\RequirePackage{fix-cm}
\documentclass{svjour3}                     
\smartqed  
\usepackage{graphicx}
\usepackage{fancyhdr,tikz}
\usepackage{epsfig,amsfonts,amsmath,amssymb}
\usepackage{enumitem}
\usepackage{longtable,mathtools}

\usepackage[font={small,it}]{caption}

\usepackage{graphicx}
\usepackage{amssymb}

\usepackage{inputenc}
\usepackage{color}
\usepackage{pdfpages}
\usepackage{url,calc,array}
\usepackage{calc,algorithmicx}
\usepackage{algorithm}
\usepackage{algpseudocode}

\usepackage{hhline}
\usepackage{arydshln}

\usepackage{multirow}

\newcommand{\dd}{\,\mathrm{d}}
\newcommand{\R}{\mathbb{R}}
\def\trans{^{\mathsf{T}}}

\newcommand{\subfigimage}[3][,]{%
	\setbox1=\hbox{\includegraphics[#1]{#3}}
	\leavevmode\rlap{\usebox1}
	\rlap{\hspace*{5pt}\raisebox{\dimexpr\ht1-2\baselineskip}{#2}}
	\phantom{\usebox1}
}

\usepackage{placeins}
\usepackage{float}
\floatstyle{plain} 

 \graphicspath{{./Figs/}}
 \renewcommand{\arraystretch}{1.2}
\usepackage{etoolbox}

\AtBeginEnvironment{tabular}{\footnotesize}

\newcommand{\ii}{\operatorname{i}}
\def\ctrans{^{\mathsf{H}}}

\usepackage{hyperref}

\usepackage[normalem]{ulem}

\begin{document}

\title{A non-perturbative approach to computing seismic normal modes in rotating planets
}

\titlerunning{Normal Modes of a Rotating Planet}        

\author{Jia Shi   \and Ruipeng Li \and Yuanzhe Xi \and Yousef Saad
        \and Maarten V. de Hoop 
}


\institute{Jia Shi \at
              Department of Earth, Environmental and Planetary Sciences, Rice University, TX, USA.  Now Shell International Exploration and Production Inc, TX, USA.\\
              \email{jia.shi.work@gmail.com}           
           \and
           Ruipeng Li \at
              Center for Applied Scientific Computing, Lawrence Livermore National Laboratory, CA, USA.  
              \and
              Yuanzhe Xi \at 
              Department of Mathematics, Emory University, Atlanta, GA, USA.
              \and 
              Yousef Saad \at 
              Department of Computer Science and Engineering, University of Minnesota, MN, USA.
              \and 
              Maarten V. de Hoop \at
             Department of Computational and Applied Mathematics, Rice University, TX, USA.
}

\date{Received: date / Accepted: date}

\maketitle

\begin{abstract}
A Continuous Galerkin method based approach is
presented to compute the seismic normal modes of rotating 
planets. 
Special care is taken to separate out the essential spectrum
in the presence of a fluid outer core using a polynomial filtering eigensolver. The relevant
elastic-gravitational system of equations, including the Coriolis
force, is subjected to a mixed finite-element method, while
self-gravitation is accounted for with the fast multipole method. 
Our discretization utilizes fully unstructured tetrahedral
meshes for both solid and fluid regions. The relevant eigenvalue
problem is solved by a combination of several highly parallel and
computationally efficient methods. 
We validate our three-dimensional results in the non-rotating case using analytical results 
for constant elastic balls, as well as numerical results for an isotropic Earth model from standard ``radial" algorithms.
We also validate the computations in the rotating case, but only in the slowly-rotating regime 
where perturbation theory applies, because no other independent algorithms are available in the general case.
The algorithm and code are used to compute the point spectra of
eigenfrequencies in several Earth and Mars models studying the effects
of heterogeneity on a large range of scales.
\end{abstract}

\section*{Declarations}
This research was supported by the Simons Foundation under the
MATH+X program, the National Science Foundation grant
DMS-1815143, the members of the Geo-Mathematical Imaging Group at Rice
University, and XSEDE research allocation TG-EAR170019. 
The work by R.L. was performed under the auspices of the
U.S. Department of Energy by Lawrence Livermore National Laboratory
under Contract DE-AC52-07NA27344 (LLNL-JRNL-780818). Y.X. and
Y.S. were supported by NSF-1812695. 

The codes are made available via \href{https://github.com/js1019/NormalModes}{https://github.com/js1019/NormalModes} and 
\href{https://github.com/eigs/pEVSL}{https://github.com/eigs/pEVSL}.
The data can be reproduced using the codes in \href{https://github.com/js1019/PlanetaryModels}{https://github.com/js1019/PlanetaryModels}. 
In addition, the Mars models can be found in \cite{agwx-jd58-20}, 
where the performance and reproducibility were studied in \cite{shi2021planetary}.

\section{Introduction}

Planetary normal modes are instrumental for studying the dynamic
response to sources including earthquakes along faults and meteorite
impacts, as well as tidal forces 
\cite{dahlen1998theoretical,lognonne2005planetary}. 
The low-angular-order eigenfrequencies
contain critical information about the planet's large-scale structure
and provide constraints on heterogeneity in composition, temperature,
and anisotropy, while rotation constrains the shapes as well as possible
density distributions of planets. The effect of rotation on the
seismic point spectrum of the Earth is well understood and has been
observed for decades \cite[Fig.1]{park2005earth}.  The observation of
spectral energy of low-frequency toroidal modes in vertical seismic
recordings of the 1998 Balleny Islands earthquake
\cite{zurn2000observation}, is a manifestation of the
three-dimensional heterogeneity and anisotropy of the mantle
structures and rotation.

For a review of Earth's free oscillations, we refer to
\cite{woodhouse2007}. Current standard approaches to computing the
seismic point spectrum and associated normal modes have several
limitations. Assuming spherical symmetry for non-rotating planets, the
problem becomes one-dimensional and the computation of normal modes in
such models using \texttt{MINEOS}
\cite{woodhouse1988calculation,masters2011mineos} is still common
practice; these are then typically used in perturbation-theory and
mode-coupling approaches to include lateral heterogeneities. Full-mode
coupling methodology utilizing normal modes in a spherically symmetric
model as a basis has been adopted to studying Earth's interior for
decades  \cite{dahlen1968normal,dahlen1969normal,woodhouse1978effect,woodhouse1980coupling,park1986synthetic,park1990subspace,romanowicz1987multiplet,lognonne1990modelling,hara1991inversion,hara1993inversion,um1991normal,lognonne1991normal,deuss2001theoretical,deuss2004iteration,al2012calculation,yang2015synthetic}.
This methodology is of
Rayleigh-Ritz type, and is justified under the assumption that the
space in which the normal modes lie contains the mentioned basis,
which requires spherically symmetric fluid-solid and surface
boundaries. Here, we remove this limitation. Moreover, a separation of
the essential spectrum needs to be carefully carried out, which has
been commonly ignored in the ``radial'' algorithms. We discuss the
mode-coupling approach and the conditions under which it applies in
Appendix \ref{app:sphharmonics}.

To simulate seismic waves in strongly heterogeneous media, the
spectral-element method (SPECFEM)
\cite{komatitsch1998spectral,komatitsch1999introduction} has been
widely used for more than two decades. We mention the software package
SPECFEM3D\_globe
\cite{komatitsch2002spectral1,komatitsch2002spectral2}, which is
capable of modeling relatively high-frequency waveforms in an entire
planet while suppressing the perturbation to the gravitational potential. 
Other implementations of SPECFEM
\cite{chaljub2003solving,chaljub2004spectral,chaljub2007spectral}
have been developed with alternative numerical approaches pertaining
to the fluid outer core. In principle, seismic eigenfrequencies show up
by taking a discrete Fourier transform of numerical solutions;
however, it is a major computational challenge to control the
accuracy at very long time scales. We note that in SPECFEM3D\_globe, the
fluid displacement is replaced by a scalar potential, which results in
a non-symmetric system of discretized equations. Moreover, 
the (square of the) Brunt-V\"{a}is\"{a}l\"{a} frequency is
assumed to be zero. Rotation in the fluid regions is unnaturally introduced
by means of an additional vector 
(cf. \cite[(16) and (17)]{komatitsch2002spectral2} and
\cite[(30)]{chaljub2007spectral}). 
In addition, current SPECFEM3D\_globe does not include the incremental gravitational 
field, which limits its usage for relatively higher frequency wave propagation. 

One may view the computational approach developed in this paper as
forming a bridge between SPECFEM3D\_globe, and the mode-coupling
approaches derived from modes in a spherically symmetric model,
involving finer scale heterogeneity and higher seismic
eigenfrequencies. Our approach facilitates the studies of the highly
heterogeneous crust models and complex three-dimensional models
through the planetary spectrum, as well as the naturally efficient
computation of seismograms from many different sources. Naturally, we
also include the Coriolis force and centrifugal potential and
formulate it as a nonlinear eigenvalue problem. 
We can accommodate arbitrarily shaped fluid-solid boundaries which becomes 
increasingly important at higher rotation rates. 
In our formulation, the rotation rate might spatially
vary, which is relevant to the future computation of normal modes in
gas giants in our solar system. 

In this paper, we revisit the work of
\cite{buland1984computation}. Buland and collaborators encountered
several complications that we overcome by characterizing and
separating the essential spectrum using a polynomial filtering eigensolver and introducing a new formulation
that properly models the elastic-gravitational system without
simplifications. In our proposed formulation,  the displacement, the
proper orthonormal condition and the symmetry of the system for
non-rotating planets are preserved.  We apply fully unstructured meshes
to model fully heterogeneous planets, and the mixed finite-element
method (FEM) to discretize the elastic-gravitational system. Our
method can handle fully heterogeneous planetary models easily, and
guarantee that accurate solutions lie in the space to which normal
modes associated with the seismic point spectrum belong.  In a previous
paper \cite{DBLP:conf/sc/ShiLXSH18}, we introduced a highly parallel
algorithm for solving the generalized eigenvalue problem resulting
from our analysis for Cowling approximation using P1 mixed FEM. 
We achieved high parallel computational and memory scalabilities with demonstrated performance on modern supercomputers. 
In the following paper \cite{shi2021planetary}, we extended our algorithm using P2 mixed FEM for better accuracy 
and discussed the reproducibility of our codes reported from several universities 
during the student cluster competion at the supercomputing conference. 

Self-gravitation manifests itself in the incremental gravitational
potential as the density changes with displacement. We utilize the
Green's solution of Poisson's equation and treat the self-gravitation as an $N$-body
problem. We then apply the fast multipole method (FMM)
\cite{greengard1997new,gimbutas2011fmmlib3d,yokota2013fmm}, which
reduces the algorithmic complexity significantly, to compute both the
reference gravitational and the incremental gravitational potentials.
Alternatively, one can apply a finite-infinite element method
\cite{zienkiewicz1983novel,burnett1994three} for modeling unbounded
domain problems to approximate the far-field of Poisson's equation.
More recently, the spectral-infinite-element method
\cite{gharti2018spectral} has been developed to incorporate 
gravity. While our eigensolver \cite{DBLP:conf/sc/ShiLXSH18} only
takes matrix-vector products, any suitable schemes, including FMM or
infinite-element methods, can be used in our computational framework.

To include rotation in the elastic-gravitational system through the
Coriolis force and the centrifugal potential, in this work, we utilize extended
Lanczos vectors computed in a non-rotating planet -- with the shapes
of boundaries of a rotating planet and accounting for the centrifugal
potential -- as a truncated basis to properly facilitate reduction to
one of the equivalent linear forms of the quadratic eigenvalue problem
(QEP). Here, the separation of the essential spectrum comes into play
again and the normal modes computed are guaranteed to lie in the
appropriate space of functions. The reduced system can be solved with
a standard eigensolver.

We present and validate our three-dimensional computations using
constant elastic balls and an isotropic preliminary reference Earth
model as non-rotating planets with standard radial codes.  The
computational accuracy for rotating planets is illustrated and tested but only 
in the regime where perturbation theory applies as no other
independent algorithms are available in the general case. We use our algorithm and code to
compute the point spectra of eigenfrequencies in several Earth and
Mars models, acknowledging relatively low rotation rates, studying the
effects of heterogeneity on a large range of scales. The Mars models
are relevant to the InSight (Interior exploration using Seismic
Investigations, Geodesy and Heat Transport)
\cite{banerdt2013insight,lognonne2019seis} mission. It is expected
that a set of eigenfrequencies is observable 
\cite{panning2017planned,bissig2018detectability}. Here, we select
one Mars model \cite{khan2016single} from the set of blind tests
\cite{clinton2017preparing,van2019preparing} and combine it with the
topography \cite{zuber1992mars,smith1999global} and a
three-dimensional crust
\cite{belleguic2005constraints,goossens2017evidence} to create a
realistic Mars model. We compute the low-angular-order
eigenfrequencies and study the general effects of rotation and
heterogeneity combined.

The outline of this paper is as follows. In
Section~\ref{sec:weakform}, we revisit the form and physics of the
elastic-gravitational system of a rotating planet and establish the
weak formulation of the system with a separation of the essential
spectrum using a polynomial filtering eigensolver. 
In Section~\ref{sec:liquidcore}, we discuss the hydrostatic
equilibrium of a rotating fluid outer core in the presence of the
gravitational and the centrifugal forces. In
Section~\ref{sec:mixedFEM}, we introduce the Continuous Galerkin mixed
FEM and obtain the corresponding matrix equations. In
Section~\ref{sec:selfG}, we study the computation of the reference
gravitational field and the perturbation of the gravitational field
using the FMM. In Section~\ref{sec:exp}, we validate the computational
accuracy of our work for non-rotating Earth models and quantify the
effect on the point spectrum from three-dimensional heterogeneity. In
Section~\ref{sec:comprotation}, we illustrate the computational
accuracy of our proposed method and show several computational
experiments for different planetary models, including standard Earth
and Mars models as well as related effects due to rotation and a
three-dimensional crust. In Section~\ref{sec:conclusion}, we discuss
the significance of our results and directions of future research.

\section{The elastic-gravitational system with rotation}
\label{sec:weakform}

In this section, we present a modified elastic-gravitational system of
equations of a rotating planet to deal with the separation of the essential spectrum 
in the weak form \cite{de2015system} (see \cite{dahlen1998theoretical} for the strong
formulation).

\subsection{Natural subdomains and computational meshes}\label{subsec:geometry}

\begin{figure}[ht!]
\centering
\includegraphics[width=0.7\linewidth]{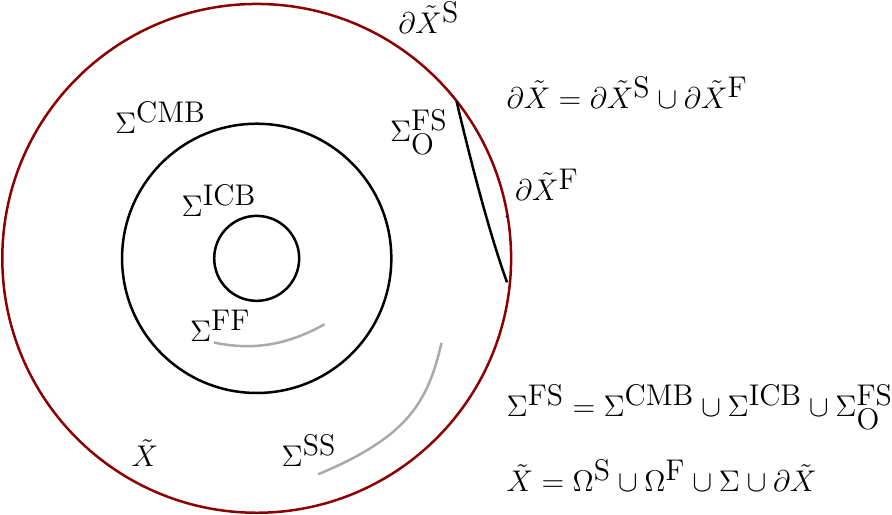}
\caption{Conceptual figure of the geometry of a planet using Earth as an example. 
The red, black and grey lines indicate the outer boundary $\partial \tilde{X}$, 
the fluid solid boundaries $\Sigma^{\text{FS}}$, and interfaces only in the solid or fluid regions. }
\label{fig:concepturalgeometry}
\end{figure}

Following the notation in \cite{de2015system}, a bounded set
$\tilde{X} \subset \mathbb{R}^3$ is used to represent the interior of
the Earth, with Lipschitz continuous exterior boundary $\partial
\tilde{X}$. The exterior boundary $\partial \tilde{X}$ contains fluid
(ocean) surfaces $\partial \tilde{X}^{\text{F}}$ and solid surfaces
$\partial \tilde{X}^{\text{S}}$. We subdivide the set
$\tilde{X}$ into solid regions $\Omega^{\text{S}}$
and fluid regions $\Omega^{\text{F}}$. The fluid regions contain the
liquid outer core $\Omega^{\text{OC}}$ and the oceans
$\Omega^{\text{O}}$. The solid regions
can be further subdivided into the crust and mantle
$\Omega^{\text{CM}}$ and the inner core $\Omega^{\text{IC}}$. We use
$\Sigma$ to represent the interfaces between these subregions. In summary,
\begin{equation*}
\tilde{X} = \,  \Omega^{\text{S}} \cup \Omega^{\text{F}} \cup \Sigma \cup \partial \tilde{X} ,\ 
\partial \tilde{X} = \partial \tilde{X}^{\text{S}}  \cup \partial \tilde{X}^{\text{F}} ,\
\Omega^{\text{S}} = \Omega^{\text{CM}} \cup \Omega^{\text{IC}} ,\
\ \Omega^{\text{F}} = \Omega^{\text{OC}} \cup \Omega^{\text{O}} . 
\end{equation*}
The interior interfaces can further be subdivided into three categories: interfaces between two fluid regions $\Sigma^{\text{FF}}$, 
interfaces between two solid regions $\Sigma^{\text{SS}}$, and interfaces between fluid and solid regions $\Sigma^{\text{FS}}$. 
We can subdivide $\Sigma^{\text{FS}}$ into two major interfaces: 
internal interfaces  $\Sigma^{\text{FS}}_{\text{int}}$ and the bottom interface $\Sigma^{\text{FS}}_{\text{O}}$ of the oceans. 
The internal interfaces include the interfaces between the lower mantle and the outer core $\Sigma^{\text{CMB}}$, 
which is known as the Core-Mantle Boundary (CMB); 
the interface between the outer core and the inner core is denoted as $\Sigma^{\text{ICB}}$, 
which is known as the Inner-Core Boundary (ICB). Thus,
\[
\Sigma = \Sigma^{\text{SS}} \cup \Sigma^{\text{FF}} \cup \Sigma^{\text{FS}} ,\  
\Sigma^{\text{FS}} = \Sigma^{\text{FS}}_{\text{int}} \cup \Sigma^{\text{FS}}_{\text{O}} ,\
\Sigma^{\text{FS}}_{\text{int}} = \Sigma^{\text{CMB}} \cup \Sigma^{\text{ICB}} . 
\]
In Fig.~\ref{fig:concepturalgeometry}, we illustrate the concepts of 
the main mathematical symbols for the geometry used in this work. 
Since a general terrestrial planet may contain multiple complex
discontinuities associated with different geological and geodynamical
features, utilization of a flexible, fully unstructured tetrahedral
mesh would be natural. We discretize the major
discontinuities using triangulated surfaces that are generated via
\texttt{distmesh} \cite{persson2004simple} and then build up the
Earth model using an unstructured tetrahedral mesh via
\texttt{TetGen} \cite{si2015tetgen}.  In Fig.~\ref{fig:mesh}, we
illustrate the interfaces and meshes with one hundred thousand and one
million elements. These techniques show great flexibility and can
provide models with multiple resolutions. 
In Figs.~\ref{fig:MIT3D}, we
illustrate a three-dimensional Earth model built on a tetrahedral mesh.  In
Fig.~\ref{fig:MIT3D} (a), we show the Moho interface that is
constructed using an unstructured triangular mesh. The color shows the
depth and the black lines are the edges of the triangles.  In
Fig.~\ref{fig:MIT3D} (b), we illustrate the three-dimensional $V_P$ model based on
MIT's mantle tomographic results \cite{burdick2017model} and crust
1.0 \cite{laske2013update}. The core model is based on the Preliminary Reference Earth Model (PREM) 
\cite{dziewonski1981preliminary}.

\begin{figure}[ht!]
	\centering
	\begin{tabular}{cc}
		\subfigimage[trim= 8cm 1cm 8cm 1cm,clip=true,width=0.40\linewidth]{(a1)}{models/Inter3L_100k-eps-converted-to.pdf} &
		\subfigimage[trim= 8cm 1cm 8cm 1cm,clip=true,width=0.40\linewidth]{(a2)}{models/mesh3L_100k-eps-converted-to.pdf}  \\
		\subfigimage[trim= 8cm 1cm 8cm 1cm,clip=true,width=0.40\linewidth]{(b1)}{models/Inter7L_1M-eps-converted-to.pdf} &
		\subfigimage[trim= 8cm 1cm 8cm 1cm,clip=true,width=0.40\linewidth]{(b2)}{models/mesh7L_1M-eps-converted-to.pdf}  
	\end{tabular}
	\captionof{figure}{Illustration of different meshes. (a1) Three triangularized surface meshes; (a2) A tetrahedral mesh with 100k elements that is generated from (a1); (b1) Seven triangularized surface meshes; (b2) A tetrahedral mesh with one-million elements that is generated from (b1). The light surfaces in (b1) and (b2) denote the CMB. }\label{fig:mesh}
\end{figure}

\begin{figure}[ht!]
\centering
\begin{tabular}{cc}
\includegraphics[trim= 4cm 0cm 0cm 0cm,clip=true, width=0.45\linewidth]{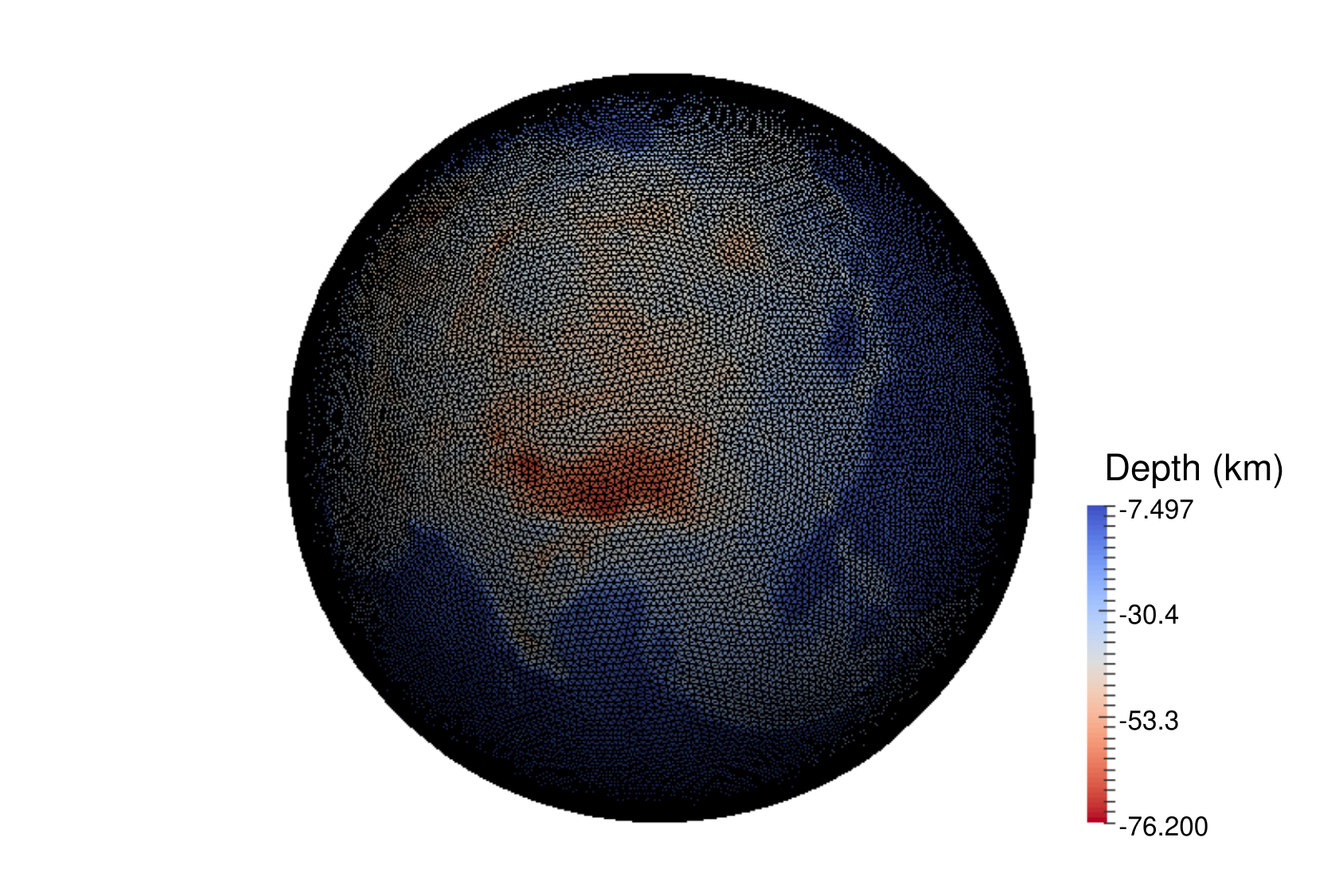} &
\includegraphics[trim= 4cm 0cm 0cm 0cm,clip=true, width=0.45\linewidth]{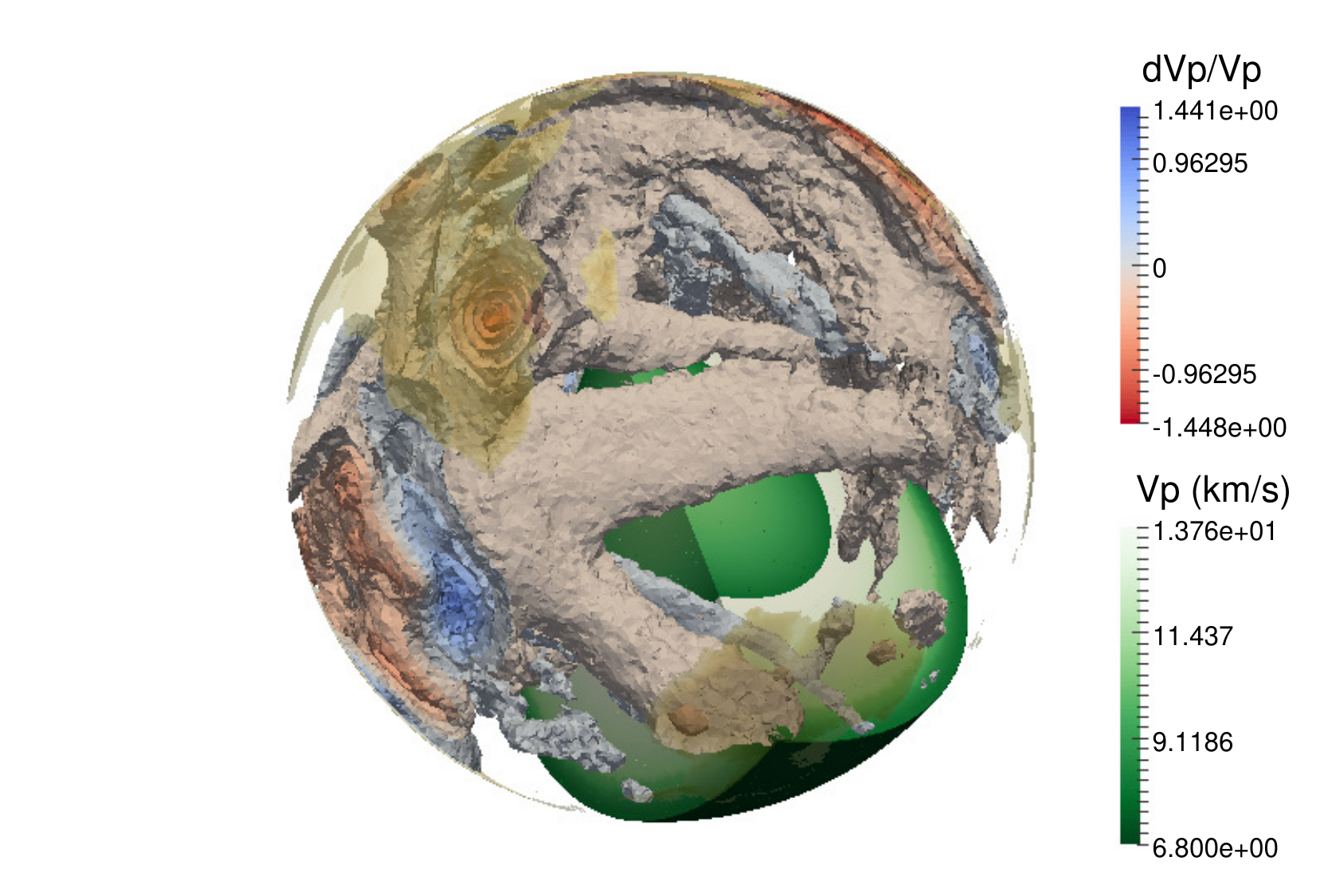} 
\\ 
(a) Moho & (b) MIT $V_P$ model
\end{tabular}
\caption{A three-dimensional Earth model built using MIT tomographic results \cite{burdick2017model} and crust 1.0 \cite{laske2013update}. (a) A triangluar mesh built for the Moho interface. The color indicates the depth below the reference surface of the Earth. The bottom of the Tibet Plateau is shown.  (b) MIT mantle $V_P$ model built on a tetrahedral mesh. The $V_P$ model and the contours of $\dd V_P/V_P$ (\%) are shown. } \label{fig:MIT3D}
\end{figure}

\begin{figure}[ht!]
\centering
\begin{tabular}{cc}
\includegraphics[width=0.45\linewidth]{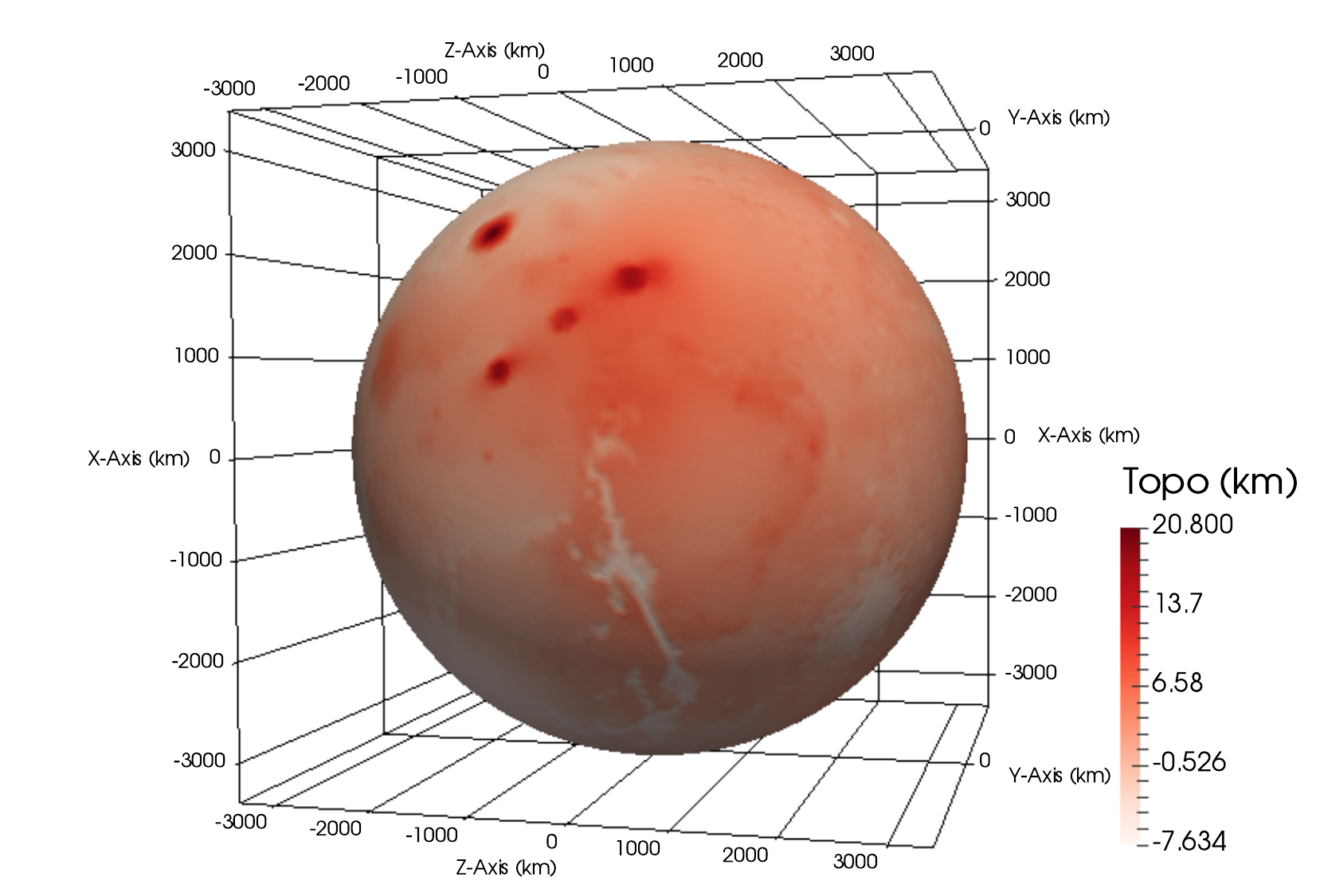}
&
\includegraphics[width=0.45\linewidth]{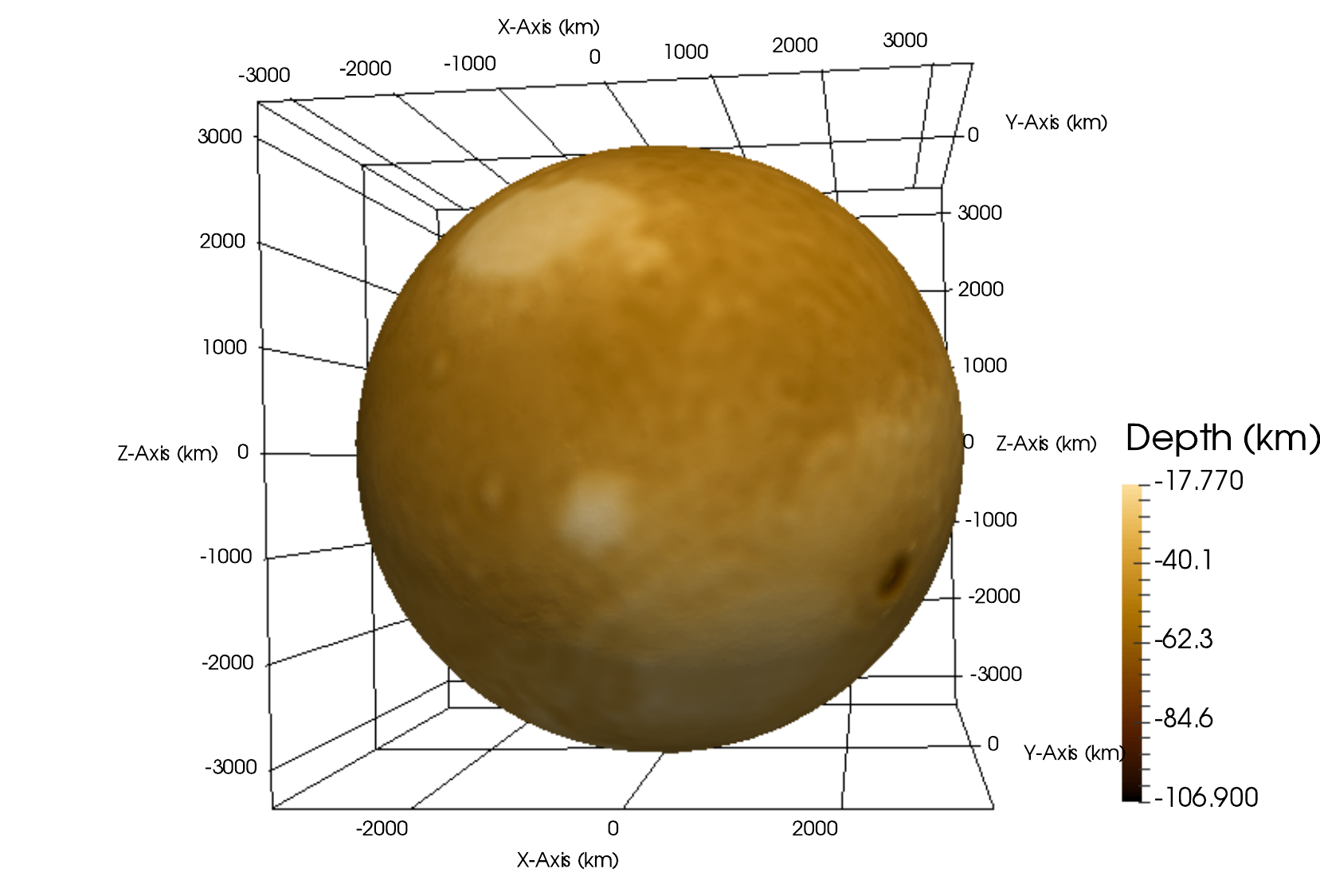}
\\
(a) Topography & (b) Crust-mantle interface
\end{tabular}
\caption{Illustration of (a) the topography and (b) the crust-mantle interface of the Mars using 
MOLA and gravity data \cite{zuber1992mars,smith1999global,goossens2017evidence}.}
\label{fig:marstopocmi}
\end{figure}

\begin{figure}[ht!]
\centering
\begin{tabular}{ccc}
\includegraphics[width=0.3\linewidth]{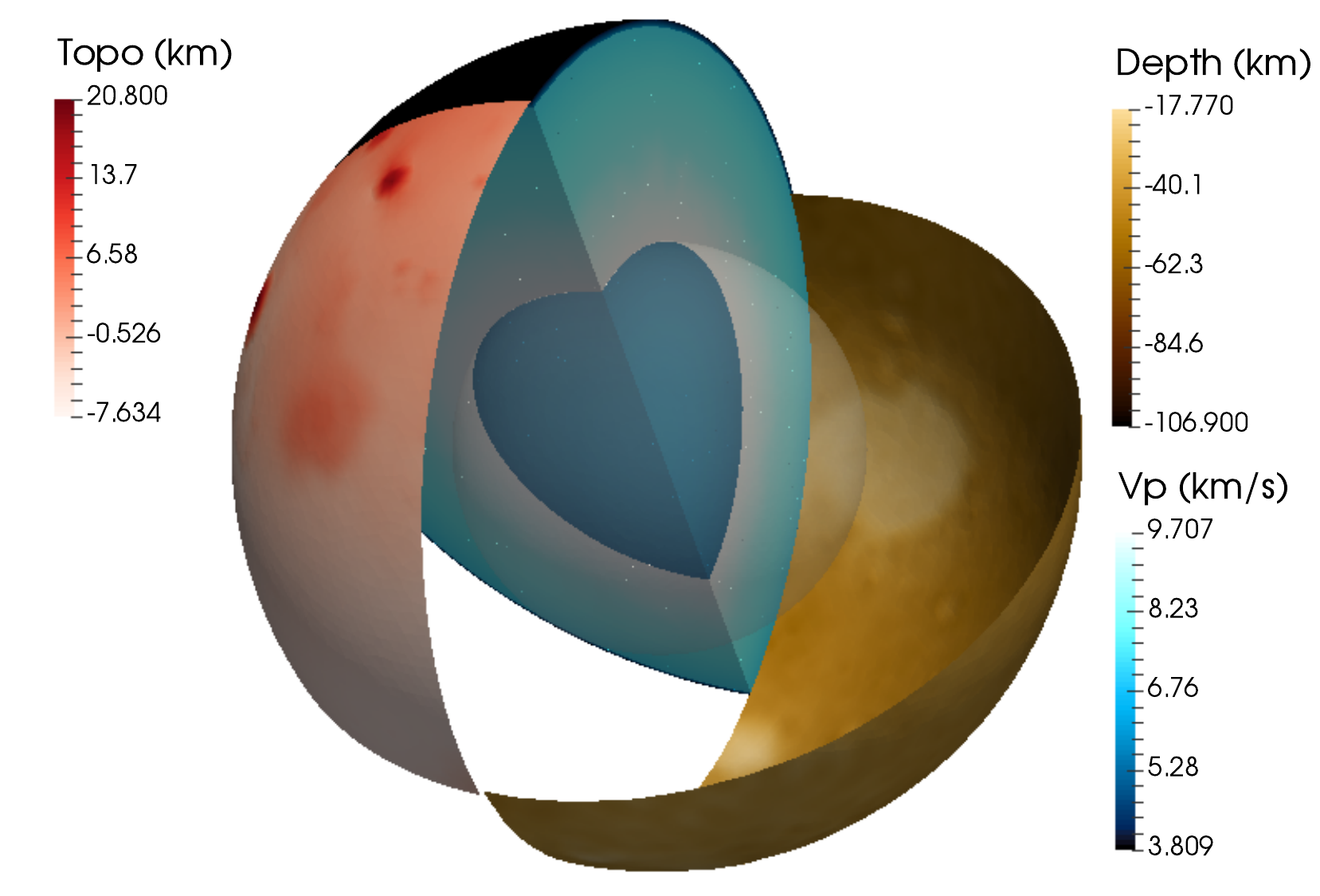} &
\includegraphics[width=0.3\linewidth]{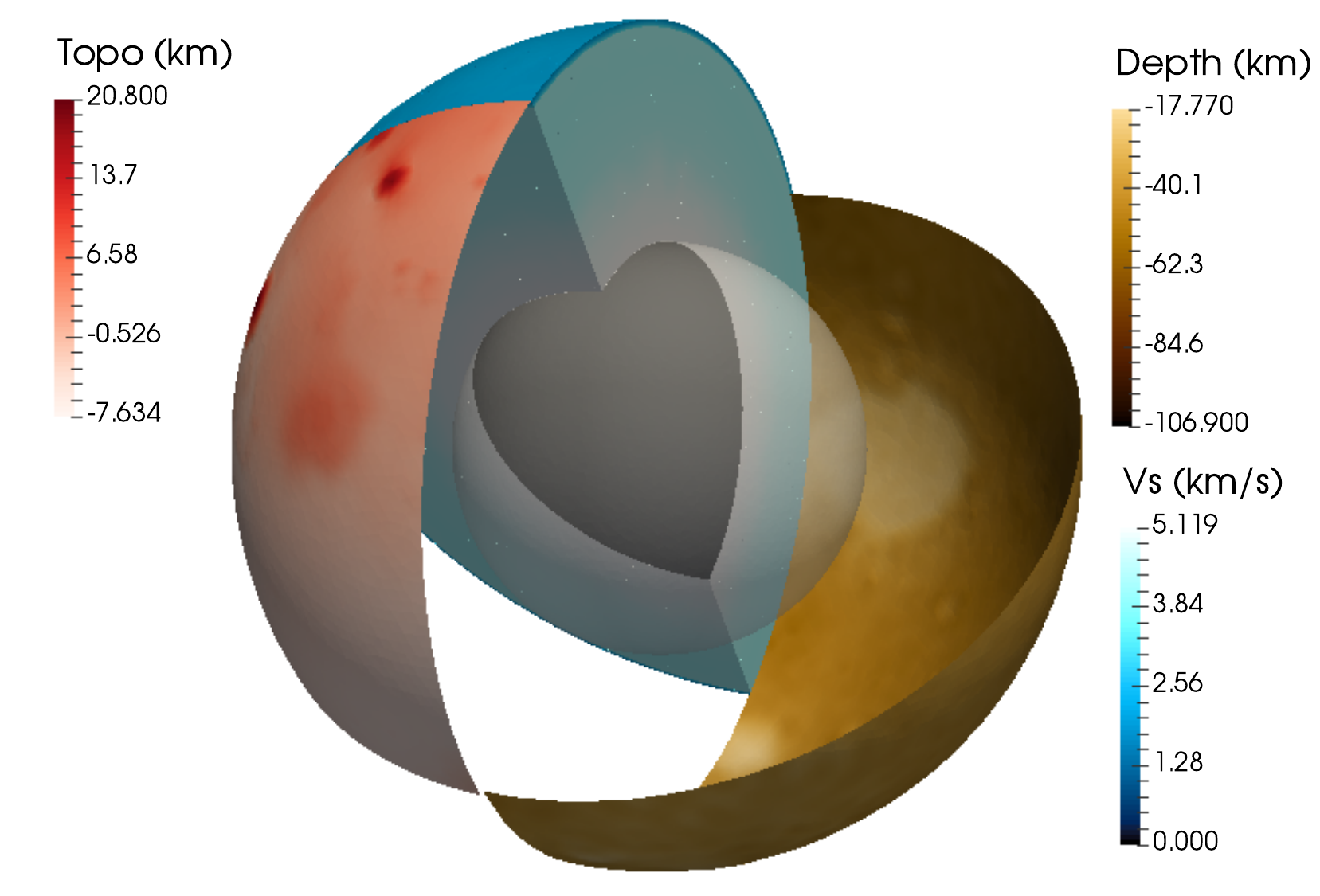} &
\includegraphics[width=0.3\linewidth]{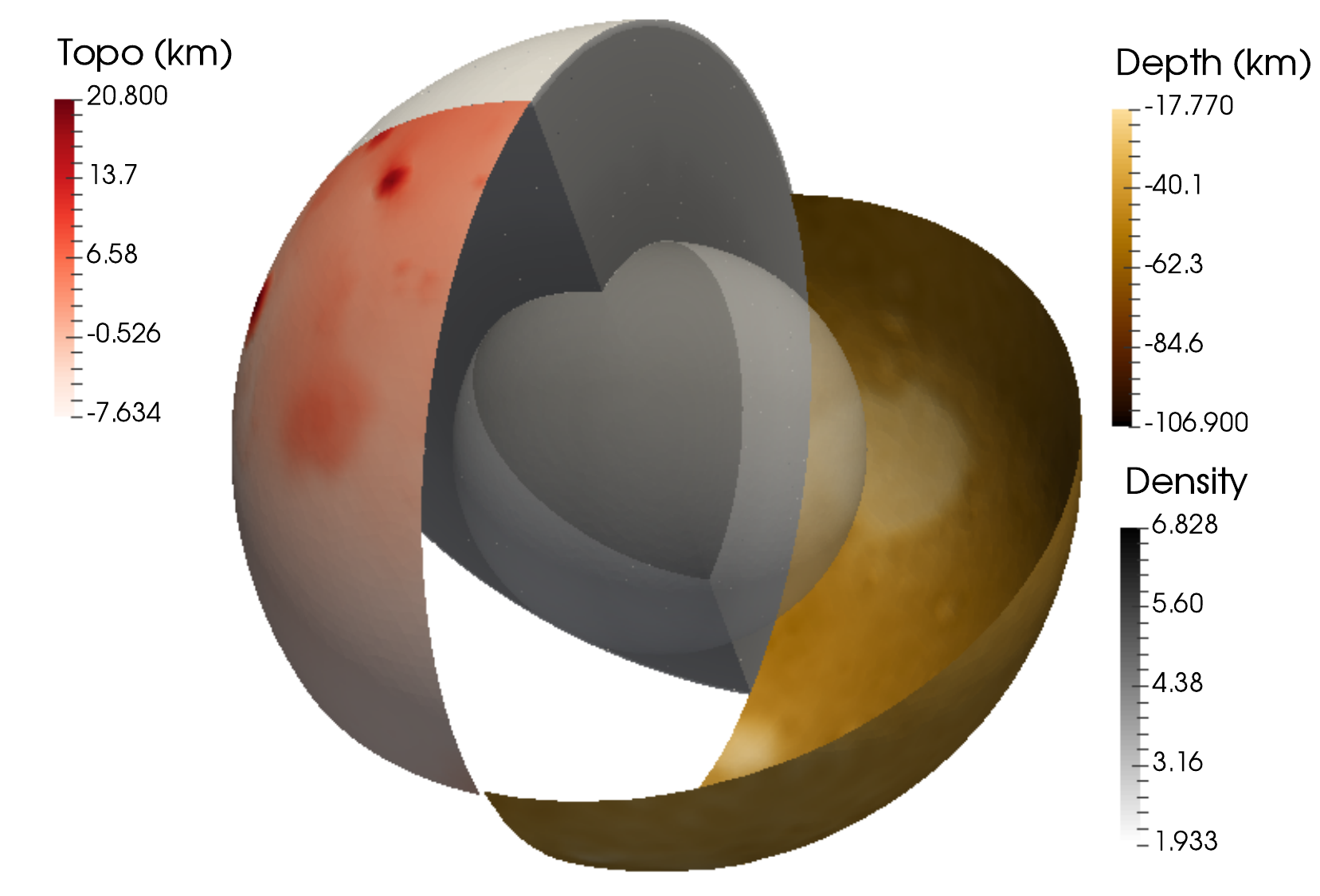} \\
(a) $V_P$ & (b) $V_S$ & (c) $\rho^0$
\end{tabular}
\caption{Illustration of (a) $V_P$, (b) $V_S$, and (c) $\rho^0$ of our
  Mars model with a three-dimensional crust shown in
  Fig.~\ref{fig:marstopocmi}.}
\label{fig:vpvsrhomars}
\end{figure}

\begin{figure}[ht!]
\centering
\begin{tabular}{cc}
\includegraphics[trim= 1cm 0cm 0cm 0cm,clip=true,width=0.45\linewidth]{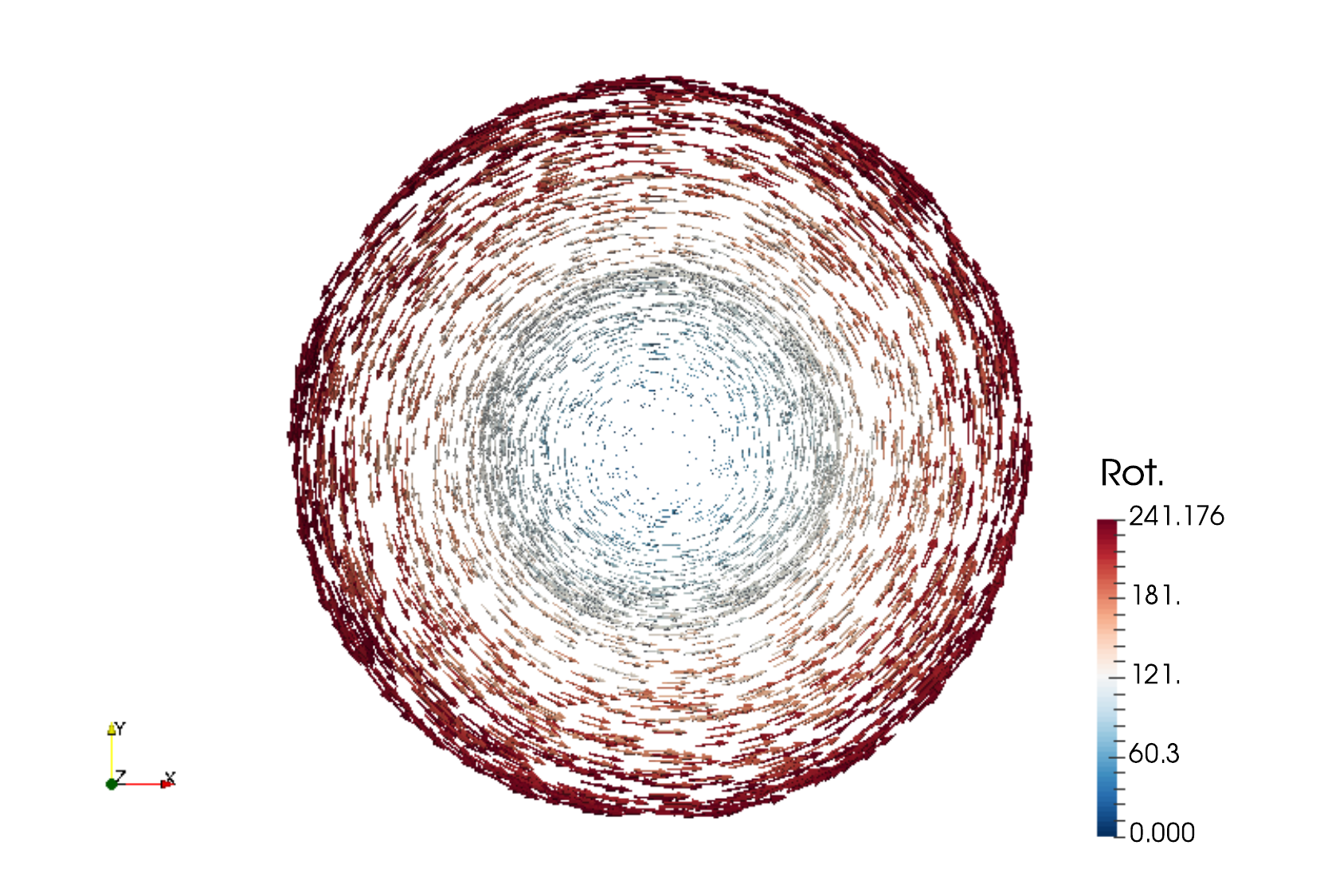} &
\includegraphics[trim= 1cm 0cm 0cm 0cm,clip=true,width=0.45\linewidth]{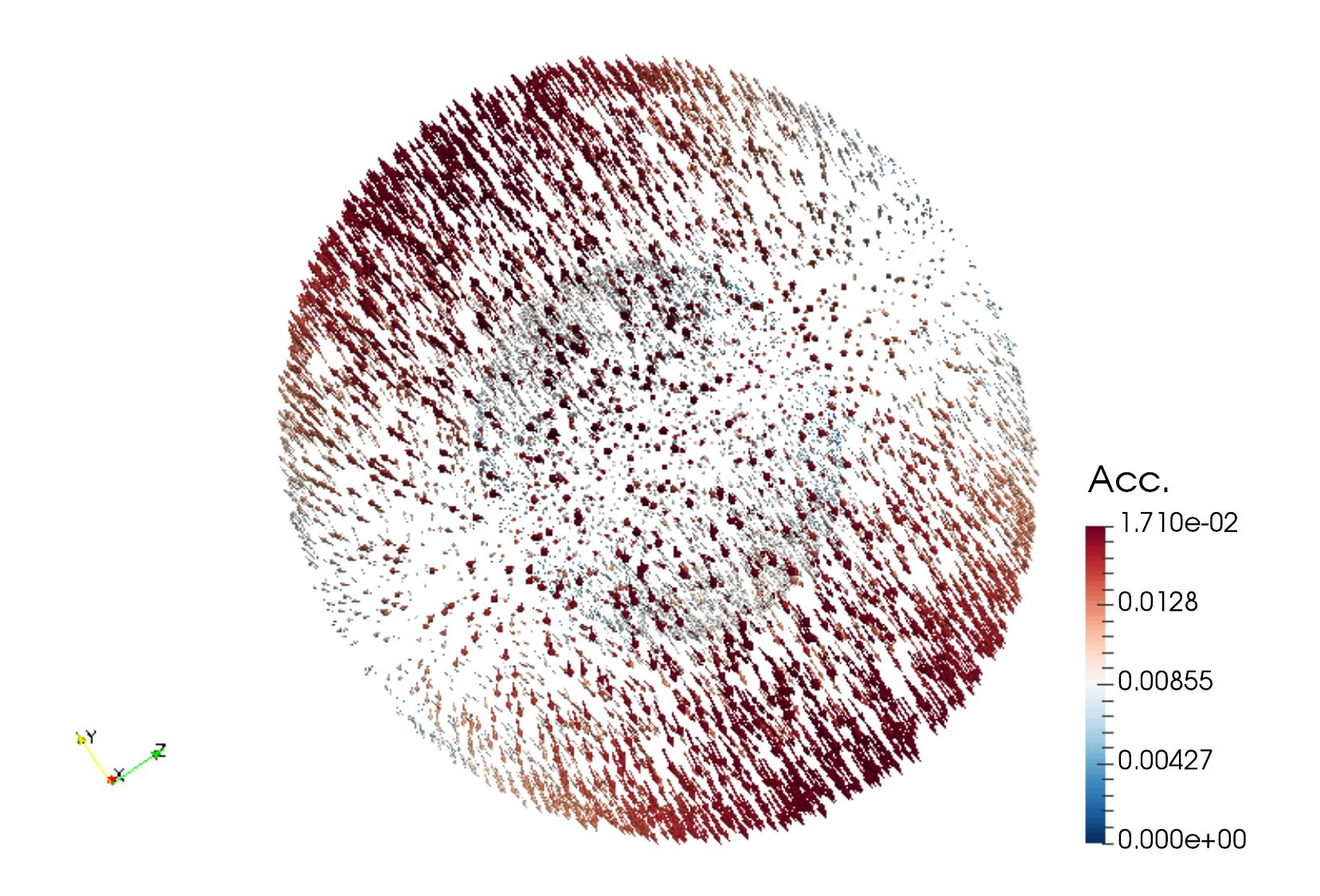} \\
(a) The axial spin mode, $\Omega \times x$ ($m/s$)  & (b) Centrifugal acceleration, $-\nabla \psi$ ($m/s^2$)
\end{tabular}
\caption{Illustration of (a) the axial spin mode, $\Omega \times x$, and (b) the centrifugal acceleration with $z$ as the rotational axis, $-\nabla \psi$, of the Mars model shown in Fig.~\ref{fig:vpvsrhomars}. } \label{fig:rotforcemars} 
\end{figure}

We also use a Mars model as an
example to illustrate our construction of a terrestrial planet. The
topography of Mars was measured by the Mars Orbiter Laser Altimeter
(MOLA) \cite{zuber1992mars,smith1999global} with high accuracy. The
thickness and density of the Martian crust were constructed with the help of the works of
\cite{belleguic2005constraints,goossens2017evidence}. In
Fig.~\ref{fig:marstopocmi} (a), we illustrate the topography of Mars
using data from MOLA \cite{smith1999global}; in
Fig.~\ref{fig:marstopocmi} (b), we show the crust-mantle interface of
Mars using data provided by \cite{goossens2017evidence}.  In
Figs.~\ref{fig:vpvsrhomars} (a)--(c), we illustrate $V_P$, $V_S$ and
$\rho^0$ of Mars integrating a radial model \cite{khan2016single}
with a three-dimensional crust as shown in
Fig.~\ref{fig:marstopocmi}. In Figs.~\ref{fig:rotforcemars} (a) and
(b), we illustrate the axial spin mode, $\Omega\times x$, and the centrifugal
acceleration, $-\nabla \psi$, of the Mars model, respectively.

\subsection{The basic equations}
\label{subsec:formulaQ}

Given the reference density $\rho^0$ and the gravitational constant
$G$, we let $\Phi^0$ denote the gravitational potential which
satisfies,
\begin{equation}
\Delta \Phi^0 = 4 \pi G \rho^0 , \label{eq:potentialQ}
\end{equation}
and $S(u)$ denote the Eulerian perturbation of the Newtonian potential
associated with the displacement $u$,  
\begin{equation}
\Delta S(u) = - 4 \pi G \nabla \cdot (\rho^0u).  \label{eq:perturbedpotentialQ}
\end{equation}
To include the centrifugal force, we introduce the centrifugal potential 
\begin{equation}\label{eq:centrifugalpotentialQ}
\psi(x) = - \frac{1}{2} \left[ \Omega^2 x^2 - \left( \Omega \cdot x\right)^2 \right] , 
\end{equation}
where $\Omega \in \R^3$ is the angular velocity of rotation. 
We form the gradient,
\begin{equation}\label{eq:definteg}
g'  = g - \nabla \psi = -\nabla ( \Phi^0 + \psi) , 
\end{equation}
where the reference gravitational field 
\begin{equation}\label{eq:referenceg}
g = - \nabla \Phi^0. 
\end{equation}
The initial stress $T^0$ satisfies the mechanical equilibrium given by the static momentum equations, 
\begin{equation}\label{eq:staticmomentum}
\nabla \cdot T^0  = - \rho^0 g' .
\end{equation}
The elastic-gravitational system of a rotating non-hydrostatic terrestrial planet has the form
\begin{equation}\label{eq:lagrangenonrotationEarth}
-\omega^2 \rho^0  u + 2 \ii \omega \rho^0 R_{\Omega} u = \nabla \cdot T^{\mathrm{L1} } - \nabla \cdot ( u \cdot \nabla T^0) 
- \rho^{\mathrm{E1}} \nabla \Phi^0 - \rho^0 \nabla S(u),
\end{equation}
where $\omega$ denotes the angular frequency; $R_{\Omega} u = \Omega \, \times \, u$; 
$\rho^{\mathrm{E1}} =  -\nabla \cdot (\rho^0 u)$ denotes 
the first-order Eulerian density perturbation and
$T^{\mathrm{L1}} = \Upsilon^{T^0} \colon \nabla u$ denotes 
the incremental Lagrangian Cauchy stress.  
The elasticity tensor, $\Upsilon^{T^0}_{ijkl}$, attains the form,
\[
\Upsilon^{T^0}_{ijkl} = c_{ijkl} + \frac{1}{2} ( - T_{ij}^0 \delta_{kl}  +T_{kl}^0 \delta_{ij}
+ T_{ik}^0 \delta_{jl} - T_{jl}^0 \delta_{ik} + T_{jk}^0 \delta_{il} - T_{il}^0 \delta_{jk}),
\] 
where $c$ denotes the elastic stiffness tensor. 
In fact, \eqref{eq:staticmomentum} does not determine the entire tensor $T^0$. 
It is common practice to invoke the hydrostatic assumption when $T_{ij}^0 = - p^0 \delta_{ij}$; then   
$\Upsilon^{T^0}_{ijkl}$ reduces to $c_{ijkl}$. 
Under the hydrostatic assumption, we reduce \eqref{eq:lagrangenonrotationEarth} into
\begin{equation}\label{eq:hydromainQ}
\omega^2 \rho^0 u - 2 \ii \omega \rho^0 R_{\Omega} u  = - \nabla \cdot (c \colon \nabla u) - \nabla (\rho^0 u \cdot g') 
+ \nabla \cdot (\rho^0 u) g' + \rho^0 \nabla S(u) .
\end{equation}
The boundary conditions for the system (\ref{eq:hydromainQ}) governing a
hydrostatic planet are summarized in Table~\ref{table:l1bcs}.
  
\begin{table}[ht!]
	\centering
	\begin{tabular}{ l @{\hskip 0.5in} r }
		\hline
		boundary types    & linearized boundary conditions  \\ \hhline{==}
		free surface, $\partial \tilde{X}$ &  $ T^0 \cdot \nu = 0; \quad \nu \cdot T^{\mathrm{L1}} = 0$ \\ \hdashline
		solid-solid interfaces $\Sigma^{\text{SS}}$ & $[\nu \cdot T^{\mathrm{L1}}]^+_- = 0; \quad [T^0 \cdot \nu]^+_- = 0; \quad [u]^+_- = 0$\\ \hdashline
		fluid-solid interfaces $\Sigma^{\text{FS}}$ & $[T^0 \cdot \nu]^+_- = 0; \quad [u\cdot \nu]^+_- = 0$  \\ 
		\& fluid-fluid interfaces $\Sigma^{\text{FF}}$ 
		& $[\nu \cdot T^{\mathrm{L1}}]^+_- = \nu [\nu \cdot T^{\mathrm{L1}} \cdot \nu]^+_- = 0$ \\ \hdashline
		all interfaces $\Sigma$ \& $\partial \tilde{X}$   &  $[S(u)]^+_- = 0; \quad [\nabla S(u)\cdot \nu + 4\pi G \rho^0 u\cdot \nu]^+_- =0$ \\ \hline  
	\end{tabular}
	\captionof{table}{Boundary conditions for a hydrostatic planet (cf. \cite[Table 3.4]{dahlen1998theoretical}).} \label{table:l1bcs}
\end{table}

\subsection{The weak formulation}\label{sec:bilinearforms}

We let $u^s$ denote displacement in the solid regions and $u^f$
denote displacement in the fluid regions.  We treat the solid and
fluid parts differently and then deal with $S(u)$ globally. We use $v$
to denote test functions and denote $v^s$ and $v^f$ for the solid and
fluid test displacements, respectively. The mass term from the first 
and the second term of \eqref{eq:hydromainQ} take the form
\begin{equation} \label{eq:authQ}
   b_{H}(u,v) = \int_{\Omega^{\text{S}}}  (\overline{v}^s \cdot u^s) \rho^0 \dd x
         + \int_{\Omega^{\text{F}}}  (\overline{v}^f \cdot u^f) \rho^0 \dd x ,
\end{equation}
and 
\begin{equation} \label{eq:rotationQ}
c_r(u,v) =  \int_{\Omega^{\text{S}}} \overline{v}^s \cdot ( \Omega \times  u^s ) \rho^0 \dd x  +  \int_{\Omega^{\text{F}}} \overline{v}^f \cdot  (\Omega \times  u^f ) \rho^0 \dd x ,
\end{equation}
respectively.  
We note that the coercivity of the original weak form of
the right-hand side of \eqref{eq:hydromainQ}, identified as
$a_{\text{original}}(u,v)$ in \cite[(3.5)]{de2015system}, is not apparent. 
 The early work by Valette \cite{valette1989spectre}, which is written in French, 
analyzed this problem in a proper mathematical space 
while the details can be found in a preprint of a book chapter \cite{de2015system}. 
In the work of \cite{de2015system}, 
it is revisited and a proper form, $a_{2}(u,v)$,
for the weak formulation is introduced. 
The coercivity of $a_{2}(u,v)$ is established in \cite[Sections 5.2 and 6]{de2015system}. 
The equivalence, that is, $a_{2}(u,v) = a_{\text{original}}(u,v)$ under
the boundary conditions (cf. \cite[Table 3.1]{dahlen1998theoretical}),
is proven in \cite[Lemma 4.1]{de2015system}.

In this work, we will study $a_2(u,v)$ under the hydrostatic assumption. 
The right hand side of \eqref{eq:hydromainQ} can be written in the form 
\begin{multline}\label{eq:a2hydro}
a_{2}(u,v) 
=\int_{\Omega^{\text{S}}}  (\nabla \overline{v}^s) \colon (c \colon \nabla u^s)   \dd x  
 + \int_{\Sigma^{\text{FS}}} \mathfrak{S}\{ (\overline{v}^s \cdot g')  (\nu^{s \rightarrow f} \cdot u^s) [\rho^0]^f \} \dd \Sigma  \\
+ \int_{\Omega^{\text{S}}} \mathfrak{S} \big\{ (\nabla \cdot \overline{v}^s)   (g' \cdot u^s)  \rho^0 
- u^s \cdot (\nabla g') \cdot \overline{v}^s  \rho^0 
- u^s \cdot (\nabla \overline{v}^s) \cdot g' \rho^0 \big\} \dd x    \\
+ \int_{\Omega^{\text{F}}} \rho^0 N^2 \frac{ (g' \cdot \overline{v}^f) (g' \cdot u^f) }{\| g' \|^2} \dd x 
+ \int_{\Sigma^{\text{FF}}} (g' \cdot \nu)  (\overline{v}^f \cdot \nu) (u^f \cdot \nu) [\rho^0]^+_- \dd \Sigma  \\
+ \int_{\Omega^{\text{F}}} \kappa (\nabla \cdot \overline{v}^f + \rho^0 \kappa^{-1} g' \cdot \overline{v}^f ) (\nabla \cdot u^f + \rho^0 \kappa^{-1} g' \cdot u^f )  \dd x \\
- \frac{1}{4\pi G}\int_{\R^3} \nabla S(\overline{v}) \cdot \nabla S(u) \dd x, 
\end{multline}
where $N^2 = (\nabla \rho^0 / \rho^0 - g'
\rho^0/\kappa )\cdot g'$ signifies the square of the Brunt-V\"{a}is\"{a}l\"{a}
frequency; $\nu^{s \rightarrow f}$ denotes the normal vector at
the fluid-solid boundary pointing from the solid to the fluid side;
the symmetrization operation $\mathfrak{S}$ 
is defined as $\mathfrak{S}\{L(u,\overline{v})\} := \frac{1}{2}
(L(u,\overline{v}) + L(\overline{v},u))$, for any bilinear form
$L(u,v)$. 
The first
integral over $\Omega^{\text{F}}$ is responsible for the inertial or
gravity modes, and the second integral over $\Omega^{\text{F}}$ yields the
acoustic modes. The integral over $\Sigma^{\text{FF}}$ generates
Kelvin modes that occur at boundaries with density jumps. 
To solve the basic equation \eqref{eq:hydromainQ}, we combine \eqref{eq:authQ} with
\eqref{eq:a2hydro} and obtain the system
\begin{equation}\label{eq:original}
a_2(u,v) = \omega^2 b_H(u,v) - \ii \omega c_r(u,v). 
\end{equation}
However, it is computationally infeasible to obtain the accurate
normal modes from the direct discretization of \eqref{eq:original} due
to the existence of spurious oscillations \cite{kiefling1976fluid}.
We discuss various approaches in Subsection~\ref{sec:reviewform} and
note that the solution needs to be restricted to the space associated with
the seismic point spectrum. 
In Subsections~\ref{sec:restrictionH1}, \ref{sec:weakfluid}, \ref{sec:weaksolid} and \ref{sec:weakformsu}, 
we present our scheme to deal with the fluid-solid and fluid surface boundary conditions, 
fluid regions, solid regions and perturbation of the gravitational potential and field, respectively.  
In the Subsection~\ref{subsec:spaceh1h2}, we introduce the mathematical
spaces associated with the seismic point and essential spectra and their separation
using a polynomial filtering eigensolver.

\subsubsection{Choice of physical variables for fluid regions without rotation}
\label{sec:reviewform}
To study planetary normal modes, we include the linear elasticity,
compressible fluids, and the fluid-solid and free-surface boundary
conditions. Discretization of the standard formulation leads to
computational difficulties, since the non-seismic modes from the
compressible fluid may pollute the computation of the point spectrum. In
this paper, we use a displacement-pressure formulation and 
later substitute the pressure term using an equivalent formula.  

Here, we review different approaches pertaining to the above-mentioned
separation of the essential spectrum for non-rotating bodies and then include the rotation.
The natural displacement formulation for a non-rotating body 
will result in a symmetric eigenvalue problem. 
However, the drawback is the existence of spurious oscillations
\cite{kiefling1976fluid}.  Several finite-element methods have been
developed for modeling the fluid regions with fluid-solid interaction:
a displacement formulation \cite{hamdi1978displacement}, a pressure
formulation \cite{zienkiewicz1969coupled,craggs1971transient}, a
displacement-pressure formulation \cite{wang1997displacement}, and a
velocity potential formulation
\cite{everstine1981symmetric,olson1985analysis}.  However, the
pressure formulation leads to a non-symmetric eigenvalue problem
\cite{zienkiewicz1969coupled,craggs1971transient}, and the velocity
potential formulation \cite{everstine1981symmetric,olson1985analysis}
leads to a quadratic eigenvalue problem.

In the engineering community, several approaches have been designed to
resolve this issue. A penalty method \cite{hamdi1978displacement} has
been applied by imposing an irrotational constraint.  However, the
study by \cite{olson1983study} has shown that this penalty method has
issues dealing with a solid vibrating in the fluid cavity, which is
the case in this paper. A four-node element with a reduced integration
using a mass matrix projection technique \cite{chen1990vibration} has
been designed to eliminate the spurious modes. A method using
different elements for solid and fluid regions was proposed for
two-dimensional \cite{bermudez1994finite} and three-dimensional
cases \cite{bermudez1999finite} when non-physical spurious modes
appear \cite{bermudez1995finite}.  The displacement/pressure
formulation \cite{wang1997displacement} has been developed via
introducing mixed elements; still, the fluid-solid coupling needs
additional consideration \cite{bermudez1994finite,bermudez1999finite}.

Compared with the above-mentioned engineering problems, we encounter a
more complicated system (\ref{eq:hydromainQ}) with different boundary
conditions (cf. Table~\ref{table:l1bcs}).  Due to the presence of the
reference gravitational field and the incremental gravitational field,
the essential spectrum of the elastic-gravitational system is more
complicated than the one of the elastic systems with fluid structures
in the engineering problems. In the geophysical community, the
pressure formulation
\cite{komatitsch2002spectral1,komatitsch2002spectral2,nissen20082}
has been commonly used, which is based on replacing the displacement
by a scalar potential in the fluid regions. It results in
non-symmetric stiffness and mass matrices for a non-rotating body. An
alternative approach
\cite{chaljub2003solving,chaljub2004spectral,chaljub2007spectral},
using several additional variables to represent the fluid
displacement, also leads to a non-symmetric system. To preserve the
necessary symmetry and guarantee the correct orthonormality condition
for the eigenfunctions or normal modes, we note that the fluid
displacement must be kept in the formulation.

\subsubsection{Fluid-solid and fluid surface boundary conditions}\label{sec:restrictionH1}

In this work, to deal with fluid-solid and fluid surface boundary conditions
we applied a similar approach \cite{wang1997displacement} with no any penalty terms
by augmenting the system of equations (cf.~\eqref{eq:a2hydro}) and introducing an
additional variable, $p$, according to
\begin{equation} \label{eq:fluidconstraint}
   -p \kappa^{-1} = \nabla \cdot u^f + \rho^0 \kappa^{-1} g' \cdot u^f
   \ \text{in}\ \Omega^{\text{F}} .
\end{equation}
Here, $\kappa$ signifies \textit{the compressibility of the fluid}. 
Imposing the fluid-solid boundary condition
$\left.\left[\nu^{f \rightarrow s} \cdot u^f - \nu^{f \rightarrow s}
  \cdot u^s \right]\right\vert_{\Sigma^{\text{FS}}} = 0$ naturally with the introduction of the 
  additional variable $p$, we obtain
the weak form for (\ref{eq:fluidconstraint}),
\begin{align} \nonumber
   0 &= -\int_{\Omega^{\text{F}}}  \overline{v}^p  p \kappa^{-1}\dd x
       + \int_{\Omega^{\text{F}}} \Big[ (\nabla \overline{v}^p) \cdot u^f
              -  \overline{v}^p (g' \cdot u^f) \rho^0 \kappa^{-1} \Big] \dd x \\
      & - \int_{\Sigma^{\text{FS}}}\overline{v}^p 
           (\nu^{f \rightarrow s} \cdot u^s) \dd \Sigma - \int_{\partial \tilde{X}^{\text{F}}}\overline{v}^p 
           (\nu \cdot u^f) \dd \Sigma, 
\label{eq:weakfluidconstraint0}
\end{align}
for all the test functions $v^p$, 
where $\nu^{f\rightarrow s}$ denotes the normal vector 
at the fluid-solid boundary pointing from the fluid
to the solid side. 
Due to the hydrostatic equilibrium, we note that $\nu|_{\partial \tilde{X}^{\text{F}}}$ is parallel to $g'$. 
Using the boundary condition, 
\begin{equation}\label{eq:pbc}
[\nu \cdot (\kappa \nabla \cdot u^f) ]|_{\partial \tilde{X}^{\text{F}}} = 0, 
\end{equation}
we have the relation
\begin{equation}\label{eq:bctrans}
(\nu \cdot u^f)|_{\partial \tilde{X}^{\text{F}}} = - \|g'\|^{-1} (g' \cdot u^f)|_{\partial \tilde{X}^{\text{F}}}
=  (\rho^0\|g'\|)^{-1} p |_{\partial \tilde{X}^{\text{F}}}. 
\end{equation}
We using \eqref{eq:bctrans} to rewrite \eqref{eq:weakfluidconstraint0}
\begin{multline}
  0 = -\int_{\Omega^{\text{F}}}  \overline{v}^p  p \kappa^{-1}\dd x
       + \int_{\Omega^{\text{F}}} \Big[ (\nabla \overline{v}^p) \cdot u^f
              -  \overline{v}^p (g' \cdot u^f) \rho^0 \kappa^{-1} \Big] \dd x \\
        - \int_{\Sigma^{\text{FS}}}\overline{v}^p 
           (\nu^{f \rightarrow s} \cdot u^s) \dd \Sigma 
       - \int_{\partial \tilde{X}^{\text{F}}}  (\rho^0\|g'\|)^{-1} \overline{v}^p p \dd \Sigma 
            =: c_g([u,p],v^p) . \label{eq:weakfluidconstraint}
\end{multline}
A short-hand notation $c_g([u,p],v^p)$ in \eqref{eq:weakfluidconstraint} 
is introduced for simplification.
In this work, since we only consider planets with a solid surface, 
the integral over $\partial \tilde{X}^{\text{F}}$ will be omitted. 
But it will be needed while including the oceans, or dealing with gas giants, such as Saturn or Jupiter. 

\subsubsection{Fluid regions}\label{sec:weakfluid}

We use (\ref{eq:fluidconstraint}) in \eqref{eq:a2hydro} and obtain
\begin{multline}\label{eq:weakfluid1}
   \int_{\Omega^{\text{F}}} \kappa (\nabla \cdot \overline{v}^f
      + \rho^0 \kappa^{-1} g' \cdot \overline{v}^f )(\nabla \cdot u^f
      + \rho^0 \kappa^{-1} g' \cdot u^f )  \dd x
\\
   = \int_{\Omega^{\text{F}}} [ (\overline{v}^f \cdot \nabla p) 
     - (\overline{v}^f \cdot g') p \rho^0 \kappa^{-1} ] \dd x
     - \int_{\Sigma^{\text{FS}}} (\overline{v}^f \cdot \nu^{f \rightarrow s}) p
            \dd \Sigma .
\end{multline}
Since 
\begin{equation}\label{eq:fsbcpressure}
   -\int_{\Sigma^{\text{FS}}} (\overline{v}^f \cdot  \nu^{f \rightarrow s}) p
            \dd \Sigma
   = \int_{\Sigma^{\text{FS}}} (\overline{v}^s \cdot \nu^{s \rightarrow f}) p
                  \dd \Sigma , 
\end{equation}
we include the right-hand side of \eqref{eq:fsbcpressure} in the
contributions from the solid regions. Thus, we obtain the contributions
to $a_2(u,v)$ in \eqref{eq:a2hydro} from the fluid regions,
\begin{multline} \label{eq:weakfluid2}
   a^f_2([u,p],v) = \int_{\Omega^{\text{F}}}
        \rho^0 N^2 \frac{(g' \cdot \overline{v}^f) (g' \cdot u^f) }{\| g' \|^2} \dd x
   + \int_{\Omega^{\text{F}}} \overline{v}^f \cdot  (\nabla p
       -  g' p    \rho^0 \kappa^{-1}) \dd x \\
       + \int_{\Sigma^{\text{FF}}} (g'\cdot \nu) (\overline{v}^f \cdot \nu) (u^f \cdot \nu)  [\rho^0]^+_- \dd \Sigma 
        .
\end{multline}

\subsubsection{Solid regions}\label{sec:weaksolid}

For the solid regions, we add the right-hand side of
\eqref{eq:fsbcpressure} to the terms related to the solid regions in
\eqref{eq:a2hydro} and obtain
\begin{multline} \label{eq:weaksolidparts}
   a^s_2(u,v)  = \int_{\Omega^{\text{S}}}
    (\nabla \overline{v}^s) \colon (c \colon \nabla u^s) \dd x  
 \\
   + \int_{\Omega^{\text{S}}} \mathfrak{S}\{
     (\nabla \cdot \overline{v}^s)  (g' \cdot u^s) \rho^0 
           -  u^s \cdot (\nabla g') \cdot \overline{v}^s \rho^0
           -  u^s \cdot (\nabla \overline{v}^s) \cdot g' \rho^0\} \dd x \\
                + \int_{\Sigma^{\text{FS}}} \mathfrak{S}\{
      (\overline{v}^s \cdot g')  (\nu^{s \rightarrow f} \cdot u^s) [\rho^0]^f  \}
                \dd \Sigma
   + \int_{\Sigma^{\text{FS}}}   (\overline{v}^s \cdot \nu^{s\rightarrow f}) p
                               \dd \Sigma . 
\end{multline}

\subsubsection{Perturbation of the gravitational potential and field}\label{sec:weakformsu}

Here, we discuss the contribution of the perturbation of the
gravitational potential $S(u)$. 
Since the test functions are divided into test functions on solid and
fluid regions, we have
\begin{multline}
a_G(u,v) = -\frac{1}{4\pi G}\int_{\R^3} \nabla S(\overline{v}) \cdot \nabla S(u) \dd x   = 
 \\ 
 - \int_{\Omega^{\text{S}}} \nabla \cdot (\rho^0 \overline{v}^s)  S(u) \dd x  
 - \int_{ \Sigma^{\text{SS}} \cup \partial \tilde{X}^{\text{S}} }  
(\nu \cdot \overline{v}^s) S(u) [\rho^0]^+_- \dd \Sigma   
   \\
- \int_{\Omega^{\text{F}}} \nabla \cdot  (\rho^0 \overline{v}^f)  S(u) \dd x 
- \int_{ \Sigma^{\text{FF}}  \cup \partial \tilde{X}^{\text{F}} }  
(\nu \cdot \overline{v}^f) S(u) [\rho^0]^+_- \dd \Sigma   \\
- \int_{\Sigma^{\text{FS}}} \left\{  (\nu^{f\rightarrow s} \cdot \overline{v}^s) S(u) [\rho^0]^s + (\nu^{s\rightarrow f} \cdot \overline{v}^f) S(u) [\rho^0]^f \right\} \dd \Sigma  ,
\label{eq:weakgravperturb1} 
\end{multline}
where $[\rho^0]^s$ denotes the solid density along the fluid-solid
boundary.  One can set up $S(u)$ as an independent variable and apply
the finite-infinite element method to approximate
\eqref{eq:perturbedpotentialQ}, but here we follow a different
approach.

Making use of Green's function \cite[Chapter 3, (3.98)]{dahlen1998theoretical}, we have
\begin{equation}\label{eq:solSu}
   S(u) = G\int_{\tilde{X}} \frac{\nabla' \cdot ( \rho^0(x') u(x'))}{\| x - x' \|} \, \dd x' + G \int_{\Sigma \cup \partial \tilde{X}} \frac{\nu(x') \cdot u(x') [\rho^0(x')]^+_-}{\| x - x' \|} \, \dd \Sigma' .
\end{equation}
Again, we separate the displacement $u$ into $u^s$ and $u^f$, 
and rewrite (\ref{eq:solSu}) as 
\begin{multline} 
S(u) = G \bigg\{ \int_{\Omega^{\text{S}}} \frac{\nabla' \cdot ( \rho^0(x') u^s(x'))}{\| x - x' \|} \, \dd x' 
+ \int_{\Omega^{\text{F}}} \frac{\nabla' \cdot ( \rho^0(x') u^f(x'))}{\| x - x' \|} \, \dd x'
\\ 
+ \int_{ \Sigma^{\text{SS}} \cup \partial \tilde{X}^{\text{S}} } \frac{\nu(x') \cdot u^s(x') [\rho^0(x')]^+_-}{\| x - x' \|} \, \dd \Sigma'  
 +  \int_{\Sigma^{\text{FF}} \cup \partial \tilde{X}^{\text{F}}  } \frac{\nu(x') \cdot u^f(x') [\rho^0(x')]^+_-}{\| x - x' \|} \, \dd \Sigma'  \\
+ \int_{\Sigma^{\text{FS}} } \frac{
[\rho^0(x')]^s \nu^{f\rightarrow s}(x') \cdot u^s(x') + [\rho^0(x')]^f  \nu^{s\rightarrow f}(x') \cdot u^f(x')}{\| x - x' \|} \, \dd \Sigma' 
 \bigg\}.
\label{eq:solSu1}
\end{multline}
Although we impose $\nu^{s\rightarrow f} \cdot u^f = \nu^{s\rightarrow
  f} \cdot u^s$ along the fluid-solid boundaries, we keep the
construction of the incremental gravitational potential $S(u)$ as
described in (\ref{eq:solSu1}). This is to preserve the symmetry of
the bilinear form as we substitute (\ref{eq:solSu1}) into
(\ref{eq:weakgravperturb1}).

Since the Green's solution is known, we apply the FMM to evaluate
$S(u)$ for a given displacement $u$ via (\ref{eq:solSu1}). The
utilization of this approach is computationally attractive, but
requires that the eigensolver can solve for the interior eigenpairs
via matrix-vector multiplications.

\subsubsection{Summary}\label{sec:summaryweakform}

To restrict the system to the computational domain, 
we can rewrite \eqref{eq:a2hydro} as
\begin{equation}\label{eq:a_sfinalQ}
a_{2}([u,p],v)  = a_2^s(u,v) + a_2^f([u,p],v) + a_G(u,v). 
\end{equation}
We obtain the complete formula for the rotating hydrostatic planetary model
\eqref{eq:authQ}, \eqref{eq:rotationQ}, \eqref{eq:a_sfinalQ} and
\eqref{eq:weakfluidconstraint}:
\begin{equation}\label{eq:finalfullQ}
\begin{cases}
a_{2}([u,p,S(u)],v) &= \omega^2 b_{H}(u,v) - 2  \ii \omega c_r(u,v), \\
c_g([u,p],v^p) & = 0. 
\end{cases}
\end{equation}
A matrix representation can be derived from \eqref{eq:finalfullQ}.  In
practice, we replace $p$ in $a_2$ by $p(u^f,u^s_{\Sigma^{\text{FS}}})$
via solving the constraint $c_g([u,p],v^p) = 0$ in
\eqref{eq:weakfluidconstraint} and obtain
\begin{equation}\label{eq:finalfullQ2}
a_{2}([u,p(u^f,u^s_{\Sigma^{\text{FS}}}),S(u)],v) = \omega^2 b_{H}(u,v) - 2  \ii \omega c_r(u,v). 
\end{equation}
The corresponding orthonormality condition is that, for an eigenpair $(\omega_{(i)}, u^{(i)})$, any other eigenpair $(\omega_{(j)}, u^{(j)})$ satisfies 
\begin{equation}\label{eq:orthonormalcond}
b_H(u^{(i)}, u^{(j)}) - 2 \ii (\omega_{(i)}+ \omega_{(j)})^{-1}  c_r(u^{(i)},u^{(j)}) = \delta_{ij}, 
\end{equation}
which is consistent with \cite[(4.82)]{dahlen1998theoretical}.


\subsection{Hilbert space for the elastic-gravitational system}
\label{subsec:spaceh1h2}
We introduce the space for the displacement field \cite[Definition 5.4]{de2015system}
\begin{equation}\label{eq:spaceE}
E = 
\left\{
u \in L^2(\tilde{X}, \rho^0 \dd x) : 
\begin{cases}
& u^s = u|_{\Omega^{\text{S}}} \in H^1(\Omega^{\text{S}}) \\
& u^f = u|_{\Omega^{\text{F}}}  \in H(\text{Div}, \Omega^{\text{F}}, L^2(\partial \Omega^{\text{F}})) \\
& [u \cdot \nu]_-^+ = 0, \, \text{along} \, \Sigma^{\text{FS}} 
\end{cases}
\right\} ,
\end{equation}
where 
\[
H(\text{Div}, \Omega^{\text{F}}, L^2(\partial \Omega^{\text{F}})) = \{u^f \in L^2(\Omega^{\text{F}}) : \nabla \cdot u^f \in L^2(\Omega^{\text{F}}) , \, u|_{\partial \Omega^{\text{F}}} \cdot \nu \in L^2(\partial \Omega^{\text{F}})  \} . 
\]
$L^2(\tilde{X}, \rho^0 \dd x)$ denotes a weighted $L^2$ Hilbert space 
with 
\begin{align*}
L^2(\tilde{X}, \rho^0 \dd x) & :=\left\{ u: \int_{\tilde{X}} |u|^2 \rho^0 \dd x < \infty \right\}; \\
\langle u, v \rangle_{L^2(\tilde{X}, \rho^0 \dd x)} & := \int_{\tilde{X}} (u \cdot v) \rho^0 \dd x. 
\end{align*}
We write $H = L^2(\tilde{X}, \rho^0 \dd x)$ subject to the constraint
$\int_{\tilde{X}} u \rho^0 \dd x = 0$ removing rigid-body
translations; $E$ is densely embedded in $H$ \cite{de2019note}.

To describe the essential spectrum, we introduce operator $T$
in \cite[Section 4]{valette1989spectre} and \cite{de2019note}, 
\begin{equation}\label{eq:Toperator}
   T u^f = \rho^0 [\nabla \cdot u^f +\rho^0 \kappa^{-1} g' \cdot u^f] .
\end{equation}
The adjoint, $T^*$, of $T$ is given by 
\begin{equation}\label{eq:Tstar}
   T^* \varphi = - \frac{1}{\rho^0} \nabla (\rho^0 \varphi)
             + \rho^0 \kappa^{-1} g' \varphi ,
\end{equation}
where $\varphi$ has the interpretation of potential. A subspace,
$H_2$, of $H$ associated with the essential spectrum is defined by the
constraints
\[
   u^s = 0 ,\ T u^f = 0\text{ and }
             u^f \cdot \nu = 0 \text{ on } \Sigma^{\text{FF}}
     \cup \Sigma^{\text{FS}} \cup \partial\tilde{X}^{\text{F}} .
\]
In fact, $u^f$ can be decomposed according to $\text{Ran}(T^*) \oplus
\text{Ker}(T)$, following the decomposition
\begin{equation}\label{eq:h1h2space}
   H = H_1 \oplus H_2 ,
\end{equation} 
where spaces $H_1$ and $H_2$ are associated with the point 
and essential spectrum, respectively. 
The space $H_2$ is designed precisely to extract, via projections, the “subseismic”
approximations to the full system of governing equations for a contained rotating, 
compressible, inhomogeneous, self-gravitating fluid. 
The rigid boundary condition, $u|_{\Omega^{\text{F}}} \cdot \nu=0$ 
on $\Sigma^{\text{FF}} \cup \partial \tilde{X}^{\text{F}}$, is consistent with a rigid mantle 
and rigid inner core as $u|_{\Omega^{\text{S}}}=0$.

In fact, $\forall u \in H_2$, we obtain $p = 0$ and for Cowling approximation, we have
\begin{equation}\label{eq:insertH2Cowling}
a_2^s(u,v) + a_2^f(u,v) = \int_{\Omega^{\text{F}}}
        \rho^0 N^2 \frac{(g' \cdot \overline{v}^f) (g' \cdot u^f) }{\| g' \|^2} \dd x ,
\end{equation}
where $a_2^s$ and $a_2^f$ are defined in \eqref{eq:weaksolidparts} and \eqref{eq:weakfluid2}, respectively. 
For the incremental gravitational potential in \eqref{eq:perturbedpotentialQ}, we have 
\begin{equation}\label{eq:insertH2Su}
     \Delta S_{H_2}(u) = - 4 \pi G \nabla \cdot (\rho^0 u^f)   =  - 4 \pi G \left[ \rho^0 N^2 \frac{(g' \cdot u^f) }{\| g' \|^2} \right]. 
\end{equation}
Combining \eqref{eq:insertH2Cowling} and \eqref{eq:insertH2Su}, we note that \eqref{eq:finalfullQ2} will be reduced to 
\begin{multline}\label{eq:insertH2}
 \int_{\Omega^{\text{F}}}
        \rho^0 N^2 \frac{(g' \cdot \overline{v}^f) (g' \cdot u^f) }{\| g' \|^2} \dd x  
        -\frac{1}{4\pi G}\int_{\R^3} \nabla S_{H_2}(\overline{v}) \cdot \nabla S_{H_2}(u) \dd x \\
        = \omega^2 b_{H}(u,v) - 2  \ii \omega c_r(u,v). 
\end{multline}
Thus, restricting $\forall u \in H_2$,  the associated spectrum of \eqref{eq:insertH2} will essentially depend on to $N^2$ and the rotating rates.

In this work, we solve for the eigenvalues and eigenfunctions of \eqref{eq:finalfullQ2} inside a target frequency interval $[f_1,f_2]$, where 
\begin{equation}\label{eq:freq}
f_2 > f_1 \gg |\Omega| + \left[ |\Omega|^2 + \max \left(0, N^2_{\sup} \right) \right]^{1/2}, 
\end{equation}
where $N^2_{\sup}$ denotes the supremum of the square of the Brunt-V\"{a}is\"{a}l\"{a} frequency. 
We note that inequality \eqref{eq:freq} holds true for most planets 
because the minimal seismic normal mode frequency is typically 
much larger than the upper bound of the associated spectrum of \eqref{eq:insertH2}, 
which is the right hand side of \eqref{eq:freq}. 
For instance, the maximum of the Brunt-V\"{a}is\"{a}l\"{a} frequency of the Earth is around 
50 $\mu$Hz and $|\Omega|$ is 7.3 $\mu$Hz while the minimal seismic normal mode frequency is around 0.3 mHz.
A well-designed polynomial filter  applied with the eigensolver, 
will have the effect of  boosting up the eigenvalues inside the interval
$[f_1,f_2]$ while lessening the rest of the spectrum, including the part
 associated with $H_2$. 
\begin{remark}
  It is important to understand the need for
  \emph{polynomial filtering} in this context. First note that eigensolvers like
  ARPACK \cite{lehoucq1998arpack} or subspace iteration, e.g.,
  \cite{saad2011numerical}, compute eigenvalues of a matrix on one end of the
  spectrum.  After discretizetion, the essential spectrum will give rise to a
  large number of eigenvalues near zero. Computing the (discrete) eigenvalues in
  the interval $[f_1, f_2]$ will be numerically challenging unless the small
  eigenvalues associated with the essential spectrum are eliminated.  In
  numerical linear algebra, this is termed an interior eigenvalue problem in
  that the target eigenvalues of the discretized problem are located well inside
  the spectrum.  If we use a standard package like ARPACK
  \cite{lehoucq1998arpack} we could compute these eigenvalues starting from the
  smallest ones until we reach the desired interval $[f_1, f_2]$, which would be
  prohibitive because of the large cluster near zero caused by the essential
  spectrum.  Alternatively, we could
  compute them from the largest ones down. This would
  also entail computing a large number of unwanted eigenpairs. Finally, we could
  also use a shift-and-invert strategy \cite{Parlett-book}
  within ARPACK. This requires  using a direct solver with a very large matrix and
  is impractical in our
  context due to the large memory requirement.  The advantage of polynomial
  filtering is that it eliminates the unwanted eigenvalues and allows
  the eigensolver to focus on those that are amplified, namely those in
  $[f_1, f_2]$. 
\end{remark}

\medskip\medskip

\noindent

In Section~\ref{sec:liquidcore}, we study the hydrostatic equilibrium 
of the liquid regions with rotation and derive a proper density distribution. 
In Section~\ref{sec:mixedFEM}, we introduce the mixed FEM
to construct the system without the perturbation of the gravitational field. 
In Section~\ref{sec:selfG}, we utilize FMM to compute the gravitational field and 
the perturbation of the gravitational field and 
then obtain the complete matrix formula for \eqref{eq:finalfullQ2}. 

\section{Hydrostatic equilibrium of the liquid core with rotation}
\label{sec:liquidcore}

In this section, we discuss the hydrostatic equilibrium with rotation
and how it constrains the shape of the boundaries and the density
distribution in planets. Rotating fluids have been extensively studied
\cite{greenspan1968theory,chandrasekhar2013hydrodynamic,zhang2017theory}. 
The outer core's properties have been studied
through seismic normal modes since the 1970s
\cite{gilbert1975application,dziewonski1975parametrically,dziewonski1981preliminary}, 
but also with body waves
\cite{morelli1993body,kennett1995constraints}. Much more recently,
an alternative radial outer core model has been proposed using the
parametrization of the equation of state for liquid iron alloys at
high pressures and temperatures, inferred from eigenfrequency
observations \cite{irving2018seismically}. Furthermore, we mention
models for
the core of Mars \cite{rivoldini2011geodesy,khan2016single} albeit
ignoring rotation.

To reach the hydrostatic equilibrium, the prepressure $p^0$ satisfies 
\begin{equation} \label{eq:hydroequilibrium}
   \nabla p^0 = \rho^0 g' ,
\end{equation}
where $g'$ is defined in \eqref{eq:definteg}. Well-posedness requires
that
\begin{equation} \label{eq:rhoparallelg}
   \nabla \rho^0 \parallel g' \parallel \nabla p^0
\quad\text{in}\ \Omega^{\text{F}}\quad \text{and}\quad
   g' \parallel \nu\quad \text{along}\
        \Sigma^{\text{FS}} \cup \partial \tilde{X}^{\text{F}} ;
\end{equation}
see \cite[Lemma 2.1]{de2015system} for details about the functional
properties of $\rho^0$, $p^0$ and $g'$.

The derivation of Clairaut's equation \cite{clairaut1743theorie}, and Radau approximation are
put in the context of a general scheme imposing
\eqref{eq:rhoparallelg} in \cite[Chapter 14.1]{dahlen1998theoretical}.
The bulk parameters of Earth and Mars are listed in
Table~\ref{tab:parametersEarthmars}. While the hydrostatic assumption
seems to apply to Earth with reasonable accuracy, 
the derivative of the ellipticity at $r_e$, $\dot{\epsilon}(r_e)$, of Mars appears to be negative, whence this
assumption fails to hold \cite{dollfus1972new,bills1978mars}.

\begin{table}[h!]
\centering
\begin{tabular}{ccccccc}\hline
parameters & $\Omega$ (s${}^{-1}$)& $r_e$ (km) & $g_{(r_e)}$ (m/s${}^{2}$) & $\dot{\epsilon}(r_e)$ & $\epsilon_{(r_e)}^{\text{hyd}}$ & $\epsilon_{(r_e)}^{\text{obs}}$ \\ \hhline{=======}
Earth & 7.2921$e{-5}$  & 6371.0  & 9.80  & 3.05$e{-5}>0$  & 3.34$e{-3}$  & 3.35$e{-3}$ \\ 
Mars  & 7.0882$e{-5}$ & 3389.5 & 3.71  & -8.98$e{-5}<0$ & N/A & 5.89$e{-3}$
\\ \hline
\end{tabular}
\caption{Bulk parameters of Earth and Mars; $\dot{\epsilon}(r_e)$
  denotes the derivative of $\epsilon$ at $a=r_e$, and
  $\epsilon_{(r_e)}^{\text{hyd}}$ and $\epsilon_{(r_e)}^{\text{obs}}$
  denote the computed hydrostatic ellipticity and observed
  ellipticity, respectively.} \label{tab:parametersEarthmars}
\end{table}

To construct models of liquid planet interiors, such as Jupiter and Saturn, equations of state 
and theory of figures are commonly used for calculating a self-consistent shape and gravity field \cite{jeans1919problems}. 
We refer to \cite{militzer2016understanding} for a review on modelling Jupiter's interior using equations of state 
and multiple mission data.  
Since Radau assumptions break down for fast rotating plants \cite[Fig.3]{wahltidal}, 
we refer to \cite{hubbard2013concentric,militzer2019models} for constructing Saturn's interior using 
the concentric Maclaurin spheroid method to match the Cassini measurements. 
The condition \eqref{eq:rhoparallelg} is satisfied along with other conditions. 
\section{The Continuous Galerkin mixed finite-element method}
\label{sec:mixedFEM}

In this section, we employ the Continuous Galerkin mixed
FEM \cite{zienkiewicz2005finite,bathe2006finite,hughes2012finite,brezzi2012mixed,ern2013theory},
for discretizing our system without the perturbation of the gravitational field. 
We thus obtain a matrix representation for the corresponding weak forms. 
The incremental gravitational potential will
be introduced in the discretization in Subsection~\ref{sec:Su}.

\subsection{The Continuous Galerkin mixed finite-element approximation}

Given a shape regular finite-element partitioning $\mathcal{T}_h$ of
the domain $\tilde{X}$, we denote an element of the mesh by $K_k \in
\mathcal{T}_h$ and a boundary element by $E_l \subset \partial K_k$
and have
\begin{equation*}
   \tilde{X} \approx \bigcup_{k=1}^{N_K} K_k ,\quad
   \Sigma \cup \partial \tilde{X} \approx \bigcup_{l=1}^{N_E} E_l
          \subseteq \bigcup_{k=1}^{N_K} \partial K_k ,
\end{equation*}
where $N_K$ denotes the total number of volume elements and $N_E$
denotes the total number of interior and exterior boundary elements.
Furthermore, we let $K_k^{\text{S}}$ and $K_k^{\text{F}}$ be elements
in the solid and fluid regions, respectively. Similarly,
$E_l^{\text{S}}$, $E_l^{\text{F}}$ and $E_l^{\text{FS}}$ denote
boundary elements on the solid $\Sigma^{\text{SS}} \cup \partial
\tilde{X}^\text{S}$, fluid $\Sigma^{\text{FF}} \cup \partial
\tilde{X}^\text{F}$ and fluid-solid $\Sigma^{\text{FS}}$
discontinuities, respectively. We have
\begin{align*}
\Omega^{\text{S}} &\approx \bigcup_{k=1}^{N_K^{\text{S}}} K_k^{\text{S}} ,
\quad
\Omega^{\text{F}} \approx \bigcup_{k=1}^{N_K^{\text{F}}} K_k^{\text{F}} ,
\\
\Sigma^{\text{FS}} & \approx \bigcup_{l=1}^{N_E^{\text{FS}}} E_l^{\text{FS}} ,
\quad 
\Sigma^{\text{SS}} \cup \partial \tilde{X}^\text{S}
\approx \bigcup_{l=1}^{N_E^{\text{S}}} E_l^{\text{S}} ,
\quad 
\Sigma^{\text{FF}} \cup \partial \tilde{X}^\text{F}
\approx \bigcup_{l=1}^{N_E^{\text{F}}} E_l^{\text{F}}
\end{align*}
with 
\[
   N_K = N_K^{\text{S}} + N_K^{\text{F}} ,\quad
   N_E = N_E^{\text{S}} + N_E^{\text{F}} + N_E^{\text{FS}} ,
\]
where $N_K^{\text{S}}$ and $N_K^{\text{F}}$ denote the total number of
volume elements in the solid and fluid regions, respectively, and
$N_E^{\text{S}}$, $N_E^{\text{F}}$ and $N_E^{\text{FS}}$ denote the
total number of boundary elements on the (interior/exterior) solid,
fluid and fluid-solid boundaries, respectively. In the above, $h$
signifies the maximum value of diameters of all the elements.

Since we separate out the fluid and solid regions, we divide the
finite-element partitioning accordingly into
\[
   \mathcal{T}_h = \mathcal{T}_h^{\text{S}}
            + \mathcal{T}_h^{\text{F}} ,\quad
   \Sigma^{\text{FS}}_h = \mathcal{T}_h^{\text{S}}
              \cap \mathcal{T}_h^{\text{F}} ,
\]
where $\mathcal{T}_h^{\text{S}}$, $\mathcal{T}_h^{\text{F}}$ and
$\Sigma^{\text{FS}}_h$ denote the partitioning of the domains
$\Omega^{\text{S}}$, $\Omega^{\text{F}}$ and boundary $\Sigma^{\text{FS}}$,
respectively.
We then introduce $E_h$ as the finite-element space corresponding with the displacement space 
$E$ in \eqref{eq:spaceE},
\begin{equation}\label{eq:spaceEh}
   E_h = \left\{ u_h : 
   \begin{cases}
   & u^s_h \in \mathbf{V}_h^s := \left\{ v^s_h
     \in H^1(\Omega^{\text{S}}) : v^s_h|_K \in \mathcal{P}_{p^s}(K) ,\
     K \in \mathcal{T}_h^{\text{S}} \right\} ,
\\
   & u^f_h \in \mathbf{V}_h^f := \Big\{ v^f_h
     \in H(\text{Div}, \Omega^{\text{F}},
           L^2(\partial \Omega^{\text{F}}) : \\
           & \quad \quad \quad \quad \quad \quad  \quad \quad \quad \quad \quad \quad 
           v^f_h|_K \in \mathcal{P}_{p^f}(K) ,\
     K \in \mathcal{T}_h^{\text{F}} \Big\} ,
\\
   & \displaystyle \int_{E^{{\text{FS}}}} [u_h \cdot \nu]_-^+ v^p_h \dd \Sigma = 0  \
     \text{for all}\ E^{{\text{FS}}} \subset \Sigma^{\text{FS}}_h ,
   \end{cases}
   \right\}
\end{equation}
and $\mathbf{V}_h^p$ as the finite-element space for $p$,
\[
   \mathbf{V}_h^p := \left\{ v^p_h \in H^1(\Omega^{\text{F}}) :
   v^p_h|_K \in \mathcal{P}_{p^p}(K) ,\
   K \in \mathcal{T}_h^{\text{F}} \right\} .
\]
Here, $\mathcal{P}_{p^s}(K)$ and $\mathcal{P}_{p^f}(K)$ are the spaces
of polynomials of degrees $p^s$ and $p^f$, respectively;
$\mathcal{P}_{p^p}(K)$ is the space of polynomials of degree $p^p$. 
Though the $u_h^f$ is discretized as $u_h^f \in H^1(\Omega^{\text{F}})$, 
the constraint equation \eqref{eq:fluidconstraint} restricts $u_h^f \in H(\text{Div}, \Omega^{\text{F}},
L^2(\partial \Omega^{\text{F}})$.
By the Galerkin method, the finite-element solutions, $u_h$, and the test
functions, $v_h$, both lie in $E_h$ and $\mathbf{V}_h^p$. We note that
the polynomial degree $p^p$ does not need to be equal to $p^f$.

We apply non-conforming finite elements across the fluid-solid
boundaries. The fluid-solid transmission condition in the definition
of $E$ has been replaced by the condition $\int_{E^{{\text{FS}}}} [u_h
  \cdot \nu]_-^+ v^p_h \dd \Sigma = 0$ in the definition of $E_h$. 
The fluid-solid transmission condition holds in the form of a boundary
integration. 
For low-degree polyomials we show, in the next subsection, that these
conditions are compatible through our formulation. Such a
compatibility was analyzed and discussed by 
\cite{bermudez1994finite,bermudez1995finite,brezzi2012mixed}. 
Several numerical studies
\cite{kiefling1976fluid,zienkiewicz1978fluid,olson1985analysis,chen1990vibration,bermudez1999finite} have been performed using
similar non-conforming schemes along the fluid-solid boundaries. For
the general theory and analysis of the mixed FEM, we refer to
\cite{brezzi2012mixed}.

\subsection{Matrix formulae}\label{sec:matrixform}

\begin{table}[!ht]
	\setlength{\tabcolsep}{3pt}
	\centering
	\begin{tabular}{ c  c  r}
		\hline
		operations  & physical meanings   & corresponding formulae \\ \hhline{===}
		&   & \(\displaystyle \int_{ \Omega^{\text{S}}} \nabla \overline{v}^s_h  : (c: \nabla u^s_h ) \dd x \) \\
		&  & \(\displaystyle + \int_{\Sigma^{\text{FS}}} \mathfrak{S} \Big\{  (\overline{v}^s_h \cdot g' )  (\nu^{s \rightarrow f} \cdot u^s_h) [\rho^0]^f  \Big\} \dd \Sigma \) \\
		\(\displaystyle (\tilde{v}^s)\ctrans A_{sg} \tilde{u}^s \)   & solid stiffness matrix with gravity  
		& \(\displaystyle  + \int_{\Omega^{\text{S}}}  \mathfrak{S} \Big\{  (\nabla \cdot \overline{v}^s_h ) ( g' \cdot u^s_h ) \rho^0 \) \\
		&        & \(\displaystyle   -  u^s_h \cdot (\nabla g') \cdot \overline{v}^s_h \rho^0
		-  u^s_h \cdot (\nabla \overline{v}^s_h) \cdot g' \rho^0 \Big\} \dd x \) \\  \hdashline
	& &
		 \(\displaystyle \int_{\Omega^{\text{F}}} \rho^0 N^2 \frac{(g' \cdot \overline{v}^f_h) (g' \cdot u^f_h) }{\| g' \|^2} \dd x \) \\      
		 \(\displaystyle (\tilde{v}^f)\ctrans A_f \tilde{u}^f \)  & buoyancy term
		& \(\displaystyle + \int_{\Sigma^{\text{FF}}} (g'\cdot \nu) (\overline{v}^f_h \cdot \nu) (u^f_h \cdot \nu)  [\rho^0]^+_- \dd \Sigma   \) \\    \hdashline  
		\(\displaystyle (\tilde{v}^p)\ctrans A_p \tilde{p} \) & fluid potential  &   \(\displaystyle - \int_{\Omega^{\text{F}}}  \overline{v}^p_h p_h \kappa^{-1}  \dd x  - \int_{\Sigma^{\text{FS}}}\overline{v}^p (\nu^{f \rightarrow s} \cdot u^s) \dd \Sigma  \) \\  
		\(\displaystyle (\tilde{v}^f)\ctrans A_{\text{dg}} \tilde{p} \) & fluid stiffness matrix with gravity  &   \(\displaystyle 
		 \int_{\Omega^{\text{F}}} \left[ \overline{v}^f_h \cdot (\nabla p_h)   -  (\overline{v}^f_h  \cdot g') p_h  \rho^0 \kappa^{-1} \right]  \dd x \) \\ 
		\(\displaystyle (\tilde{v}^p)\ctrans A_{\text{dg}}\trans \tilde{u}^f \) & constraint with gravity &   \(\displaystyle 
		 \int_{\Omega^{\text{F}}} \Big[ (\nabla \overline{v}^p_h ) \cdot u^f_h - \overline{v}^p_h (g' \cdot u^f_h)  \rho^0 \kappa^{-1} \Big] \dd x \) \\  
		\(\displaystyle (\tilde{v}^s)\ctrans E_{\text{FS}}  \tilde{p} \) & fluid-solid boundary condition & 
		\(\displaystyle \int_{\Sigma^{\text{FS}}} (\overline{v}^s_h \cdot \nu^{s \rightarrow f} )  p_h \dd \Sigma \) \\ 
		\(\displaystyle (\tilde{v}^p)\ctrans E_{\text{FS}}\trans \tilde{u}^f \) & fluid-solid boundary condition   & 
		\(\displaystyle 
		\int_{\Sigma^{\text{FS}}}  -\overline{v}^p_h  ( \nu^{f \rightarrow s} \cdot u^s_h ) \dd \Sigma\) \\ 
		\hdashline
		  \(\displaystyle (\tilde{v}^s)\ctrans R_s \tilde{u}^s\) & Coriolis force in $\Omega^{\text{S}}$ & \( \displaystyle  \int_{\Omega^{\text{S}}} \overline{v}^s_h \cdot  \big( \Omega \times u^s_h \big) \rho^0 \dd x  \) \\
		  \(\displaystyle (\tilde{v}^f)\ctrans R_f \tilde{u}^f\) & Coriolis force in $\Omega^{\text{F}}$ & \( \displaystyle  \int_{\Omega^{\text{F}}} \overline{v}^f_h \cdot \big(\Omega \times u^f_h \big) \rho^0  \dd x  \) \\
		  \hdashline
		\(\displaystyle (\tilde{v}^s)\ctrans M_s \tilde{u}^s \)  & solid mass matrix & \(\displaystyle  \int_{\Omega^{\text{S}}}  (\overline{v}^s_h \cdot  u^s_h) \rho^0 \dd x    \) \\  
		\(\displaystyle (\tilde{v}^f)\ctrans M_f \tilde{u}^f \)  & fluid mass matrix & \(\displaystyle   \int_{\Omega^{\text{F}}}   (\overline{v}^f_h \cdot u^f_h) \rho^0 \dd x    \) \\ 
			\hline    
\end{tabular}
\captionof{table}{Implicit definition of the matrices. 
  In the above, $\int_{\Omega^{\text{S}}} = \sum_{k=1}^{N^{\text{S}}_K} \int_{K_k^{\text{S}}}$, 
  $\int_{\Omega^{\text{F}}} = \sum_{k=1}^{N^{\text{F}}_K} \int_{K_k^{\text{F}}}$ and 
  $\int_{\Sigma^{\text{FS}}} = \sum_{l=1}^{N^{\text{FS}}_E} \int_{E_l^{\text{FS}}}$.}
\label{table:matrixcomponentsquadratic}
\end{table}

We introduce nodal-based Lagrange
polynomials, $\{ \ell^s_i \}$, $\{ \ell^f_i \}$, $\{ \ell^p_i\}$, on
the respective volume elements $K \in \mathcal{T}^{\text{S}}_h$,
$\mathcal{T}^{\text{F}}_h$. 
We set $N_{p^s} =(p^s+1) (p^s+2) (p^s+3)/6$, where $N_{p^s}$ is the number of nodes on
a tetrahedron for the $p^s$-th order polynomial approximation. We have
similar expressions for $N_{p^f}$ and $N_{p^p}$. 
We write
\begin{align}
   (u_h^s)_j(x) &= \sum_{i=1}^{N_{p^s}} (u_h^s)_j(x_i) \ell^s_i(x) ,
\\
   (u_h^f)_j(x) &= \sum_{i=1}^{N_{p^f}} (u_h^f)_j(x_i) \ell^f_i(x) ,
\\
   p_h(x) &= \sum_{i=1}^{N_{p^p}} p(x_i) \ell^p_i(x), 
\end{align}
for $x \in K$; similar representations hold for $v^s_h$, $v^f_h$,
$v^p_h$, respectively. We collect the values of $u^s_h$, $u^f_h$,
$p_h$ and $v^s_h$, $v^f_h$, $v^p_h$ at all the nodes, $\{x_i\}$, in
the vectors $\tilde{u}^s$, $\tilde{u}^f$, $\tilde{p}$ and
$\tilde{v}^s$, $\tilde{v}^f$, $\tilde{v}^p$, respectively.  We can
then construct the corresponding submatrices, $A_{sg}$, $A_f$, $A_p$,
$A_{\text{dg}}$, $A_{\text{dg}}\trans$, $E_{\text{FS}}$,
$E_{\text{FS}}\trans$, $R_s$, $R_f$, $M_s$ and $M_f$, see
Table~\ref{table:matrixcomponentsquadratic}, in a standard way
summarized in Appendix \ref{sec:localmatrices}.

\section{Self-gravitation as an N-body problem}\label{sec:selfG}

Self-gravitation can be treated as the solution of an N-body
problem. We discretize the entire planet into many elements and
consider them as individual bodies. The gravitational potential and
field are then computed through the interaction between these
bodies. We note that FMM is an ideal candidate for solving an N-body
problem.  FMM reduces the complexity of the N-body problem from
$O(N^2)$ to $O(N\log N)$ or even $O(N)$ \cite{greengard1987fast}. We
apply the FMM \cite{greengard1997new,gimbutas2011fmmlib3d} to
calculate the reference gravitational potential in
Subsection~\ref{sec:refgravity}. We employ \texttt{ExaFMM}
\cite{yokota2013fmm}, a massively parallel N-body problem solver, 
to solve for the perturbation of the gravitational potential.

\subsection{Reference gravitational potential and gravitational field}

For calculating the reference gravitational potential and field, we
need to evaluate two integrals \cite[(3.2) and (3.3)]{dahlen1998theoretical}. 
The N-body problem of gravitation
requires the evaluation of
\begin{equation}
   \Phi^0(\mathbf{x}_i) = - G \sum_{k=1}^{N_K}
        \frac{1}{\|\mathbf{x}_i - \mathbf{r}_k \|}
                 \int_{K_k} \rho_k^0 \dd x
\end{equation}
for the potential in \eqref{eq:potentialQ} and 
\begin{equation}
   g(\mathbf{x}_i) = - G \sum_{k=1}^{N_K}
       \frac{\mathbf{x}_i - \mathbf{r}_k}{
         \|\mathbf{x}_i - \mathbf{r}_k\|^{3/2}}
                 \int_{K_k} \rho_{k}^0 \dd x
\end{equation}
for the field in \eqref{eq:referenceg}. Here, $\mathbf{x}_i$ denotes the location of the target
vertex and $\mathbf{r}_k$ denotes the barycenter of element $K_k$.

\subsection{Incremental gravitational potential}
\label{sec:Su}

For calculating the incremental gravitational potential, we need to
evaluate (\ref{eq:solSu}) containing both the volume and boundary integral
terms. Given the finite-element partitioning, $\mathcal{T}_h$, we
approximate $S(u_h)$ in \eqref{eq:perturbedpotentialQ} via
\begin{multline} \label{eq:ph1body0}
S_{k_2}(u_h) =  G  \int_{K_{k_2}} \frac{\nabla \cdot (\rho_{k_2}^0(x) u_h(x)) }{\|\mathbf{r}_{k_2} - x \|} \dd x   +  \sum\limits_{\substack{k_1=1\\k_1 \neq k_2}}^{N_K} \frac{G}{\|\mathbf{r}_{k_2} - \mathbf{r}_{k_1} \|} \int_{K_{k_1}} \nabla \cdot (\rho_{k_1}^0 u_h) \dd x  \\
      +   \sum_{l_1 = 1}^{N_E} \frac{G}{\| \mathbf{r}_{k_2}  - \mathbf{r}_{l_1}\|} \int_{E_{l_1}} (\nu \cdot u_h) [\rho_{l_1}^0]^+_- \dd \Sigma
\end{multline}
and 
\begin{multline} \label{eq:ph1surface0}
S_{l_2}(u_h) = G \int_{E_{l_2}}  \frac{\nu(x) \cdot u_h(x)  [\rho_{l_2}^0(x)]^+_- }{\| \mathbf{r}_{l_2}  - x \|} \dd \Sigma 
+ \sum_{\substack{l_1 = 1\\l_1\neq l_2}}^{N_E} \frac{G}{\| \mathbf{r}_{l_2}  - \mathbf{r}_{l_1}\|} \int_{E_{l_1}} (\nu \cdot u_h) [\rho_{l_1}^0]^+_-\dd \Sigma \\
  + \sum\limits_{\substack{k_1=1}}^{N_K} \frac{G}{\|\mathbf{r}_{l_2} - \mathbf{r}_{k_1} \|} \int_{K_{k_1}} \nabla \cdot (\rho_{k_1}^0 u_h) \dd x ,
\end{multline}
where $k_1$ and $k_2$ label the elements $K_{k_1}$ and
$K_{k_2}$, $S_{k_2}(u_h)$ is the incremental gravitational potential
$S(u_h)$ at the barycenter of $K_{k_2}$, $l_1$ and $l_2$ label the
triangular elements $E_{l_1}$ and $E_{l_2}$, $\mathbf{r}_{l_1}$ and
$\mathbf{r}_{l_2}$ denote the barycenters of $E_{l_1}$ and $E_{l_2}$.  
The first terms in \eqref{eq:ph1body0} and \eqref{eq:ph1surface0} indicate the self-contribution.

Since the variation of $\nabla \cdot (\rho_{k_2}^0(x) u_h(x))$ is small on element $K_{k_2}$, 
we simplify the first term in (\ref{eq:ph1body0}) according to    
\[
	G  \int_{K_{k_2}} \frac{\nabla \cdot (\rho_{k_2}^0(x) u_h(x)) }{\|\mathbf{r}_{k_2} - x \|} \dd x  \simeq G \frac{\int_{K_{k_2}} \nabla \cdot (\rho_{k_2}^0 u_h) \dd x}{|K_{k_2}|} \int_{K_{k_2}}   \frac{1}{\|\mathbf{r}_{k_2} - x \|} \dd x  , 
\]
where $|K_{k_2}|$ denotes the volume of element $K_{k_2}$. 
We let 
\begin{equation*}
	\frac{1}{R_{k_2}}  = \frac{1}{|K_{k_2}|} \int_{K_{k_2}}   \frac{1}{\|\mathbf{r}_{k_2} - x \|} \dd x ,
\end{equation*}
and obtain
\begin{equation} \label{eq:k2local}
G  \int_{K_{k_2}} \frac{\nabla \cdot (\rho_{k_2}^0(x) u_h(x)) }{\|\mathbf{r}_{k_2} - x \|} \dd x  \simeq 
\frac{G}{R_{k_2}} \int_{K_{k_2}} \nabla \cdot (\rho_{k_2}^0 u_h) \dd x . 
\end{equation}
Similarly, we simplify the first term in \eqref{eq:ph1surface0} according to
\begin{equation}\label{eq:l2local}
	 G \int_{E_{l_2}}  \frac{\nu(x) \cdot u_h(x) [\rho_{l_2}^0(x)]^+_- }{\| \mathbf{r}_{l_2}  - x \|} \dd \Sigma \simeq \frac{G}{R_{l_2}} \int_{E_{l_2}}  (\nu \cdot u_h) [\rho_{l_1}^0]^+_- \dd \Sigma, 
\end{equation}
with
\begin{equation*}
   \frac{1}{R_{l_2}}  = \frac{1}{|E_{l_2}|} \int_{E_{l_2}}
            \frac{1}{\| \mathbf{r}_{l_2}  - x \|} \dd \Sigma ,
\end{equation*}
where $|E_{l_2}|$ denotes the area of the boundary element $E_{l_2}$. 
Note that $R_{k_2}$ in (\ref{eq:k2local}) and $R_{l_2}$ in
\eqref{eq:l2local} can be precomputed on each element and surface.
The second and third terms in \eqref{eq:ph1body0} and
\eqref{eq:ph1surface0} may be evaluated via FMM.

\begin{table}
	\setlength{\tabcolsep}{3pt}
	\centering
	\begin{tabular}{ c  c  r}
		\hline
		operations  & physical meanings   & corresponding formulae \\ \hhline{===}
		  & & \(\displaystyle \int_{\Omega^{\text{S}}} \nabla \cdot (\rho^0 u^s_h) \dd x,   \)  \\
		  & & \(\displaystyle  \int_{\Sigma^{\text{FS}}} (\nu^{f\rightarrow s} \cdot u^s_h) \left[\rho^0\right]^s \dd x ,  \)  \\
		  \(\displaystyle C_s \tilde{u}^s \)   & N bodies in $\overline{\Omega^{\text{S}}}$ & \(\displaystyle \int_{ \Sigma^{\text{SS} \cup \partial \tilde{X}^{\text{S}}} } (\nu \cdot u^s_h) \left[\rho^0\right]^+_- \dd x  \) \\ \hdashline 
		    & & \(\displaystyle \int_{\Omega^{\text{F}}} \nabla \cdot (\rho^0 u^f_h) \dd x , \)  \\
		    & & \(\displaystyle  \int_{\Sigma^{\text{FS}}} (\nu^{s\rightarrow f} \cdot u^f_h) \left[\rho^0\right]^f \dd x  ,  \)  \\ 
		  \(\displaystyle C_f \tilde{u}^f \)   & N bodies in $\overline{\Omega^{\text{F}}}$ & \(\displaystyle  \int_{\Sigma^{\text{FF} \cup \partial \tilde{X}^{\text{F}}} } (\nu \cdot u^f_h) \left[\rho^0\right]^+_- \dd x   \) \\ \hdashline 
		   & & \(\displaystyle  G\int_{\tilde{X}} \frac{\nabla' \cdot ( \rho^0(x') u_h(x'))}{\| x - x' \|} \, \dd x'  \) \\
		   \(\displaystyle S (C\tilde{u}) \)   & solution for Poisson's equation  & \(\displaystyle  + G \int_{\Sigma \cup \partial \tilde{X}} \frac{\nu(x') \cdot u_h(x') [\rho^0(x')]^+_-}{\| x - x' \|} \, \dd x'  \) \\ \hdashline 
		  &   & \(\displaystyle  \int_{\Omega^{\text{S}}} \nabla \cdot (\rho^0 \overline{v}^s_h)  S(u_h) \dd x   \) \\
		  & incremental gravitational field & \(\displaystyle  +  \int_{\Sigma^{\text{FS}}} (\overline{v}^s_h \cdot \nu^{f\rightarrow s}) S(u_h) [\rho^0]^s \dd x   \) \\
		  \(\displaystyle (\tilde{v}^s)\ctrans C_s\trans (S C\tilde{u}) \)  & in $\overline{\Omega^{\text{S}}}$  & \(\displaystyle  +   \int_{ \Sigma^{\text{SS}} \cup \partial \tilde{X}^{\text{S}} }  (\overline{v}^s_h \cdot \nu ) S(u_h) [\rho^0]^+_-  \dd x  \)  \\ \hdashline 
		   & & \(\displaystyle    \int_{\Omega^{\text{F}}} \nabla \cdot (\rho^0 \overline{v}^f_h) S(u_h) \dd x  \) \\
		   & incremental gravitational field & \(\displaystyle + \int_{\Sigma^{\text{FS}}} (\overline{v}^f_h \cdot \nu^{s\rightarrow f} ) S(u_h) [\rho^0]^f  \dd x  \) \\
		  \(\displaystyle (\tilde{v}^s)\ctrans C_f\trans (S C \tilde{u}) \)  &  in $\overline{\Omega^{\text{F}}}$ &  \(\displaystyle   + \int_{\Sigma^{\text{FF}} \cup \partial \tilde{X}^{\text{F}}  }  (\overline{v}^f_h \cdot \nu ) S(u_h) [\rho^0]^+_- \dd x \)  \\ 
		\hline    
	\end{tabular}
     \captionof{table}{Implicit definition of the submatrices for perturbation to the gravitational potential.}
      \label{table:matrixcomponentsselfG}
\end{table}

\subsubsection{Solid planets} 

For solid planets, we substitute (\ref{eq:k2local}) and
(\ref{eq:l2local}) into (\ref{eq:ph1body0}) and
(\ref{eq:ph1surface0}), respectively. To evaluate
\eqref{eq:weakgravperturb1} for a solid planet, we need to compute
\begin{multline}\label{eq:phi1solid}
	a_G^s(u^s_h,v^s_h)= 
	-  \sum\limits_{k_2=1}^{N_K} \int_{K_{k_2}^{\text{S}} } \left(\nabla \cdot (\rho^0_{k_2} \overline{v}^s_h) \right) S_{k_2}(u^s_h) \dd x   \\
	-  \sum\limits_{l_2=1}^{N_E}  \int_{E_{l_2}^{\text{S}}  } (\nu \cdot \overline{v}^s_h) S_{l_2}(u^s_h)  [\rho^0_{l_2}]^+_-  \dd \Sigma . 
\end{multline}
We add \eqref{eq:phi1solid} into the matrix representation and obtain
\begin{equation}\label{eq:solideigfull}
	\omega^2 M_s \tilde{u}^s - 2 \ii \omega R_s \tilde{u}^s - \big( A_{sg} - C_s\trans S_s C_s \big) \tilde{u}^s = 0,
\end{equation}
where $C_s \tilde{u}^s$ evaluates $S_{k_2}(u^s_h)$ and
$S_{l_2}(u^s_h)$, $S_s$ solves the N-body problem for the solid
planet, and $C_s\trans S_s C_s \tilde{u}^s$ evaluates
\eqref{eq:phi1solid};  the submatrix $A_{sg}$ and its corresponding 
weak formula is shown in Table~\ref{table:matrixcomponentsquadratic}, and 
the submatrices $C_s, C_s\trans, S$ and their
corresponding weak formulae are shown in
Table~\ref{table:matrixcomponentsselfG}. Here, of course, $A_{sg}$, $C_s$ and
$C_s\trans$ do not include terms related the fluid-solid boundaries
$\Sigma^{\text{FS}}$. 

\subsubsection{Planets with fluid regions}

For a planet with fluid regions, we also substitute (\ref{eq:k2local}) and (\ref{eq:l2local}) into (\ref{eq:ph1body0}) and (\ref{eq:ph1surface0}), respectively. 
To ensure the Hermitian property of the system, 
we carefully treat the fluid-solid boundary terms and evaluate the incremental gravitational potential $S(u_h)$ via (\ref{eq:solSu1}) and obtain the volume integral contributions 
\begin{multline}\label{eq:ph1body2}
S_{k_2}(u_h) = \frac{G}{R_{k_2}}  \int_{K_{k_2}} \nabla \cdot (\rho_{k_2}^0 u_h) \dd x    
  \\  
   +  \sum\limits_{\substack{k_1=1\\k_1 \neq k_2}}^{N_K^{\text{S}}} \frac{G}{\|\mathbf{r}_{k_2} - \mathbf{r}_{k_1} \|} \int_{K_{k_1}^{\text{S}} } \nabla \cdot (\rho_{k_1}^0 u^s_h) \dd x  
    +   \sum_{l_1 = 1}^{N_E^{\text{S}}} \frac{G}{\| \mathbf{r}_{k_2}  - \mathbf{r}_{l_1}\|} \int_{E_{l_1}^{\text{S}} } (\nu \cdot u^s_h) [\rho_{l_1}^0]^+_- \dd \Sigma \\
   +  \sum\limits_{\substack{k_1=1\\k_2 \neq k_2}}^{N_K^{\text{F}}} \frac{G}{\|\mathbf{r}_{k_2} - \mathbf{r}_{k_1} \|} \int_{K_{k_1}^{\text{F}}  } \nabla \cdot (\rho_{k_1}^0 u^f_h) \dd x    
 +   \sum_{l_1 = 1}^{N_E^{\text{F}}} \frac{G}{\| \mathbf{r}_{k_2}  - \mathbf{r}_{l_1}\|} \int_{E_{l_1}^{\text{F}} } (\nu \cdot u^f_h) [\rho_{l_1}^0]^+_-\dd \Sigma   \\ 
  +   \sum_{l_1 = 1}^{N_E^{\text{FS}}} \frac{G}{\| \mathbf{r}_{k_2}  - \mathbf{r}_{l_1}\|} \int_{E_{l_1}^{\text{FS}}  } \left\{ (\nu^{f\rightarrow s} \cdot u^s_h) [\rho_{l_1}^0]^s  +  (\nu^{s\rightarrow f} \cdot u^f_h) [\rho_{l_1}^0]^f  \right\} \dd \Sigma , 
\end{multline}
and boundary integral contributions
\begin{multline}\label{eq:ph1surface2}
	S_{l_2}(u_h) = \frac{G}{R_{l_2}} \int_{E_{l_2}}  (\nu \cdot u_h) [\rho_{l_2}^0]^+_- \dd \Sigma 
	   \\
     + \sum\limits_{\substack{k_1=1}}^{N_K^{\text{S}}} \frac{G}{\|\mathbf{r}_{l_2} - \mathbf{r}_{k_1} \|} \int_{K_{k_1}^{\text{S}}  } \nabla \cdot (\rho_{k_1}^0 u^s_h) \dd x 
     + \sum_{\substack{l_1 = 1\\l_1\neq l_2}}^{N_E^{\text{S}}} \frac{G}{\| \mathbf{r}_{l_2}  - \mathbf{r}_{l_1}\|} \int_{E_{l_1}^{\text{S}}  } (\nu \cdot u^s_h) [\rho_{l_1}^0]^+_- \dd \Sigma \\
      + \sum\limits_{\substack{k_1=1}}^{N_K^{\text{F}}} \frac{G}{\|\mathbf{r}_{l_2} - \mathbf{r}_{k_1} \|} \int_{K_{k_1}^{\text{F}}  } \nabla \cdot (\rho_{k_1}^0 u^f_h) \dd x  
      + \sum_{\substack{l_1 = 1\\l_1\neq l_2}}^{N_E^{\text{F}}} \frac{G}{\| \mathbf{r}_{l_2}  - \mathbf{r}_{l_1}\|} \int_{E_{l_1}^{\text{F}} } (\nu \cdot u^s_h) [\rho_{l_1}^0]^+_- \dd \Sigma \\
       +  \sum_{l_1 = 1}^{N_E^{\text{FS}}} \frac{G}{\| \mathbf{r}_{l_2}  - \mathbf{r}_{l_1}\|} \int_{E_{l_1}^{\text{FS}}  } \left\{  (\nu^{f\rightarrow s} \cdot u^s_h) [\rho_{l_1}^0]^s +  (\nu^{s\rightarrow f} \cdot u^f_h) [\rho_{l_1}^0]^f \right\} \dd \Sigma  
	   .
\end{multline}
With (\ref{eq:ph1body2}) and (\ref{eq:ph1surface2}), we have the full solution for the incremental gravitational potential. 
To evaluate (\ref{eq:weakgravperturb1}) for a planet with fluid regions, we need to compute 
\begin{multline}\label{eq:phi1fluidsolid}
a_G(u_h,v_h) = \\
-  \sum\limits_{k_2=1}^{N_K^{\text{S}}} \int_{K_{k_2}^{\text{S}} } \left(\nabla \cdot (\rho^0_{k_2} \overline{v}^s_h) \right) S_{k_2}(u_h) \dd x
-  \sum\limits_{l_2=1}^{N_E^{\text{S}} }  \int_{E_{l_2}^{\text{S}}   }  (\nu \cdot \overline{v}^s_h) S_{l_2}(u_h) [\rho^0_{l_2}]^+_- \dd \Sigma 
  \\
-  \sum\limits_{k_2=1}^{N_K^{\text{F}}} \int_{K_{k_2}^{\text{F}} } \left(\nabla \cdot (\rho^0_{k_2} \overline{v}^f_h) \right) S_{k_2}(u_h) \dd x   
-  \sum\limits_{l_2=1}^{N_E^{\text{F}} }  \int_{E_{l_2}^{\text{F}}  }  (\nu \cdot \overline{v}^f_h) S_{l_2}(u_h) [\rho^0_{l_2}]^+_-  \dd \Sigma \\
- \sum_{l_2 = 1}^{N_E^{\text{FS}}}  \int_{E_{l_2}^{\text{FS}}  } \left\{  (\nu^{f\rightarrow s} \cdot \overline{v}^s_h) S_{l_2}(u_h) [\rho^0_{l_2}]^s +  (\nu^{s\rightarrow f} \cdot \overline{v}^f_h) S_{l_2}(u_h) [\rho^0_{l_2}]^f  \right\}  \dd \Sigma   
 . 
\end{multline}
We derive the matrix representation with \eqref{eq:phi1fluidsolid} and obtain
\begin{equation}\label{eq:eigfull}
	\omega^2 M \tilde{u}  - 2 \ii \omega \tilde{R}_{\Omega} \tilde{u} - \big(A_{G}-E_G A_p^{-1} E_G\trans -  C\trans S C \big) \tilde{u} =  0, 
\end{equation}
with 
\begin{align*}
 A_{G} &= \left(\begin{array}{cc}
   A_{sg} & 0   \\
   0      & A_f                   
   \end{array} \right) ,\ 
    \tilde{R}_{\Omega} = \left(\begin{array}{cc}
   R_s & 0   \\
   0   & R_f                           
   \end{array}\right), \
   M = \left(\begin{array}{cc}
   M_s & 0   \\
   0   & M_f                           
   \end{array}\right), 
\\
  E_G\trans &= \left(\begin{array}{cc}
   E_{\text{FS}}  &  A_{\text{dg}}                                   
   \end{array}\right) ,\
	C = \big( \begin{array}{cc}
	C_s       &
    C_f                                  
	\end{array} \big) 	,
\end{align*}
where $C \tilde{u} = C_s \tilde{u}^s + C_f \tilde{u}^f$ evaluates
(\ref{eq:ph1body2}) and (\ref{eq:ph1surface2}) to get $S_{k_2}(u_h)$
and $S_{l_2}(u_h)$, $S$ solves the N-body problem, and $C\trans S C
\tilde{u}$ evaluates (\ref{eq:phi1fluidsolid}); 
the submatrices $A_{sg}$, $A_f$, $A_p$, $R_s$, $R_f$, $M_s$, $M_f$, $E_{\text{FS}}$, 
$A_{\text{dg}}$ and their corresponding weak formulae are shown in 
Table~\ref{table:matrixcomponentsquadratic} and the submatrices $C_s$,
$C_s\trans$, $C_f$, $C_f\trans$, $S$ and their corresponding weak formulae
are shown in Table~\ref{table:matrixcomponentsselfG}. 
The construction of submatrices $C_s$, $C_s\trans$, $C_f$, $C_f\trans$ can 
be found in \ref{sec:matrixselfG}. 
We note that $A_p$ is always symmetric positive definite since $\kappa$ is always positive. 
We note that \eqref{eq:eigfull} is the discretization of \eqref{eq:finalfullQ2}. 

\section{Computational experiments for non-rotating planets}\label{sec:exp}

In this section, we first show the computational accuracy of our
algorithm for the reference gravitational field using FMM in
Subsection~\ref{sec:refgravity}.  We then illustrate computational
experiments yielding planetary normal modes with or without
perturbation of the gravitational potential.  In this section and
Section~\ref{sec:comprotation}, two supercomputers, Stampede2 (an
Intel cluster) at the Texas Advanced Computing Center and Abel (a Cray
XC30 cluster) at Petroleum Geo-Services are utilized for the
computational experiments.


\subsection{Computational accuracy for the reference gravitational field}\label{sec:refgravity}

In this subsection, we illustrate the computational accuracy 
for the reference gravitational field using FMM. 
We begin with a simple constant-density ball. 
In Table~\ref{table:errGconstball}, we show the FMM solution 
for a gravitational field of a constant density ball and a comparison with the closed-form solution. 
We note that FMM provides an accurate solution for this example.

\begin{table}[ht!]
	\centering
 \begin{tabular}{cccccc} \hline
   \# of elements & 116,085 & 1,136,447 & 2,019,017 & 3,081,551 & 4,035,022  \\ \hhline{======}
   MSE of $\Phi^0$ & 2.133e-6 & 7.452e-8 & 1.784e-8 & 1.545e-8 & 1.430e-8 \\ 
   MSE of $g$   &  1.102e-3  & 1.848e-4 & 1.156e-4 & 8.781e-5 & 7.363e-5  \\ \hline
 \end{tabular}
  \captionof{table}{Errors in the gravitational calculation of a constant density ball. } \label{table:errGconstball}
\end{table}

We use PREM to build our Earth models on unstructured meshes with different sizes. 
In Table~\ref{table:errG3layers}, we show the approximation errors 
of different three-layer models, which contain two major discontinuities (CMB and ICB) when compared  with the semi-analytical solution. 
In Fig.~\ref{fig:referenceg}, we show the comparison of the gravitational field computed via FMM with the semi-analytical solution in PREM.

\begin{table}[ht!]
	 \setlength{\tabcolsep}{3pt}
	\centering
 \begin{tabular}{cccccccc} \hline
   \# of elements & 5,800 & 57,490 & 503,882 & 1,136,447 & 2,093,055 & 5,549,390 & 7,825,918  \\ 
   \hhline{========}
   MSE of $\Phi^0$ & 3.604e-3 & 2.635e-4 & 4.071e-5 & 2.092e-5 & 1.354e-5 & 4.059e-6 & 2.396e-9  \\ 
   MSE of $g$   & 5.805e-2 & 5.479e-3 & 7.320e-4 & 3.218e-4 & 2.068e-4 & 9.524e-5 & 5.609e-5  \\ \hline
 \end{tabular}
 \captionof{table}{Errors of three-layer approximations in the gravitational calculation. } \label{table:errG3layers}
\end{table}

\begin{figure}[ht!]
 \setlength{\tabcolsep}{3pt}
	\centering
  \begin{tabular}{cc}
  	\includegraphics[trim= 8cm 0cm 4cm 0cm,clip=true,width=0.42\linewidth]{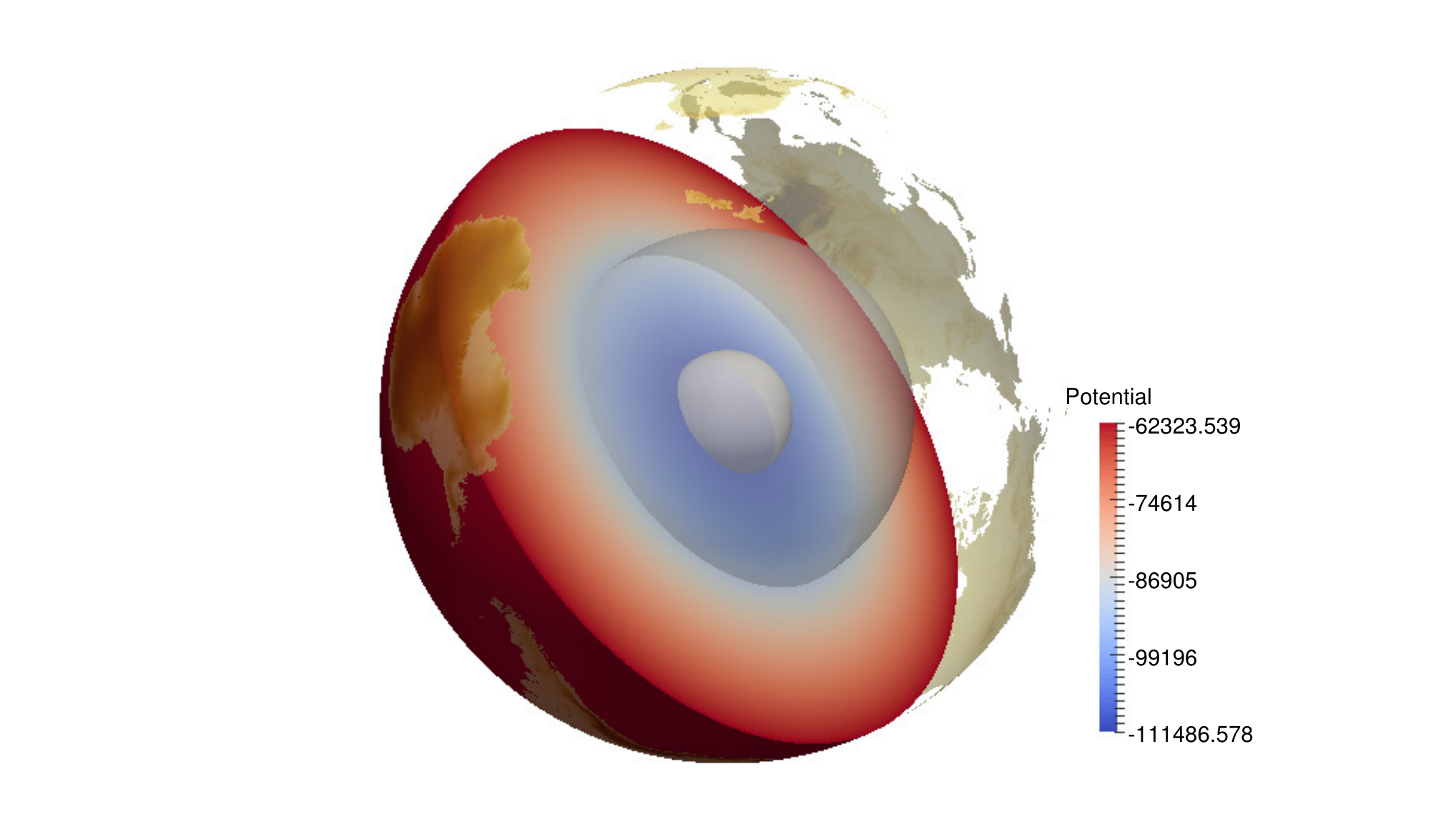} &
  	\includegraphics[trim= 0cm 0cm 0cm 0cm,clip=true,width=0.42\linewidth]{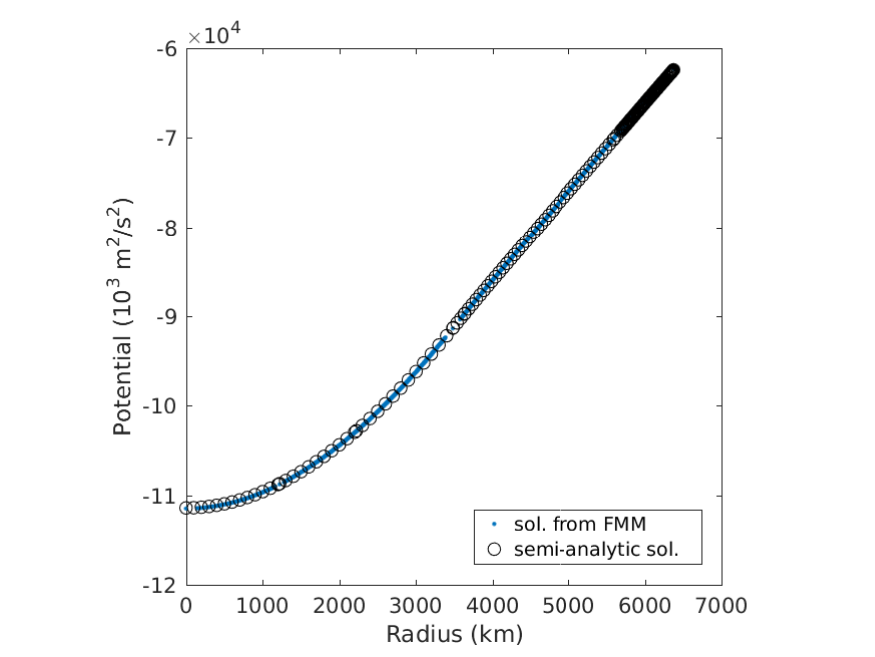}  \\
  	(a1) & (a2) \\
  	\includegraphics[trim= 8cm 0cm 4cm 0cm,clip=true,width=0.42\linewidth]{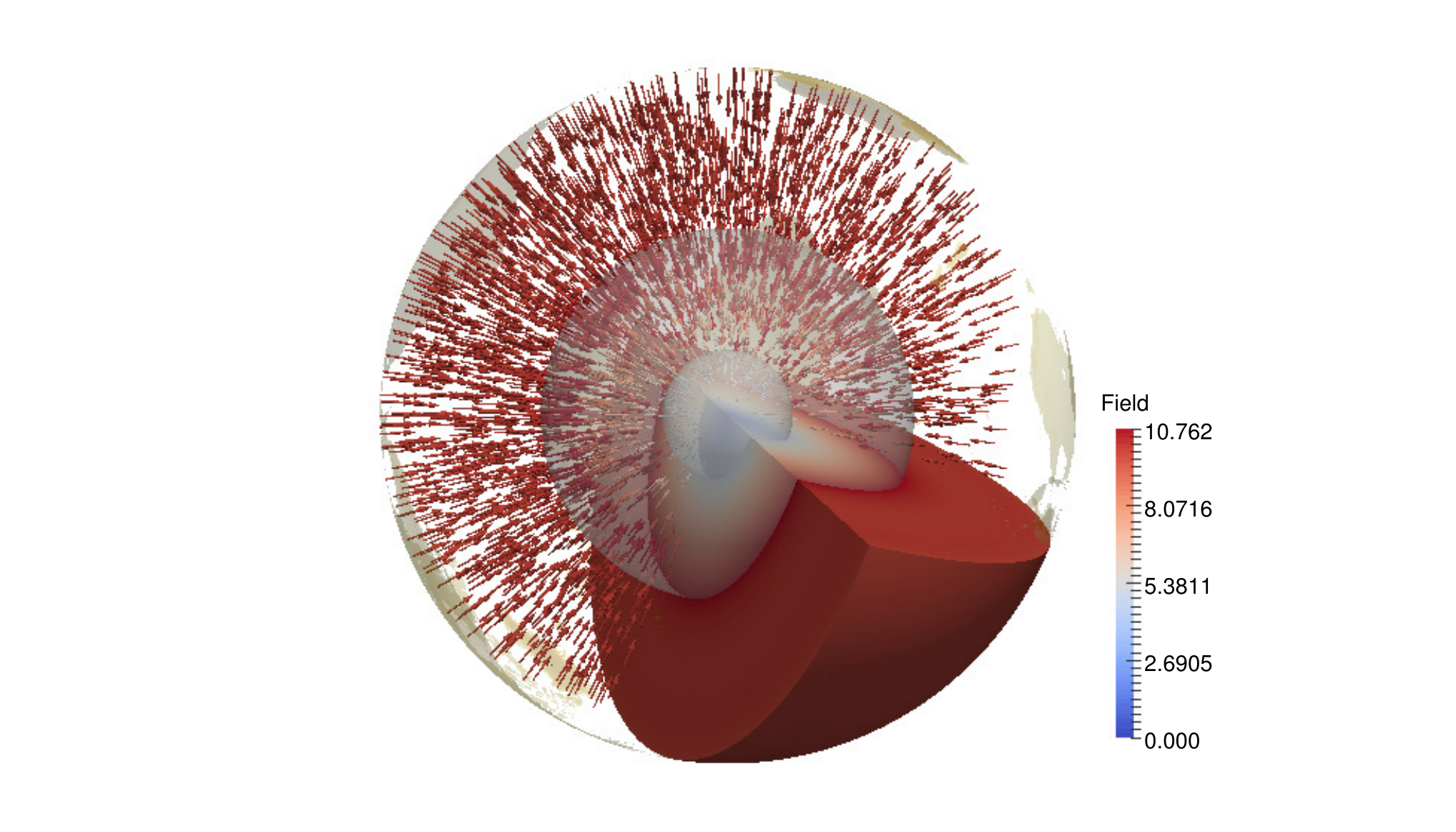} &
  	\includegraphics[trim= 0cm 0cm 0cm 0cm,clip=true,width=0.42\linewidth]{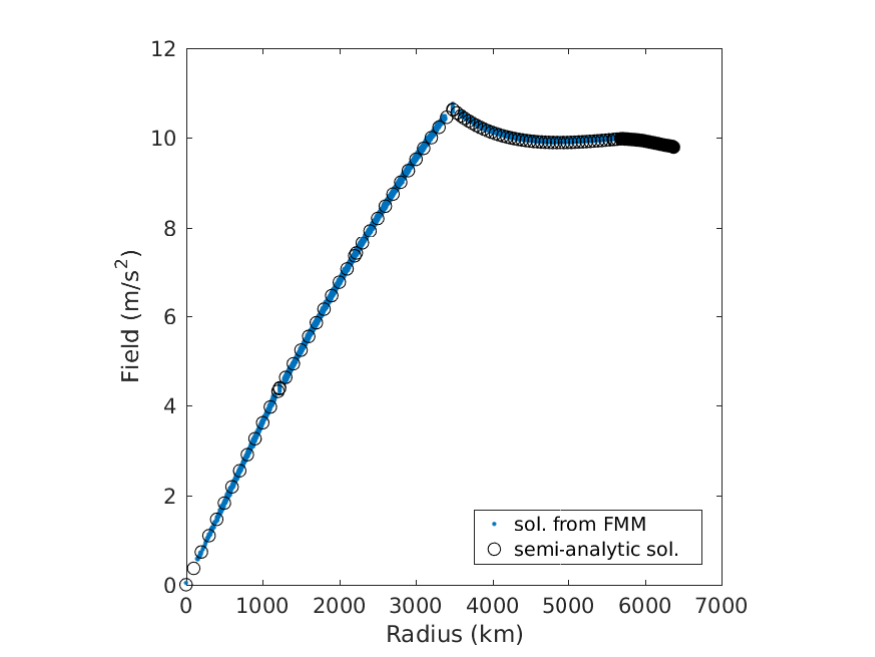}
  	\\
  	(b1) & (b2)
  \end{tabular}
  \captionof{figure}{Comparison between the semi-analytical and FMM
    solutions: (a1) FMM gravitational potential; (a2) comparison in
    the radial direction; (b1) FMM gravitational field; (b2)
    comparison in the radial direction.}\label{fig:referenceg}
\end{figure}

In Table~\ref{table:errG7layers}, we show the approximation errors of different seven-layer models
which contain six major discontinuities (Moho, top of Low Velocity Zone (LVZ), bottom of LVZ, 660, CMB and ICB) with the semi-analytical solution. 	
	
\begin{table}[ht!]
	 \setlength{\tabcolsep}{3pt}
	\centering
	\begin{tabular}{cccccc} \hline
		\# of elements & 2,031,729 & 5,018,249 & 8,043,617 & 12,479,828 & 16,560,615  \\ 
		\hhline{======}
		MSE of $\Phi^0$ & 2.333e-7 & 4.485e-8  & 1.286e-8 & 9.785e-9 & 5.548e-9 \\ 
		MSE of $g$   & 1.926e-4 & 8.606e-5 & 5.186e-5 & 4.036e-6 & 3.394e-5  \\ \hline
	\end{tabular}
	\captionof{table}{Errors of seven-layer approximations in the gravitational calculation. } \label{table:errG7layers}
\end{table}	

\subsection{Computational accuracy for non-rotating planets}\label{sec:compnonrotation}

In this subsection, we do not consider rotation and study the computational accuracy with 
existing algorithms for spherically-symmetric planets. 
Let the angular velocity of rotation $\Omega=0$, without loss of generality, 
we write \eqref{eq:eigfull} and its
pure solid planet version \eqref{eq:solideigfull} in the form of
generalized eigenvalue problems:
\begin{equation}\label{eq:gep}
   A \tilde{u} = \omega_{N}^2  M \tilde{u} , 
\end{equation}
where $A$ represents $A_{sg} - C_s\trans S_s C_s$ in \eqref{eq:solideigfull} or 
$A_{G}-E_G A_p^{-1} E_G\trans -  C\trans S C$ in \eqref{eq:eigfull} and 
$\omega_{N}$ denotes the frequency for the non-rotation planets. 
Since the explicit formation of $A$ with
self-gravitation requires excessive storage, it is necessary to solve
\eqref{eq:gep} via a matrix-free scheme, where $A$, $M$ and $M^{-1}$
are only accessed through matrix-vector multiplications. We combine
several efficient parallel approaches to solve (\ref{eq:gep}) with a
matrix-free scheme.

In this work, we utilize polynomial filtering techniques 
\cite{saad:filtered-cr06,Filtlan-paper,spectrumslicing} as these do not
involve solving linear systems with the indefinite matrices. 
Here, the bulk of computations are carried out in the form of matrix-vector
products. The polynomial filtering technique is
ideally suited for solving large-scale three-dimentional interior eigenvalue problems
because it significantly enhances the memory and computational
efficiency without any loss of accuracy \cite{DBLP:conf/sc/ShiLXSH18}. In
this paper, we adopt the polynomial filtering algorithms recently
developed in \cite{spectrumslicing,DBLP:conf/sc/ShiLXSH18,li2019evsl} 
due to their simplicity and robustness on 
  a prescribed interval $[f_1,f_2]$ mHz. The
details about our parallel algorithms and their performance can be
found in \cite{DBLP:conf/sc/ShiLXSH18}.

We show the convergence of our numerical formulation and approach for constant
elastic balls and PREM. 
The constant balls have a radius of 6,371 km,
density $\rho^0=5.51\times 10^3$ kg/$m^3$, \textsl{P}-wave speed $V_P$ = 10.0
km/s and \textsl{S}-wave speed $V_S$ = 5.7735 km/s. The PREM used in our tests
is modified in an isotropic model without attenuation, with $V_P =
(V_{PV} + V_{PH})/2$ and $V_S = (V_{SV} + V_{SH})/2$. The ocean layer
in PREM is replaced by crust.  
In the work of \cite{Matchette-Downes2021}, a good agreement of 
the one-dimensional solution based on the classical approach \texttt{MINEOS} \cite{masters2011mineos}
and a radial FEM \cite{jingchen2018revisiting} is demonstrated. 
The discretization of the radial FEM code is described in Appendix~\ref{app:sphharmonics}, 
In this work, we show our three-dimensional results are in a good agreement with the one-dimensional solutions. 

\subsubsection{Solid models with self-gravitation}

We present our results for purely solid models with self-gravitation.
In Tables~\ref{tab:solidfull} and \ref{tab:solidfullp2}, we list
the number of elements `\#elm.' as well as the problem sizes (labeled
as `size of $A$' for the solid cases and `size of $A_G$' and `size of
$A_p$' for the Earth examples), the number of surfaces `\#surf.', the
size of $S_s$ or $S$, and the target frequency interval in milliHertz (labeled as $[f_1,f_2]$ (mHz)), 
the degree of the
polynomial filter `deg', the number of the Lanczos iterations required
`\#it', and the number of the normal modes computed `\#eigs'.
\begin{table}[ht!]
    \centering
    \begin{tabular}{cccccccccc} \hline
        Exp.   & \#elm. & size of $A$ &  \#surf.  & size of $S_s$ & $[f_1,f_2]$ (mHz)  & (deg,\#it) & \#eigs \\ \hhline{========}
        C1p1   & 5,123      & 2,727       & 392        & 5,515         & [0.1,1.0]             & (14,192)   & 70 \\
                C2p1   & 21,093     & 10,644      & 956        & 22,049        & [0.1,1.0]              & (25,232)  & 92 \\
                C3p1   & 39,273     & 19,131      & 956        & 40,229        & [0.1,1.0]           & (34,252)  & 92 \\  
                C4p1   & 105,115    & 51,933      & 3,608      & 108,723       & [0.1,1.0]  & (50,252) &  92 \\  
                C5p1   & 495,099    & 242,721     & 14,888     & 509,987       & [0.1,1.0]  & (108,272) & 92 \\ \hline
    \end{tabular}
    \caption{Test cases with self-gravitation for different solid models using P1 elements  for the frequency range $[0.1,1.0]$ mHz.}\label{tab:solidfull}
\end{table}

Since the pure solid models do not generate any essential spectra, we
can directly compute the lowest-frequency normal modes.  We note that
the length ($\lambda_{\max}-\lambda_{\min}$) of the spectrum grows
with the size of the problem determined by the discretization.

\begin{table}[ht!]
    \centering
        \begin{tabular}{cccccccccc} \hline
           Exp.  & ${}_0T_2$ & ${}_0S_2$ & ${}_1S_1$ & ${}_0S_0$ & ${}_0T_3$ & ${}_0S_3$ & ${}_1S_2$ & ${}_0T_4$ & ${}_0S_4$ \\
           \hhline{==========} 
           C1p1  & 0.3724    & 0.4178    & 0.4600    & 0.5105    & 0.5881    & 0.6322    & 0.6900    & 0.7973    & 0.8359  \\
           C2p1  & 0.3653    & 0.4112    & 0.4511    & 0.5053    & 0.5692    & 0.6052    & 0.6708    & 0.7587    & 0.7791  \\
           C3p1  & 0.3643    & 0.4103    & 0.4502    & 0.5053    & 0.5665    & 0.6017    & 0.6680    & 0.7527    & 0.7721  \\
           C4p1  & 0.3622    & 0.4089    & 0.4472    & 0.5035    & 0.5612    & 0.5932    & 0.6622    & 0.7424    & 0.7526  \\
           C5p1  & 0.3612    & 0.4086    & 0.4460    & 0.5035    & 0.5587    & 0.5899    & 0.6596    & 0.7374    & 0.7445  \\ \hline
             1D    & 0.3607    & 0.4087    & 0.4456    & 0.5040    & 0.5574    & 0.5885    & 0.6582    & 0.7348    & 0.7406  \\ \hline  
        \end{tabular}
    \caption{Convergence tests with self-gravitation for different solid models in Table~\ref{tab:solidfull} with self-gravitation for P1 elements. }\label{tab:solidfullconv}
\end{table}

In Table~\ref{tab:solidfullconv}, we show the convergence results for
different solid models using P1 elements, 
that is, the finite-element polynomial orders $p^s=p^f=p^p=1$ are used throughout this work.  
Through comparison with 1D results,
we observe that our computational results do converge. We accept
relative differences of about 0.1\%.

\begin{table}[ht!]
    \centering
    \begin{tabular}{cccccccc} \hline
        Exp. & \# of elm. & size of $A$ &  \#surf.  & size of $S_s$ & $[f_1,f_2]$ (mHz)  & (deg,\#it)  & \#eigs \\ \hhline{========}
        C1p2   & 19,073      & 75,888     & 956        & 20,029        & [0.1,1.0]           & (44,512)  & 92\\ 
        C2p2   & 40,378      & 170,025     & 3,608        & 43,986        &  [0.1,1.0]         & (58,492) &  92 \\ 
        C3p2   & 80,554      & 335,103     & 5,924        & 86,478        & [0.1,1.0]           & (81,492) & 92 \\ 
        C4p2   & 152,426    & 645,687     & 14,888     & 167,314      &  [0.1,1.0]             & (129,492) & 92 \\ 
        C5p2   & 334,193    & 1,360,140    & 14,888     & 349,081      & [0.1,1.0]              & (200,492)  & 92 \\ \hline
    \end{tabular}
    \caption{Test cases with self-gravitation for different solid models using P2 elements  for the frequency range $[0.1,1.0]$ mHz.}\label{tab:solidfullp2}
\end{table}

\begin{table}[ht!]
    \centering
    \begin{tabular}{cccccccccc} \hline
        Exp. & ${}_0T_2$ & ${}_0S_2$ & ${}_1S_1$ & ${}_0S_0$ & ${}_0T_3$ & ${}_0S_3$ & ${}_1S_2$ & ${}_0T_4$ & ${}_0S_4$ \\
        \hhline{==========}
        C1p2  & 0.3619    & 0.4100    & 0.4473    & 0.5094    & 0.5594    & 0.5908   & 0.6605    & 0.7376   & 0.7439  \\
        C2p2  & 0.3610    & 0.4090    & 0.4459    & 0.5042    & 0.5579    & 0.5889   & 0.6587    & 0.7355   & 0.7413  \\
        C3p2  & 0.3609    & 0.4089    & 0.4463    & 0.5042    & 0.5577    & 0.5888   & 0.6585    & 0.7352    & 0.7410  \\
        C4p2  & 0.3608    & 0.4088    & 0.4456    & 0.5041    & 0.5575    & 0.5886   & 0.6583    & 0.7349    & 0.7408  \\
        C5p2  & 0.3608    & 0.4087    & 0.4456    & 0.5041    & 0.5575    & 0.5885    & 0.6583    & 0.7349    & 0.7407  \\ \hline
           1D  & 0.3607    & 0.4087    & 0.4456    & 0.5040    & 0.5574    & 0.5885    & 0.6582    & 0.7348    & 0.7406  \\ \hline  
    \end{tabular}
    \caption{Convergence tests with self-gravitation for the solid models in Table~\ref{tab:solidfullp2} using P2 elements. }\label{tab:solidfullconvp2}
\end{table}

In Table~\ref{tab:solidfullp2}, we list test cases for different solid
models using P2 elements, 
that is, the finite-element polynomial orders $p^s=p^f=p^p=2$ are used throughout this work. 
From experiments C1p2 to C5p2, we double
the number of elements and obtain proper convergence results in
Table~\ref{tab:solidfullconvp2}.  We show that even with about 330,000
elements, we are able to achieve four-digit agreement.

\subsubsection{PREM with self-gravitation}

Here, we include a liquid outer core using PREM and the presence of
the essential spectrum.  
In Table~\ref{tab:premfull}, we show test cases for PREM.  We roughly
double the number of elements from E1p1 to E7p1.  In
Table~\ref{tab:premfullconv}, we argue convergence by comparing with
1D results.  For PREM with self-gravitation, we accept relative differences
that are less than 0.1\%.


\begin{table}[ht!]
    \centering
    \begin{tabular}{ccccccccc} \hline
        Exp. & \# of elm. & size of $A_G$ & size of $A_p$ &  \#surf. & size of $S$ & $[f_1,f_2]$ (mHz) & (deg,\#it) & \#eigs
                \\ \hhline{=========}
                E1p1   &  9,721      & 7,590         & 887            & 2,304     & 12,025      & [0.1,1.0]             & (187,392) & 64\\
                E2p1   & 20,466     & 14,736        & 974          & 4,956     & 25,422     & [0.1,1.0]             & (182,372) & 72 \\
                E3p1   & 42,828     & 30,384        & 3,171         & 8,172     & 51,000     &[0.1,1.0]             & (342,452) & 83\\
                E4p1   & 83,354     & 63,225        & 5,298         & 22,104    & 105,458    & [0.1,1.0]             & (745,452) & 88 \\ 
                E5p1   & 157,057     & 96,852        & 6,771         & 22,104    & 179,161    & [0.1,1.0]            & (747,492) & 88 \\ 
                E6p1   & 303,218     & 164,673        & 10,077         & 22,104    & 325,322    &[0.1,1.0]             & (685,492) & 88 \\ 
                E7p1   & 639,791     & 361,587        & 21,824         & 60,288    & 700,079    & [0.1,1.0]            & (685,492) & 88 \\ 
                E8p1   & 1,972,263     & 1,086,702     &  70,429        & 150,288    & 2,122,551    & [0.1,1.0]          & (1565,492) & 88 \\ 
                E8p2   & 1,972,263   & 8,400,630    & 522,705 & 150,288 & 2,122,551  & [0.3,1.5]    & (1185,1051) &  268 \\ 
                \hline
    \end{tabular}
    \caption{Test cases with self-gravitation for different Earth models E1p1 - E8p1 using P1 elements for the frequency range $[0.1,1.0]$ mHz and Earth model E8p2 using P2 elements for the frequency range $[0.3,1.5]$ mHz.  }\label{tab:premfull}
\end{table}

\begin{table}[ht!]
\setlength{\tabcolsep}{2pt}
    \centering
        \begin{tabular}{cccccc} \hline
           Exp. & ${}_0S_2$ & ${}_0T_2$ & ${}_2S_1$ & ${}_0S_3$ & ${}_0T_3$   \\
           \hhline{======}
           E1p1  & 0.3284    & 0.3953    & 0.4179    & 0.5242   & 0.6241      \\ 
           E2p1  & 0.3229    & 0.3921    & 0.4149    & 0.5077    & 0.6146      \\ 
           E3p1  & 0.3177   & 0.3884    & 0.4113    & 0.4932    & 0.6062      \\ 
           E4p1  & 0.3166    & 0.3842    & 0.4090    & 0.4903    & 0.5980      \\   
           E5p1  & 0.3137    & 0.3845    & 0.4085    & 0.4863    & 0.5962      \\   
           E6p1  & 0.3126    & 0.3840    & 0.4080    & 0.4768    & 0.5945      \\   
           E7p1  & 0.3116    & 0.3834    & 0.4073    & 0.4742    & 0.5933      \\ 
           E8p1  & 0.3112    & 0.3829    & 0.4067    & 0.4721    & 0.5920      \\     
           E8p2  & 0.3106    & 0.3826    & 0.4063    & 0.4708    & 0.5912      \\     
            \hline
           1D &  0.3110   & 0.3826    & 0.4063    & 0.4713     & 0.5912   \\ 
   \hline  
        \end{tabular}
        \caption{Convergence tests with self-gravitation for different Earth models in Table~\ref{tab:premfull}. }\label{tab:premfullconv}
\end{table}

\subsection{Fully heterogeneous models}\label{sec:heterogeneity}

Here, we study the effects of heterogeneity on the normal modes. 
In Subsection~\ref{subsec:crustmit} and Subsection~\ref{subsec:shapecmb}, 
we study the effects of the crust and upper mantle, and shape of the CMB, respectively.

\subsubsection{Shape of the CMB}\label{subsec:shapecmb}

\begin{table}[ht!]
    \centering
    \begin{tabular}{ccccccccc} \hline
        Exp. & \# of elm. & size of $A_G$ & size of $A_p$ &  \#surf. & size of $S$ & $[f_1,f_2]$ (mHz) & (deg,\#it) & \#eigs
                \\ \hhline{=========}
                CMB8   & 2,007,479   &  8,711,940   & 633,358 & 177,352 & 2,184,831  & [1.5,2.0]    & (3591,1251) &  350 \\ 
                \hline
    \end{tabular}
    \caption{Test case with self-gravitation for an Earth model with a non-spherically symmetric CMB using P2 elements. }\label{tab:premcmb}
\end{table}

Here, we study the effects of the CMB. 
Long-wavelength topography of the CMB was proposed by \cite{creager1986aspherical,morelli1987topography}. 
Many studies 
\cite{bataille1988inhomogeneities,doornbos1989models,pulliam1993bumps,rodgers1993inference,obayashi1997p,earle1997observations,earle1998observations,garcia2000amplitude,sze2003core,lassak2010core,tanaka2010constraints,colombi2014seismic,schlaphorst2015investigation} were later performed to model the topography of the CMB. 

\begin{figure}[ht!]
\centering
\includegraphics[trim= 1cm 2cm 1cm 2cm,clip=true,width=0.7\linewidth]{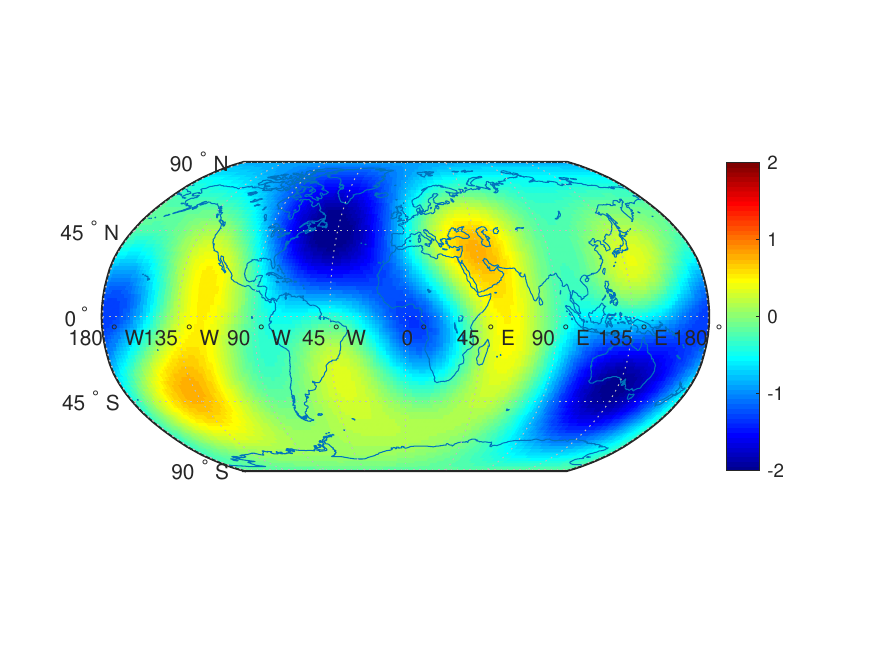}
\caption{Shape of the CMB using the result of \cite{tanaka2010constraints}. The values in the color bar indicate the variations in kilometers.}
\label{fig:shapeofCMB}
\end{figure}

\begin{figure}
\centering
\begin{tabular}{cc}
\includegraphics[trim= -1cm 0cm -1cm 0cm,width=0.40\linewidth]{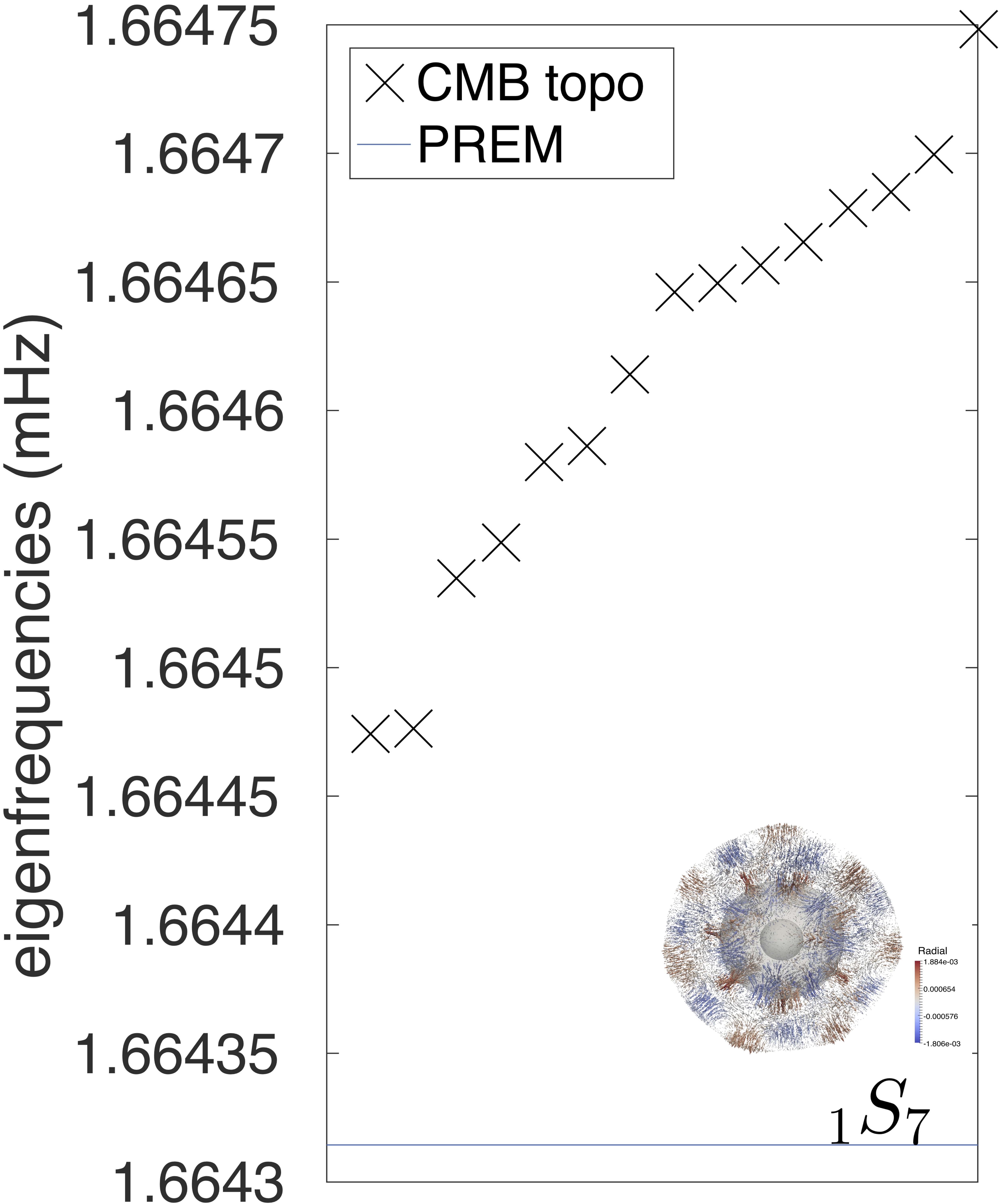} & 
\includegraphics[trim= -1cm 0cm -1cm 0cm,width=0.40\linewidth]{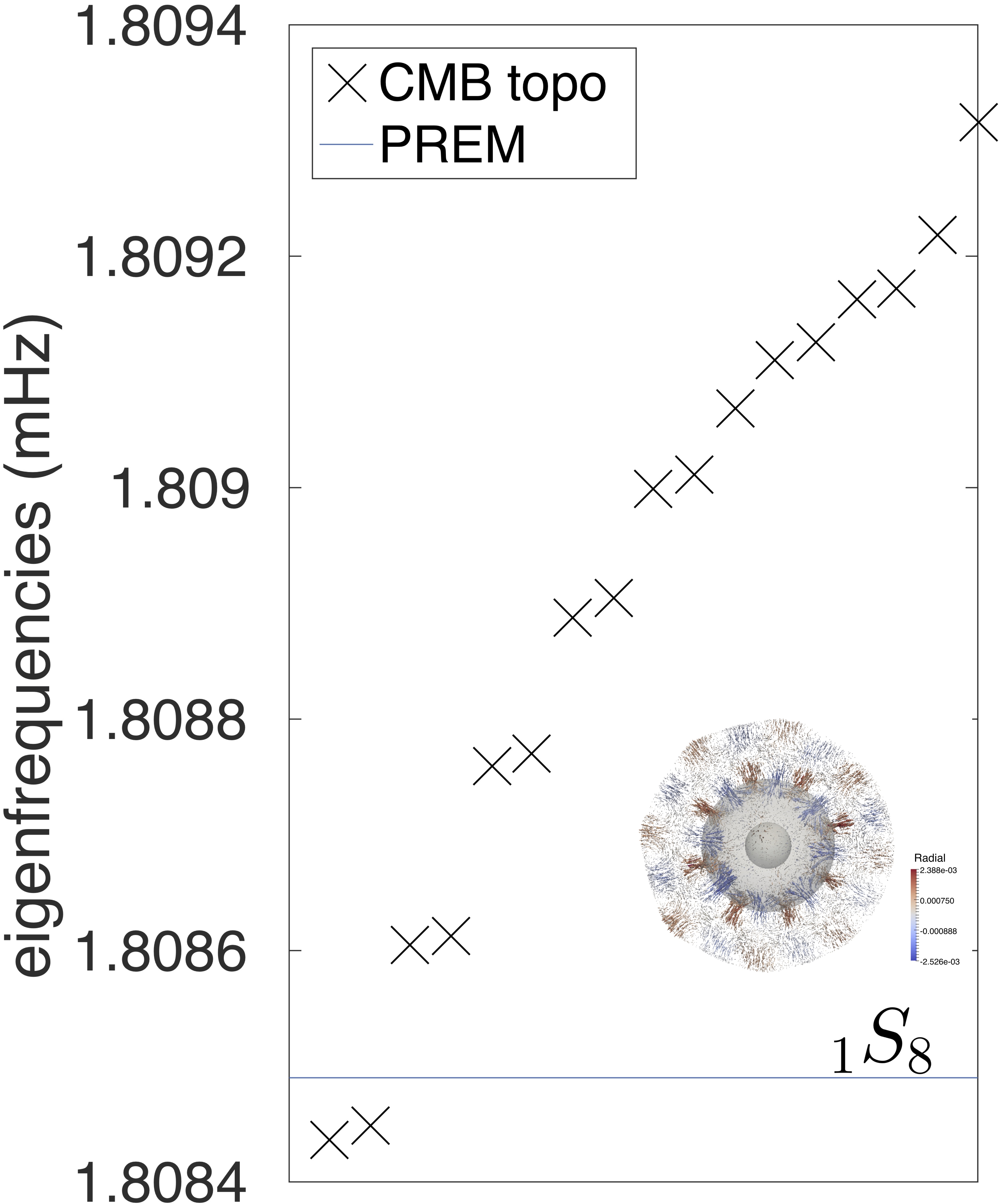} 
\end{tabular}
\caption{Splittings of the modes ${}_1S_7$ and ${}_1S_8$ due to the non-spherically symmetric CMB, which is shown in Figs.~\ref{fig:shapeofCMB}. }
\label{fig:cmbcomparison}
\end{figure}

In Fig.~\ref{fig:shapeofCMB}, we show the topography of the CMB from the result by  \cite{tanaka2010constraints}. 
We use a triangular mesh to model the shape with ellipticity combined. 
In Table~\ref{tab:premcmb}, we show the information of the experiment CMB8, 
which indicates a PREM-like model with the mentioned CMB embedded.
In Fig.~\ref{fig:cmbcomparison}, we illustrate the splittings of modes ${}_1S_7$ and ${}_1S_8$ due to the non-spherically symmetric CMB. 
Since the modes ${}_1S_7$ and ${}_1S_8$ are sensitive to the change of the CMB, the splittings of 
these modes are quite clear.  

\subsubsection{Heterogeneity of the crust and upper mantle}\label{subsec:crustmit}
Self-gravitation is important for the normal modes with frequencies lower than 5.0 mHz or so \cite{kennett1998density}. 
However, in this subsection, we restrict ourselves to models without perturbation of the gravitational potential 
for computational efficiency. 
We reduce the full generalized eigenvalue problem \eqref{eq:gep} into Cowling approximation
\begin{equation}\label{eq:cowling}
(A_{G}-E_G A_p^{-1} E_G\trans ) \tilde{u} = \omega_{C}^2  M \tilde{u} , 
\end{equation}
where $\omega_C$ is the frequency for Cowling approximation. 

\begin{table}[ht!]
    \centering
    \setlength{\tabcolsep}{1.pt}
    \begin{tabular}{cccccccc} \hline
        Exp. & \# of elm. & size of $A_G$ & size of $A_p$  & $[f_1,f_2]$ (mHz) & (deg,\#it) & \#eigs 
        \\ \hline  \hhline{========}
          E9p2   & 4,094,031 & 17,469,666    & 1,181,103   & $[2.0,2.5]$            & (4054,1892) & 528 \\  
          MIT\_2016May     & 4,048,932  & 16,578,945   & 879,067  & $[2.0,2.5]$            & (2674,1912) & 520 \\  
          MIT+crust 1.0  & 4,044,225  & 16,550,922   & 878,808  & $[2.0,2.5]$            & (6984,1912) & 550 \\  
        \hline
    \end{tabular}
    \caption{Test cases for four different Earth models using the Cowling approximation.}\label{tab:threemodels}
\end{table}

\begin{figure}
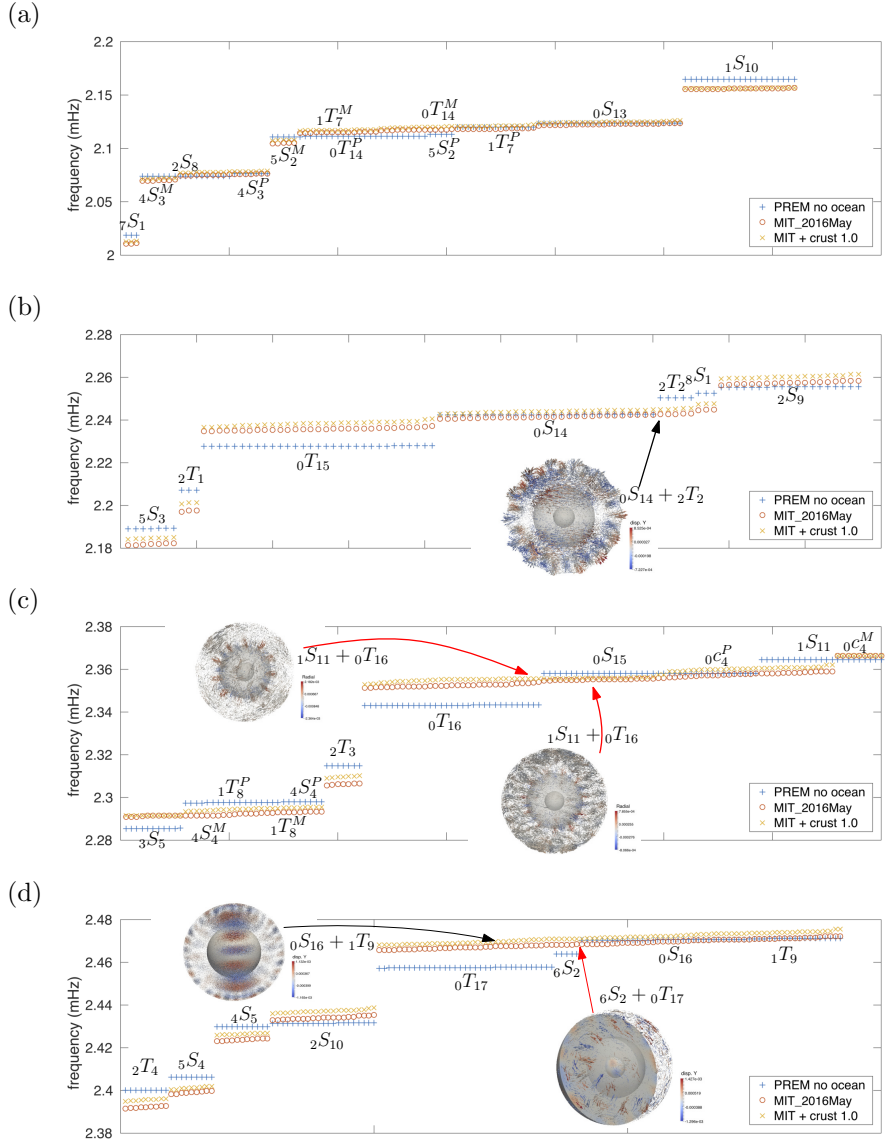

\centering
\begin{tabular}{c}
\subfigimage[width=0.95\textwidth]{(a)}{MIT/freqs/MIT_freqs_1-124_edit-eps-converted-to.pdf}
\\
\subfigimage[width=0.95\textwidth]{(b)}{MIT/freqs/MIT_freqs_125-221_edit-eps-converted-to.pdf}
\\
\subfigimage[width=0.95\textwidth]{(c)}{MIT/freqs/MIT_freqs_222-361_edit-eps-converted-to.pdf}
\\
\subfigimage[width=0.95\textwidth]{(d)}{MIT/freqs/MIT_freqs_362-503_edit-eps-converted-to.pdf}
\end{tabular}
\captionof{figure}{Comparisons between different Earth models in the
  Cowling approximation. The results from PREM without ocean, the MIT
  model, and the MIT model with the three-dimensional crust are shown
  using blue $+$, red $\circ$ and yellow $\times$, respectively. The
  superscripts $P$, $M$ on the mode symbols denote PREM and MIT
  models, respectively. (a-d) Comparison for different modes in
  $[2.0,2.18]$, $[2.18,2.28]$, $[2.28,2.38]$ and $[2.38,2.48]$ mHz,
  respectively. The mode in (b) couples ${}_0S_{14}$ with
  ${}_2T_2$. The two modes in (c) couple ${}_1S_{11}$ with
  ${}_0T_{16}$. The left mode in (d) couples ${}_0S_{16}$ with
  ${}_1T_9$. The right mode in (d) couples ${}_6S_{2}$ with
  ${}_0T_{17}$.} \label{fig:modesMIT0}
\end{figure}

In Table~\ref{tab:threemodels}, we show three different Earth models 
using the Cowling approximation. 
We construct two three-dimensional Earth models using MIT's mantle
tomographic results \cite{burdick2017model} and crust 1.0
\cite{laske2013update}.  The core model is based on PREM.  The
mantle seismic reference wave speeds are based on AK135
\cite{kennett1995constraints}. One model is obtained by combining
MIT's mantle tomographic model and PREM for the core and
density. The other one replaces PREM's crust by crust 1.0, which is
shown in Fig.~\ref{fig:MIT3D}. In the first three rows of Table~\ref{tab:threemodels}, we
show the information of three different tests for these three
different Earth models. Since with similar degrees of freedom, the
largest eigenvalue of the MIT model with the three-dimensional crust is much larger
than these of the other two models, we expect that significant
mode coupling and splitting occur 
\cite{deuss2001theoretical,romanowicz2008computation,beghein2008signal,irving2009normal,koelemeijer2012normal,nader2015normal,yang2015synthetic,akbarashrafi2017exact,al2018hamilton}.

We visualize different modes. The normal modes computed in the two
MIT models are non-degenerate. In Fig.~\ref{fig:modesMIT0}, we compare
different modes computed in the three models in the frequency range
$[2.0,2.5]$ mHz. Since the background models have only slight
differences, some of the eigenfrequencies are similar amongst PREM and
the MIT models. We illustrate most of the modes computed in PREM.  In
Fig.~\ref{fig:modesMIT0} (a), we observe that, even at low
frequencies, weak mode splitting occurs for surface wave modes,
including ${}_2S_8$, ${}_0S_{13}$, ${}_0T_{14}$ and ${}_1T_7$. We also
report that no coupled modes are observed in $[2.0,2.18]$ mHz. In
Figs.~\ref{fig:modesMIT0} (b-d), we show the different modes in
$[2.18,2.28]$, $[2.28,2.38]$ and $[2.38,2.48]$ mHz, respectively. The
splitting of most surface wave modes becomes larger with increasing
frequency. However, since modes like ${}_1S_{10}$ (strong at the
core-mantle boundary) in Fig.~\ref{fig:modesMIT0} (a), ${}_0c_4$ (an
inner core toroidal mode) and ${}_3S_5$ (an ICB Stoneley mode) in
Fig.~\ref{fig:modesMIT0} (c), are not sensitive to the crust and upper
mantle structure, no clear splitting is observed. We observe coupled
modes in Figs.~\ref{fig:modesMIT0} (b-d) computed in the MIT model
with the three-dimensional crust. The eigenfunction of one mode in
Fig.~\ref{fig:modesMIT0} (b) shows that ${}_0S_{14}$ and ${}_2T_2$ are
coupled. The ${}_0T_{15}$ and ${}_8S_1$ near ${}_0S_{14}$ and
${}_2T_2$ are isolated multiplets. The eigenfunctions of the two modes
in Fig.~\ref{fig:modesMIT0} (c) show that ${}_1S_{11}$ and
${}_0T_{16}$ are coupled. The ${}_0S_{15}$ near ${}_1S_{11}$ and
${}_0T_{16}$ is an isolated multiplet. These coupled modes are
interesting because ${}_1S_{11}$ is clearly sensitive to the
core-mantle boundary and the fundamental Love mode ${}_0T_{16}$
illustrated can be measured at the surface. The left mode in
Fig.~\ref{fig:modesMIT0} (d) is a ${}_0S_{16}$ and ${}_1T_9$ coupled
mode.  The right mode in Fig.~\ref{fig:modesMIT0} (d) is a ${}_6S_{2}$
and ${}_0T_{17}$ coupled mode. This mode is also very interesting
because ${}_6S_{2}$ illustrated is an inner core mode and the
fundamental Love mode ${}_0T_{17}$ illustrated can be detected at the
surface. Since the relative wave speed variations of the MIT
tomographic model vary roughly from -1.4\% to 1.4\% in the upper
mantle and the crust's thickness is small, strong mode coupling occurs
only to two modes. In this frequency range $[2.0, 2.5]$ mHz, the width
of each multiplet is small and no significant coupling between three
or more modes is observed.

\section{Computational experiments for rotating planets}
\label{sec:comprotation}

In this section, we include the rotation and study its effects on normal modes. 
To simplify \eqref{eq:eigfull} and \eqref{eq:solideigfull}
without any loss of the generality, we extend \eqref{eq:gep} 
and derive a standard form for the QEP,
\begin{equation} \label{eq:simplequad}
   \omega^2 M \tilde{u} - 2 \ii \omega \tilde{R}_{\Omega} \tilde{u}
                        - A \tilde{u} = 0 . 
\end{equation}
We note that $\tilde{R}_{\Omega} = - \tilde{R}\trans_{\Omega}$, that
is, $2 \ii \tilde{R}_{\Omega}$ is Hermitian.  The eigenfrequencies are
real and come in pairs $(\omega,-\omega)$.

To solve the QEP of the original form, the QEP is often projected onto
a properly chosen low-dimensional subspace to facilitate the reduction
to a QEP directly of lower dimension, such as in the Jacobi–Davidson
method \cite{sleijpen1996jacobi,sleijpen1996quadratic}. The reduced
QEP can then be solved by a standard dense matrix technique. Both
Arnoldi- and Lanczos-type processes \cite{hoffnung2006krylov} have
been developed to build such projections of the QEP. A subspace
approximation method \cite{holz2004subspace} was presented via
applying perturbation theory to the QEP. A second-order Arnoldi
procedure \cite{bai2005soar} was developed to generate an orthonormal
basis for solving a large-scale QEP directly. We note that the above
mentioned methods typically utilize a shift-and-invert scheme for
solving the interior eigenpairs. These techniques become
impractical for eigenvalue problems of the size of ours due to the high memory costs.

Instead, we can utilize extended Lanczos vectors from  
solving the generalized eigenvalue problem \eqref{eq:gep} through the
polynomial filtering method. We then approximate the solution
$\tilde{u}$ using the basis computed from
\begin{equation} \label{eq:linearpb}
   A X_e = M X_e \Lambda_e , 
\end{equation}
where $X_e$ stands for the Ritz vectors of the linear system and
$\Lambda_e$ denotes a diagonal matrix whose diagonal is a collection
of $\omega_{N}^2$ in \eqref{eq:gep}. We take $m_e$ eigenvectors
spanning a subspace and let $\tilde{u}_{e} = X_e y_e$ approximate
$\tilde{u}$ in \eqref{eq:simplequad}, where $y_e$ is complex.  We
apply $$\left( \begin{array}{cc} X_e\trans & 0 \\0 &
  X_e\trans \end{array}\right)$$ to an equivalent form of
\eqref{eq:simplequad},
\[
\left( \begin{array}{cc}
			0         &  	A     \\
			A       & 2\ii \tilde{R}_{\Omega}                 
\end{array} \right) 
\left( \begin{array}{c}
\tilde{u} \\
\omega \tilde{u}
\end{array}	\right)	= \omega 
\left( \begin{array}{cc}
		   A    &   0      \\
			0       &  M                 
\end{array} \right) 	
\left( \begin{array}{c}
\tilde{u} \\
\omega \tilde{u}
\end{array}	\right)	. 
\]  
Making use of $X_e\trans A X_e = \Lambda_e$, we obtain
\begin{equation} \label{eq:sublinear0}
\left( \begin{array}{cc}
			0         &  \Lambda_e       \\
			\Lambda_e          & 2\ii X_e\trans \tilde{R}_{\Omega} X_e                 
\end{array} \right) 
\left( \begin{array}{c}
y_e \\
\omega_e y_e
\end{array}	\right)			= \omega_e 
\left( \begin{array}{cc}
\Lambda_e         &   0      \\
			0       &  I                 
\end{array} \right) 	
\left( \begin{array}{c}
y_e \\
\omega_e y_e
\end{array}	\right) . 
\end{equation}
It is apparent that if $\tilde{R}_{\Omega}=0$, we have $\omega_e=
\omega_{N} = \Lambda_e^{1/2}$. The system \eqref{eq:sublinear0} can
be solved with a standard eigensolver such as the one implemented in 
\texttt{LAPACK} \cite{anderson1999lapack}.
Here, we study the spectra of two models: Earth 1066A
\cite{gilbert1975application} and a Mars model
\cite{khan2016single}. We use 23.9345 hours
\cite{allen1973astrophysical} and 24.6229 hours
\cite{lodders1998planetary} as Earth's and Mars' rotation periods, respectively.
With a large $m_e$ and a relatively small $\Omega$, 
the numerical solution $\omega_e$ is close to $\omega$ in \eqref{eq:simplequad}. 
The numerical accuracy can further be improved via solving \eqref{eq:simplequad} exactly.

\subsection{Computational accuracy}
\label{sec:accuracy}

\begin{table}[ht!]
\centering
\begin{tabular}{cccccc}\hline
Exp. & \# of elm. & size of $A$ & size of $A_p$  & size of $S$ & $[f_1,f_2]$ (mHz)   \\ \hhline{======}
Constant (C3kp1) & 3,129  & 1,821 & 0      & 3,521   & [0.35,0.85] \\
Earth (E3kp1)    & 3,330  & 2,760 & 392    & 4,242   & [0.3,0.86]   \\ 
Mars (M2kp1)     & 1,887  & 1,677 & 145    & 2,539   & [0.4,1.14]    \\
Mars (M8kp1)     & 8,020  & 7,557 & 152    & 12,436  & [0.4, 1.14]  \\
Earth (E40kp1)   & 42,828 & 30,384& 3,171  & 51,000  & [0.1,1.5]   \\ 
\hline
\end{tabular}
\caption{Numerical parameter values pertaining to the testing of
  computational accuracy and estimating the cost in different models.}
\label{tab:accuracy0}
\end{table}

For small models, we are able to compute the full mode expansion
associated with the point spectrum using \eqref{eq:sublinear0}. In
Table~\ref{tab:accuracy0}, we list the numerical parameter values
pertaining to the testing of computational accuracy and estimating the
cost in different models: The number of elements (labeled as \# of
elm.), size of $A_p$, size of $A$, size of $S$ and the target
frequency interval in milliHertz (labeled as $[f_1,f_2]$ (mHz)).

\begin{figure}[ht!]
\centering
\begin{tabular}{ccc}
\includegraphics[trim= 0cm 1cm 0cm 0cm,clip=true,width=0.30\textwidth]{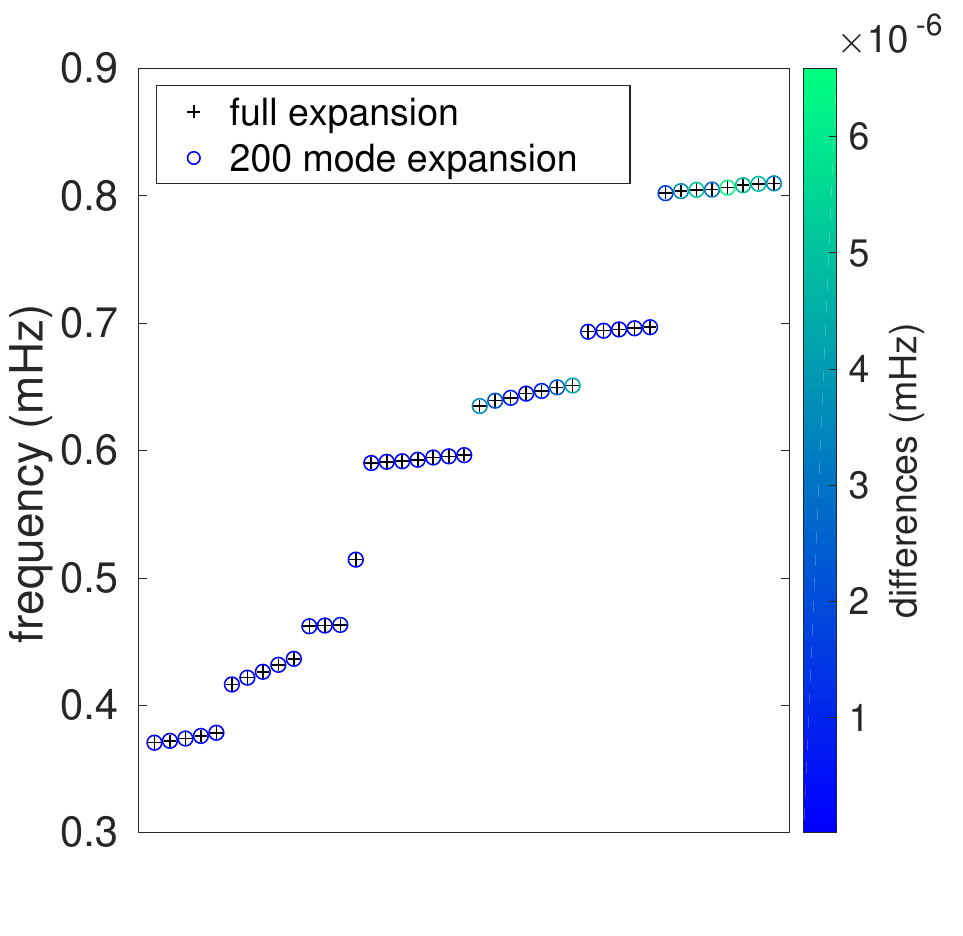} &
\includegraphics[trim= 0cm 1cm 0cm 0cm,clip=true,width=0.30\textwidth]{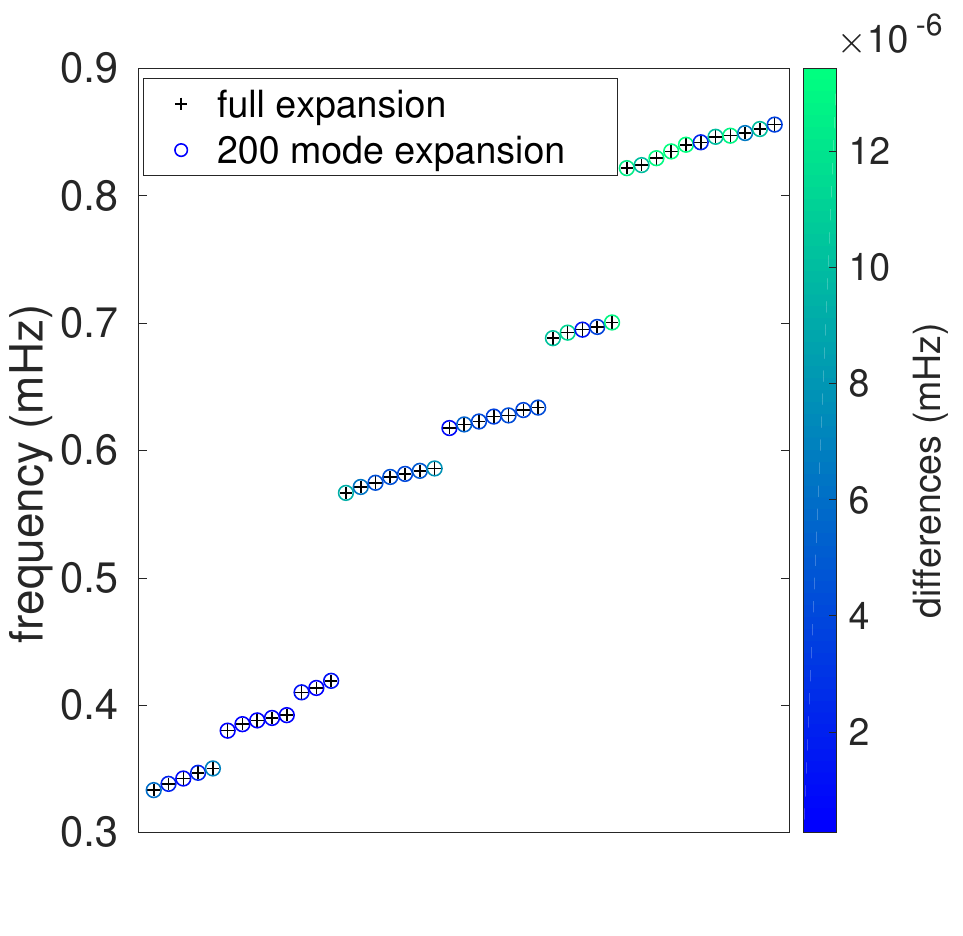} &
\includegraphics[trim= 0cm 1cm 0cm 0cm,clip=true,width=0.30\textwidth]{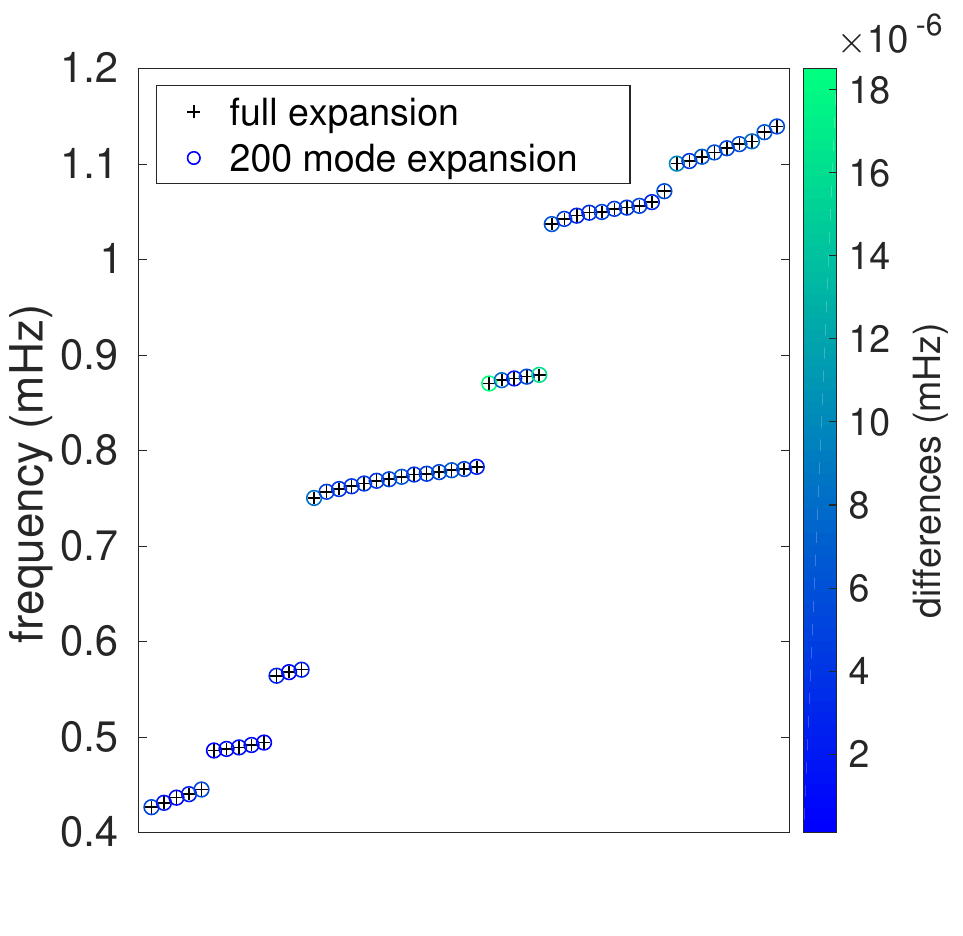} 
\\
(a) Constant (C3kp1) & (b) Earth (E3kp1) & (c) Mars (M2kp1)
\end{tabular}
\caption{Tests with three different small models for the low-frequency
  seismic eigenfrequencies. The numerical parameters of the tests are
  given in Table~\ref{tab:accuracy0}.}
\label{fig:threemodelsaccury}
\end{figure}

In Figs.~\ref{fig:threemodelsaccury} (a)--(c), we illustrate the
computational accuracy of tests in three different models, C3kp1, E3kp1
and M2kp1, respectively, on the lowest seismic eigenfrequencies using
P1 elements. We compare the differences in the eigenfrequencies
between the full mode expansion and a 200 mode expansion. 
The differences are about $5\times 10^{-6}$ mHz, which is two digits below the accuracy of common
normal mode measurements.

\begin{figure}[ht!]
\centering
\begin{tabular}{cc}
\includegraphics[trim= 0cm 1cm 0cm 0cm,clip=true,width=0.36\textwidth]{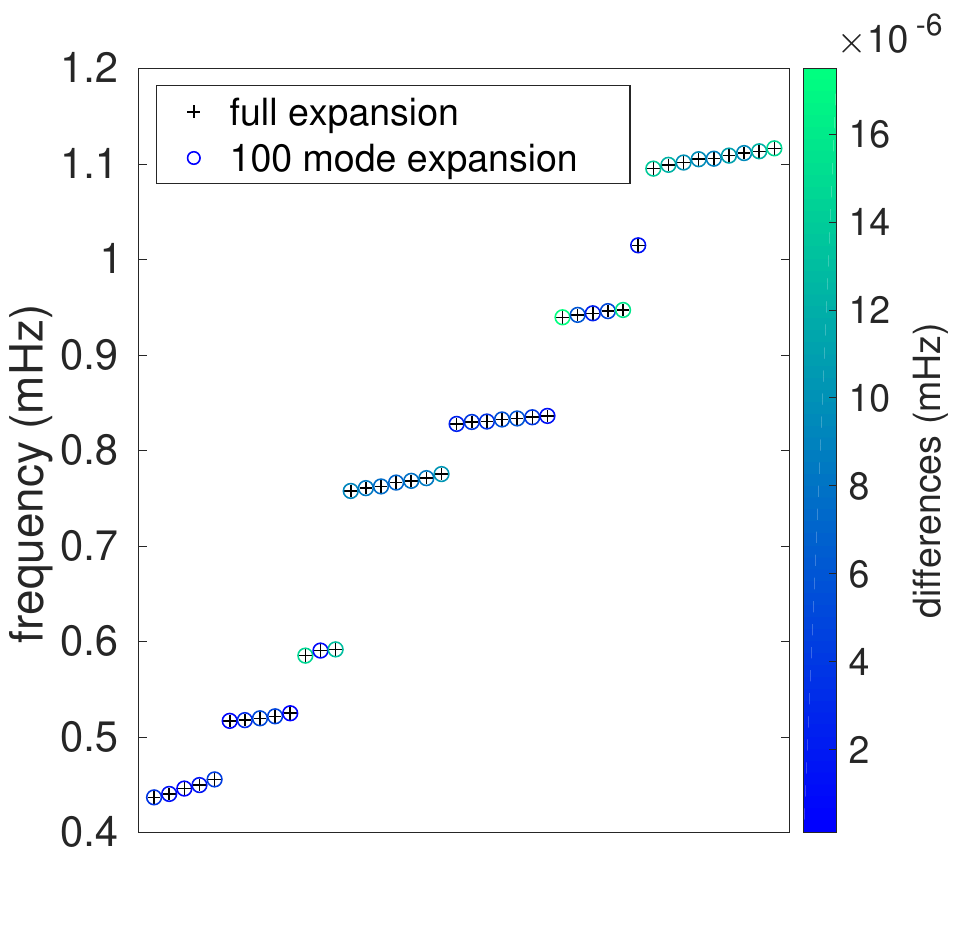} &
\includegraphics[trim= 0cm 1cm 0cm 0cm,clip=true,width=0.36\textwidth]{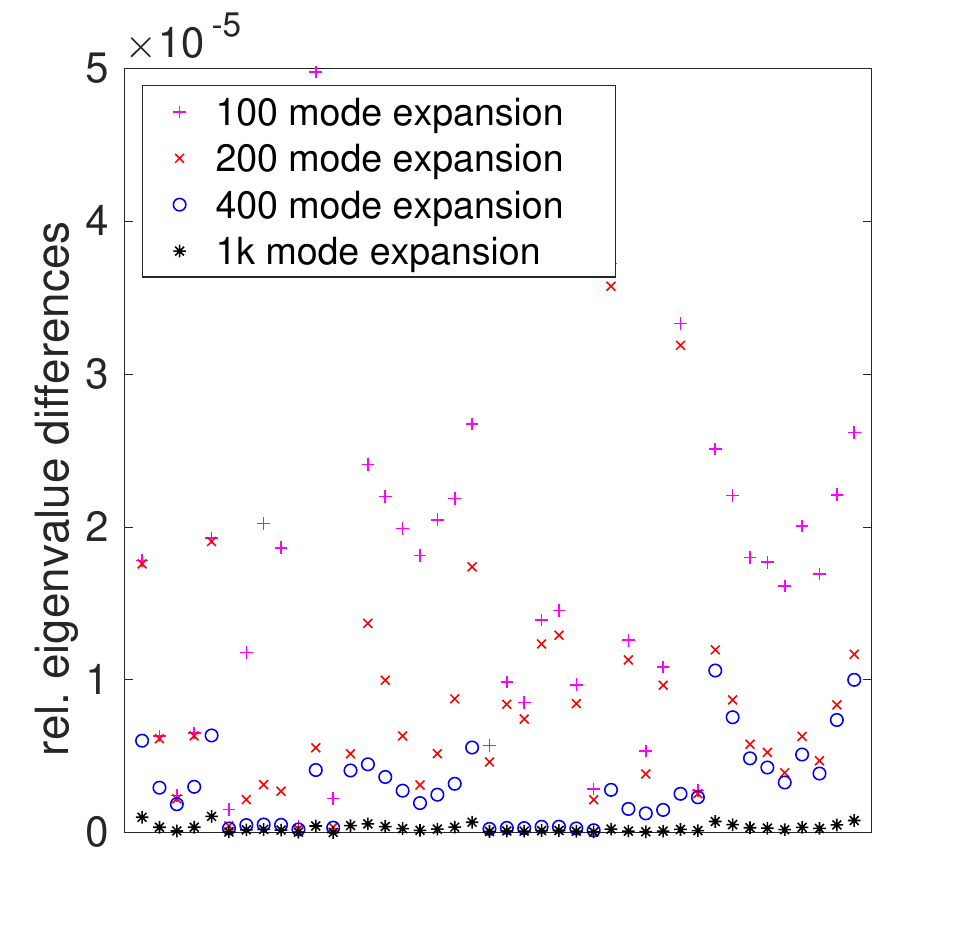} \\ 
(a) M8kp1 on [0.4, 1.14] mHz & (b) Errors of (a)  \\
\end{tabular}
\caption{Tests for computational accuracy of a Mars model using
  different numbers of mode expansion.}
\label{fig:mars8kaccuracy}
\end{figure}

In Figs.~\ref{fig:mars8kaccuracy} (a) and (b), we show the
computational accuracy of M8kp1 on [0.4, 1.14] mHz as well as the
error distribution. In Fig.~\ref{fig:mars8kaccuracy} (a), we show that
even with a 100 mode expansion, the differences are as low as $1
\times 10^{-5}$ mHz. In Fig.~\ref{fig:mars8kaccuracy} (b), we show
that with a 1000 mode expansion, the differences are further reduced
to about $1 \times 10^{-6}$ mHz.

\subsection{Benchmark experiments for Earth models with rotation}
\label{sec:earth}

\begin{figure}[ht!]
\centering
\begin{tabular}{c}
\includegraphics[trim= 0cm 0cm 0cm 0cm,clip=true,width=0.80\linewidth]{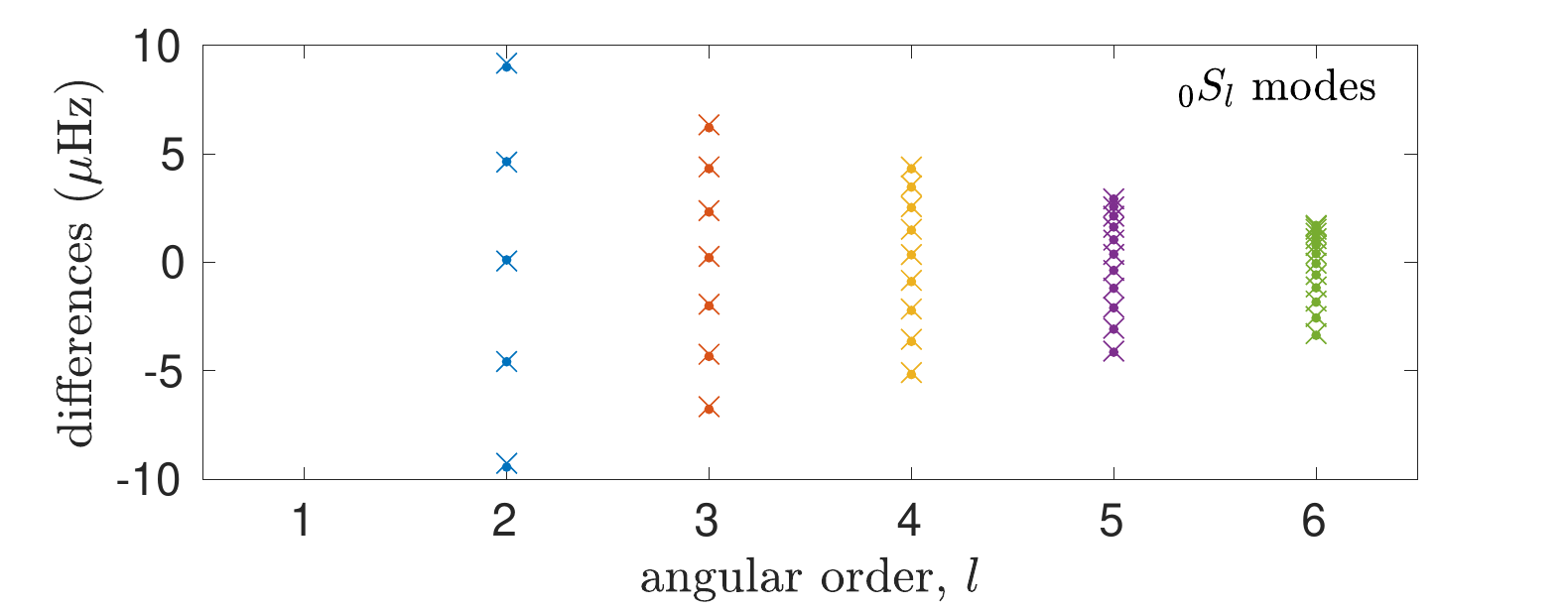} \\
(a) Comparison of ${}_0S_l$ modes \\
\includegraphics[trim= 0cm 0cm 0cm 0cm,clip=true,width=0.80\linewidth]{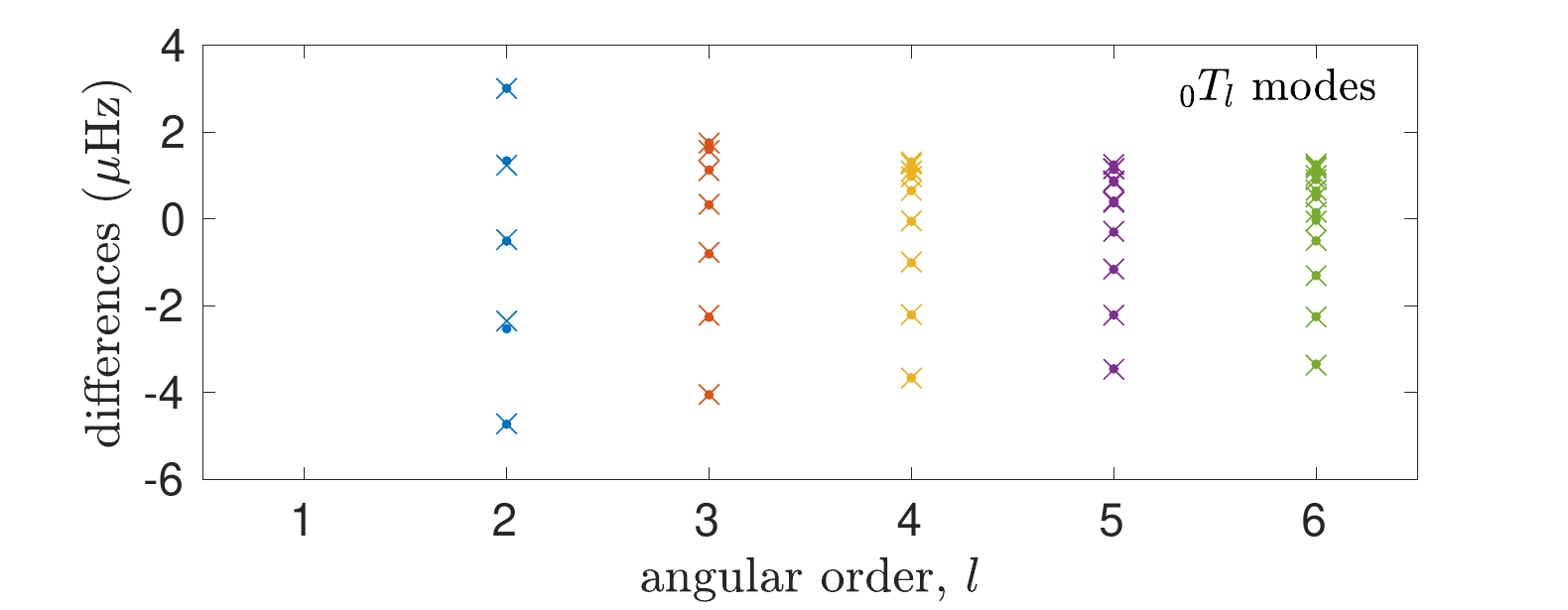} \\ 
(b) Comparison of ${}_0T_l$ modes \\
\includegraphics[trim= 0cm 0cm 0cm 0cm,clip=true,width=0.80\linewidth]{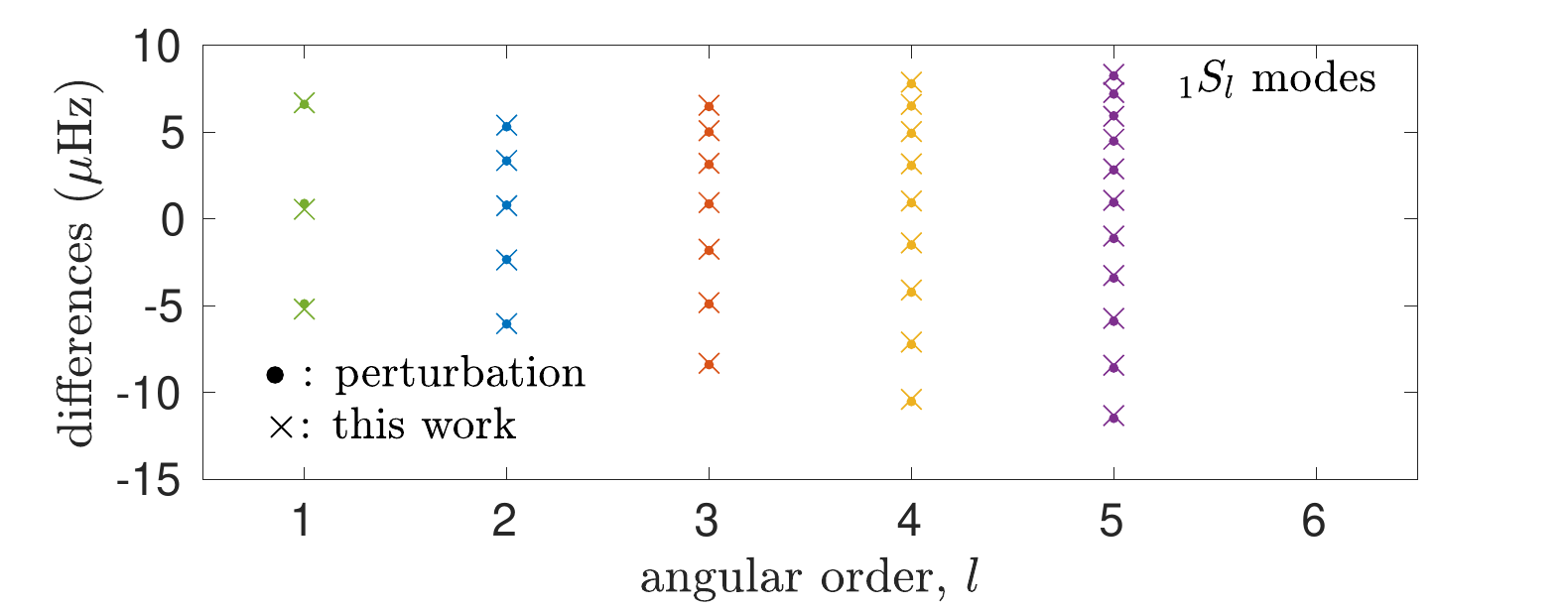} \\
(c) Comparison of ${}_1S_l$ modes 
\end{tabular}
\caption{Comparison of the results from a perturbation calculation and
  our proposed method, which are shown using symbols $\bullet$ and
  $\times$, respectively.  (a), (b) and (c) illustrate comparisons of
  ${}_0S_l$, ${}_0T_l$ and ${}_1S_l$ modes, respectively.}
\label{fig:e1066acomparison}
\end{figure}

Over the past two decades, a significant number of observational
studies have been carried out to the rotation effects on the Earth's normal modes 
\cite{zurn2000observation,millot2003normal,park2005earth,roult2010observation,nader2015normal,schimmel2018low}. 
Our computational approach can aid and complement such studies through
accurate and consistent simulations generating even relatively high
eigenfrequencies. Here, we perform a benchmark experiment of Earth
model 1066A \cite{gilbert1975application} against a perturbation
calculation \cite{dahlen1979rotational}. In the perturbation
calculation, the eigenfrequency perturbations $\delta \omega_m$ have
the following form
\begin{equation} \label{eq:pertbQuadform}
   \delta \omega_m = \omega_0 (a+bm+cm^2), \quad -l \leq m \leq l ,
\end{equation}
where $\omega_0$ denotes the eigenfrequency of the unperturbed
spherically symmetric model, $l$ denotes the angular order in the
spherical harmonic expansion, and $a$, $b$ and $c$ are the relevant
coefficients. The values of $a$, $b$ and $c$ for different radial
modes can be found in \cite[Table 14.1]{dahlen1998theoretical}. In
Table~\ref{tab:bmtests}, we list the numerical parameters of the Earth
models in the benchmark test. The models E1Mp1 and E2Mp2 used to
compute $\omega_0$ represent spherically symmetric ones without
rotation. Experiments EE1Mp1 and EE2Mp2 represent elliptic Earth
models and are used to compute eigenfrequencies with our proposed
method. The ellipticities of the Earth models are computed by solving
Clairaut's equation (cf.~Section~\ref{sec:liquidcore}). Since the
eigenfrequencies of the Slichter modes \cite{slichter1961fundamental}
are close to the upper bound of the essential spectrum and the
convergence of the proposed algorithm is relatively slow, we set
$f_1=$ 0.04 mHz and use experiments E1Mp1 and EE1Mp1 to compute the
Slichter modes using P1 elements. Experiments E2Mp2 and EE2Mp2 are
used to compute other modes using P2 elements. In
Fig.~\ref{fig:e1066acomparison}, we show the comparison between the
perturbation and our methods. The values of the computed
eigenfrequenies of our method agree with the perturbation results in
as much as that the relative differences are commonly less than 0.3
$\mu$Hz. The degree of agreement is, of course, model dependent. The
eccentricity in the Earth model is so small that the second-order
perturbation is accurate within the typical error of our numerical
computations. Higher rotation rates would increase the eccentricity
and let the second-order perturbation loose accuracy.

\begin{table}[ht!]
\centering
\begin{tabular}{ccccccc}\hline
Exp. & \# of elm. & size of $A$ & size of $A_p$ & size of $S$ & $[f_1,f_2]$ (mHz)  
\\ \hhline{=======}
Earth (E1Mp1)  & 1,011,973 & 537,198   & 31,849    & 1,074,577 & [0.04,1.5]   \\ 
Earth (E2Mp2)  & 2,015,072 & 8,569,197 & 530,721   & 2,165,360 & [0.2,1.5]   \\ 
Earth (EE1Mp1) & 1,003,065 & 533,064   & 31,688    & 1,065,629 & [0.04,1.5]   \\ 
Earth (EE2Mp2) & 2,002,581 & 8,520,432 & 528,124   & 2,153,109 & [0.2,1.5]   \\ 
\hline
\end{tabular}
\caption{Numerical parameters of the Earth models used in the
  benchmark experiments.} \label{tab:bmtests}
\end{table}

\subsection{Mars models}
\label{sec:mars}

Here, we present our computational results for Mars models. The
interiors of the Mars models are based on mineral physics calculations
\cite{khan2016single}. In Table~\ref{tab:marstests}, we list three
Mars models labeled as M2Mp2, EM2Mp2 and TM2Mp2 which represent a
spherically symmetric Mars model without rotation, a spheroidal Mars
model with rotation, and a spheroidal Mars model with a
three-dimensional crust and rotation using P2 elements. The shape of
the spheroidal Mars model's core-mantle boundary is computed by
solving Clairaut's equation. Since Mars presumably is not hydrostatic
as discussed in Section~\ref{sec:liquidcore}, its solid region is
estimated via a linear interpolation using the ellipticities of the
core-mantle boundary ($\varepsilon=4.19\times 10^{-3}$) and the
surface ($\varepsilon=5.89\times 10^{-3}$). Model TM2Mp2 is illustrated
in Fig.~\ref{fig:vpvsrhomars}.

\begin{table}[ht!]
\centering
\begin{tabular}{cccccc}\hline
Exp. & \# of elm. & size of $A$ & size of $A_p$  & size of $S$ & $[f_1,f_2]$ (mHz)  \\ \hhline{======}
Mars (M2Mp2)  & 1,996,773 & 8,967,684 & 579,338  & 2,257,801  & [0.2,2.0]   \\ 
Mars (EM2Mp2) & 2,001,619 & 8,984,532 & 579,667  & 2,262,143  & [0.2,2.0]   \\ 
Mars (TM2Mp2) & 2,008,654 & 8,289,927 & 323,810  & 2,158,366  & [0.2,2.0]   \\ 
\hline
\end{tabular}
\caption{Numerical parameters for the Mars
  models.} \label{tab:marstests}
\end{table}

\begin{figure}[ht!]
\centering
\begin{tabular}{cc}
\includegraphics[trim= 0cm 1.5cm 1cm 0.5cm,clip=true,width=0.45\linewidth]{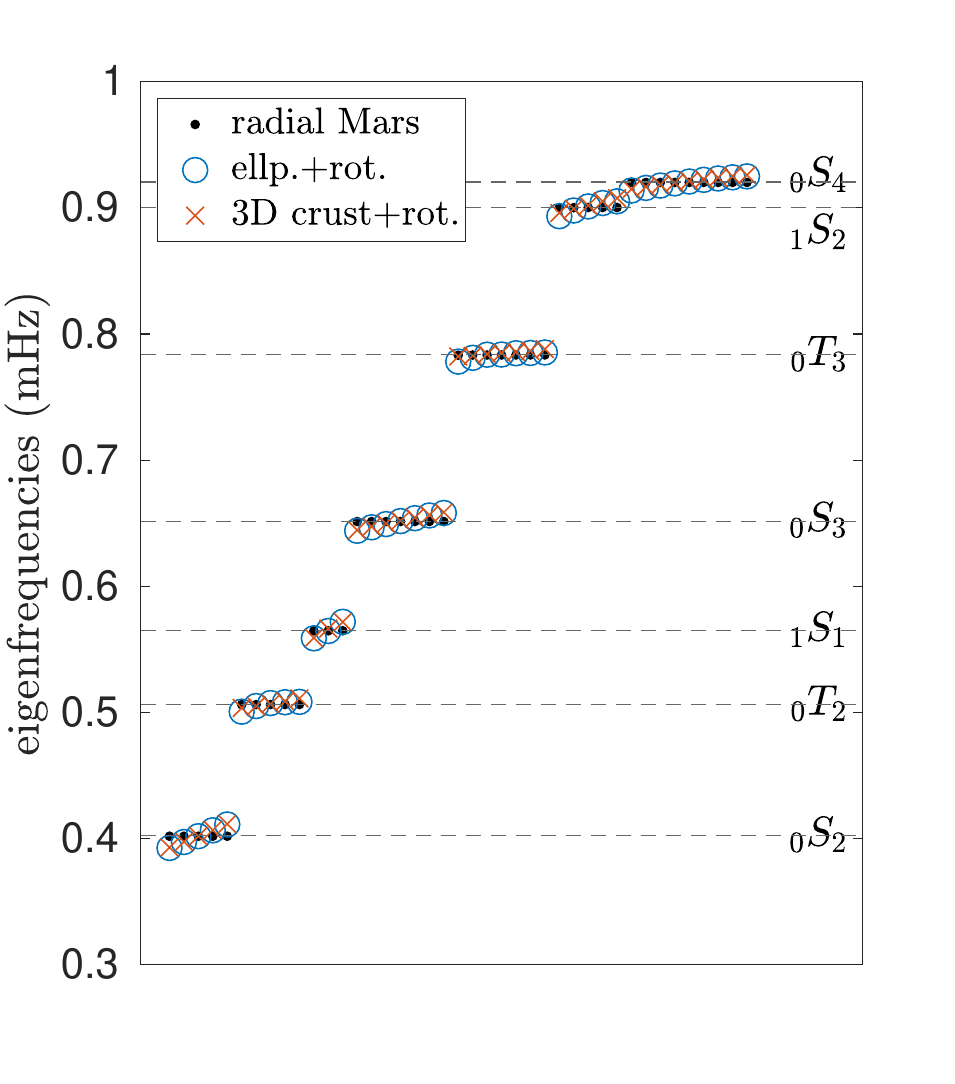} &
\includegraphics[trim= 0cm 1.5cm 1cm 0.5cm,clip=true,width=0.45\linewidth]{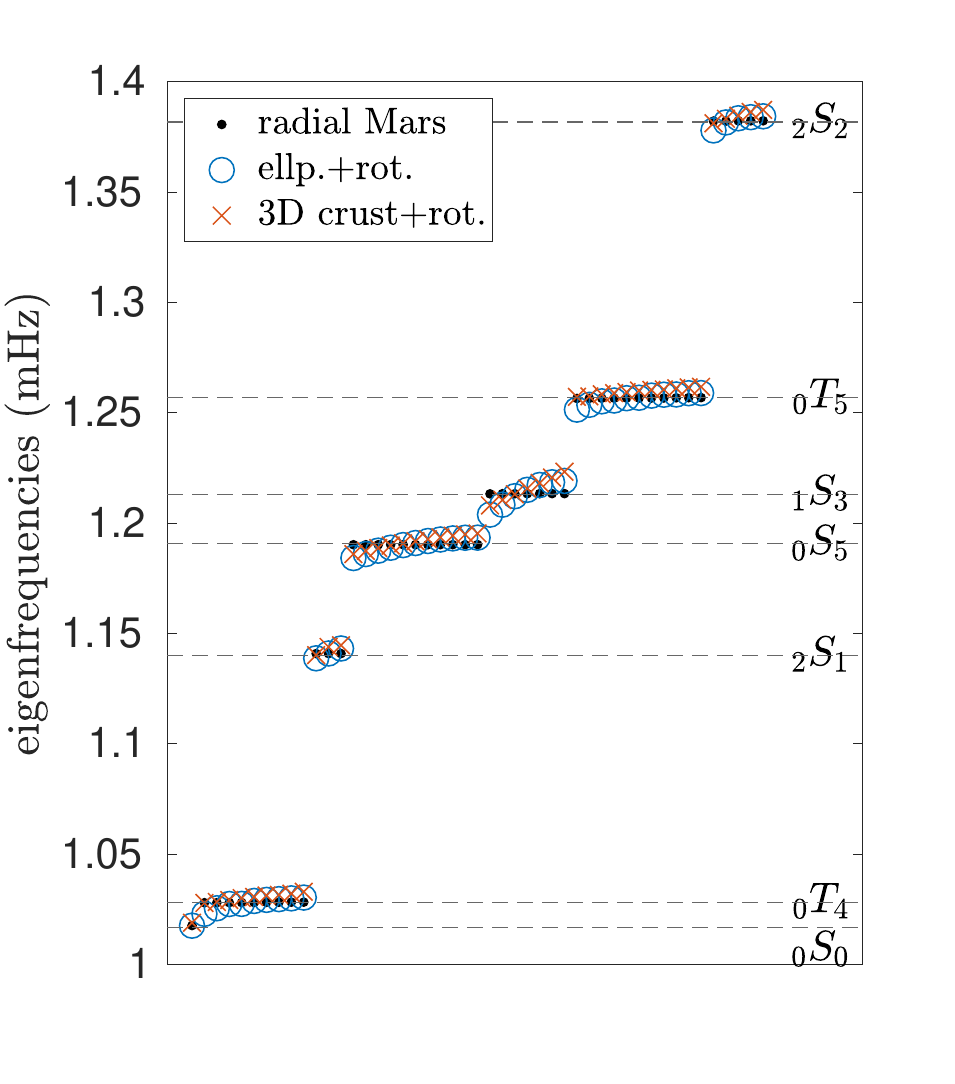} \\
(a) modes in [0.3, 1.0]mHz &  (b) modes in [1.0, 1.4]mHz \\ 
\includegraphics[trim= 0cm 1.5cm 1cm 0.5cm,clip=true,width=0.45\linewidth]{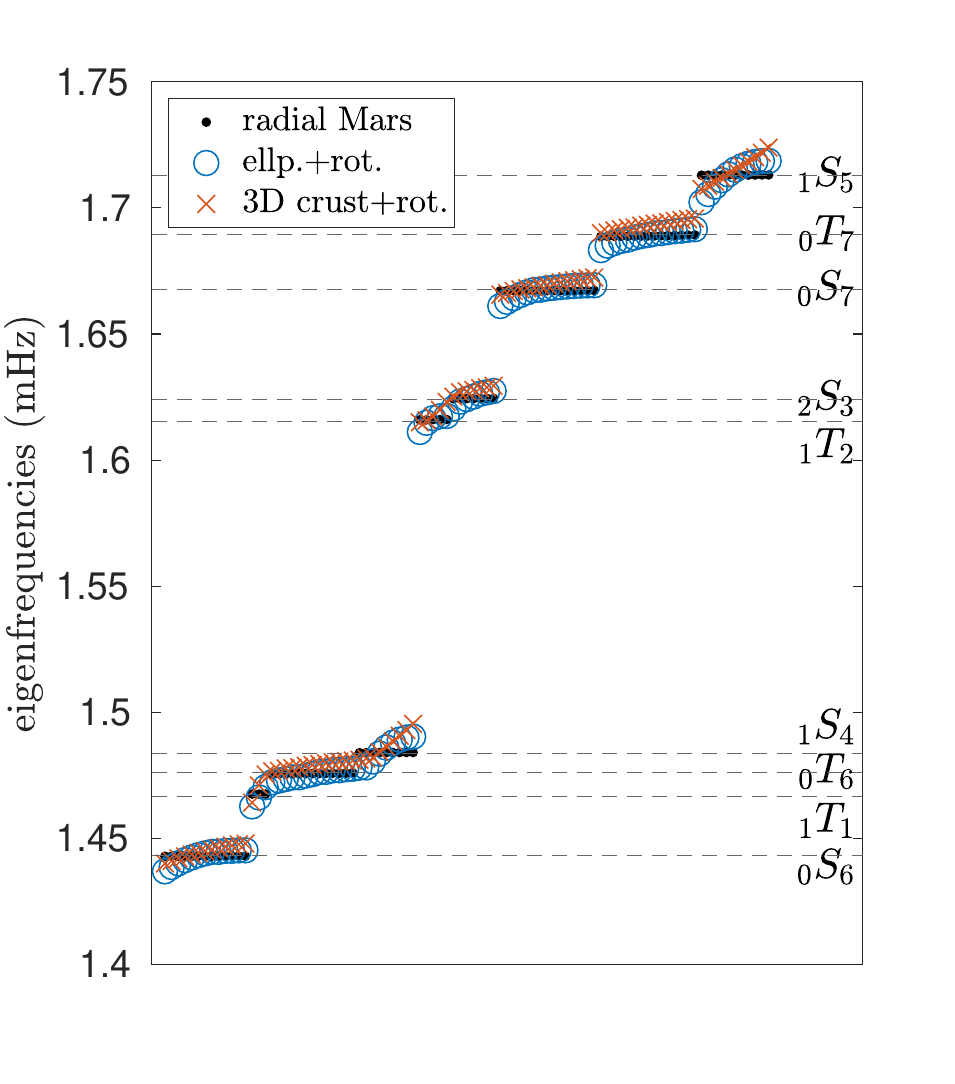} &
\includegraphics[trim= 0cm 1.5cm 1cm 0.5cm,clip=true,width=0.45\linewidth]{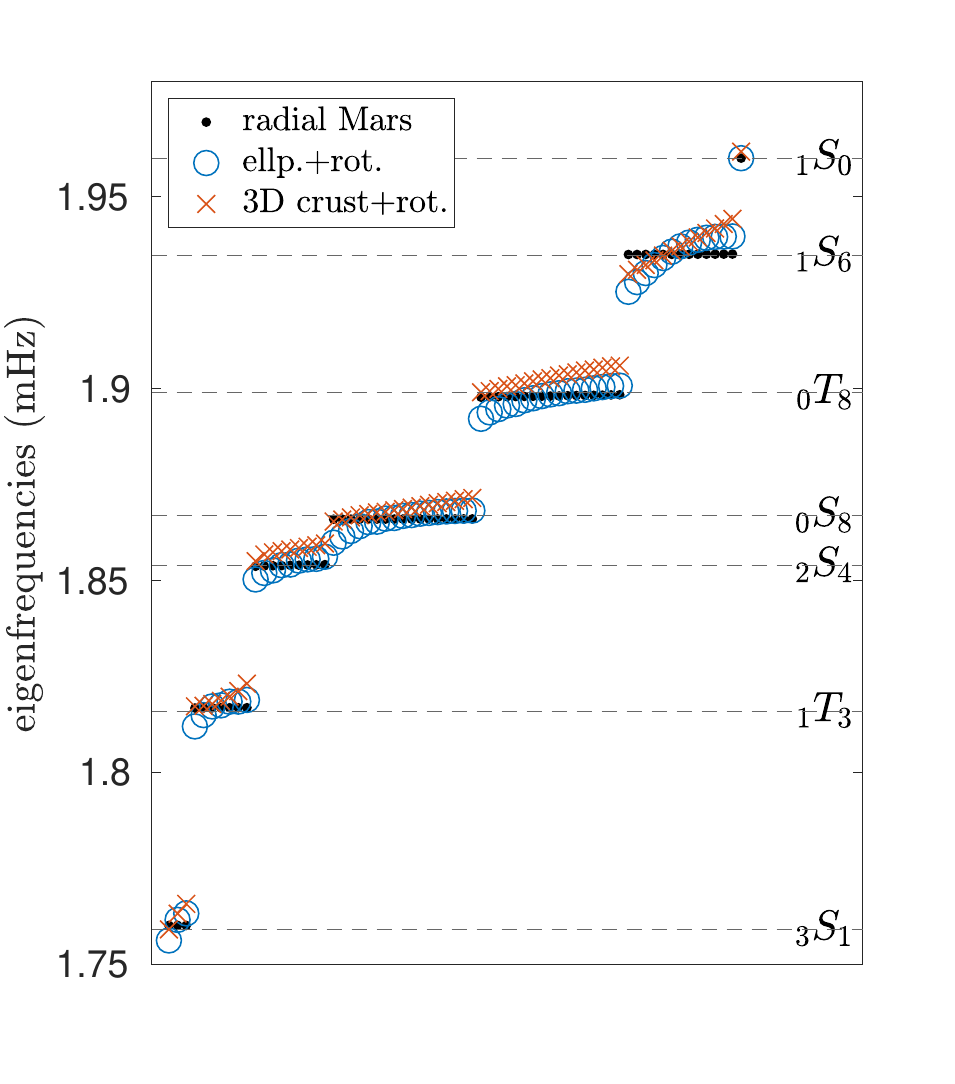} \\
(c) modes in [1.4, 1.75]mHz &  (d) modes in [1.75, 1.95]mHz
\end{tabular}
\caption{Eigenfrequencies of different Mars models. (a), (b), (c) and
  (d) illustrate eigenfrequencies in different frequency windows.
  Symbols $\bullet$, $\circ$ and $\times$ represent the
  eigenfrequencies computed from Mars models M2Mp2, EM2Mp2 and TM2Mp2
  in Table~\ref{tab:marstests}, respectively. 
  The x-axis indicates the indexes of eigenfrequencies with ascending order. 
  The horizontal dashed lines represent the eigenfrequencies of a spherically symmetric Mars
  model computed with a one-dimensional solver.} \label{fig:marsresults0}
\end{figure}

\begin{figure}[ht!]
\centering
\includegraphics[trim= 0cm 0.5cm 1cm 0.5cm,clip=true,width=0.8\linewidth]{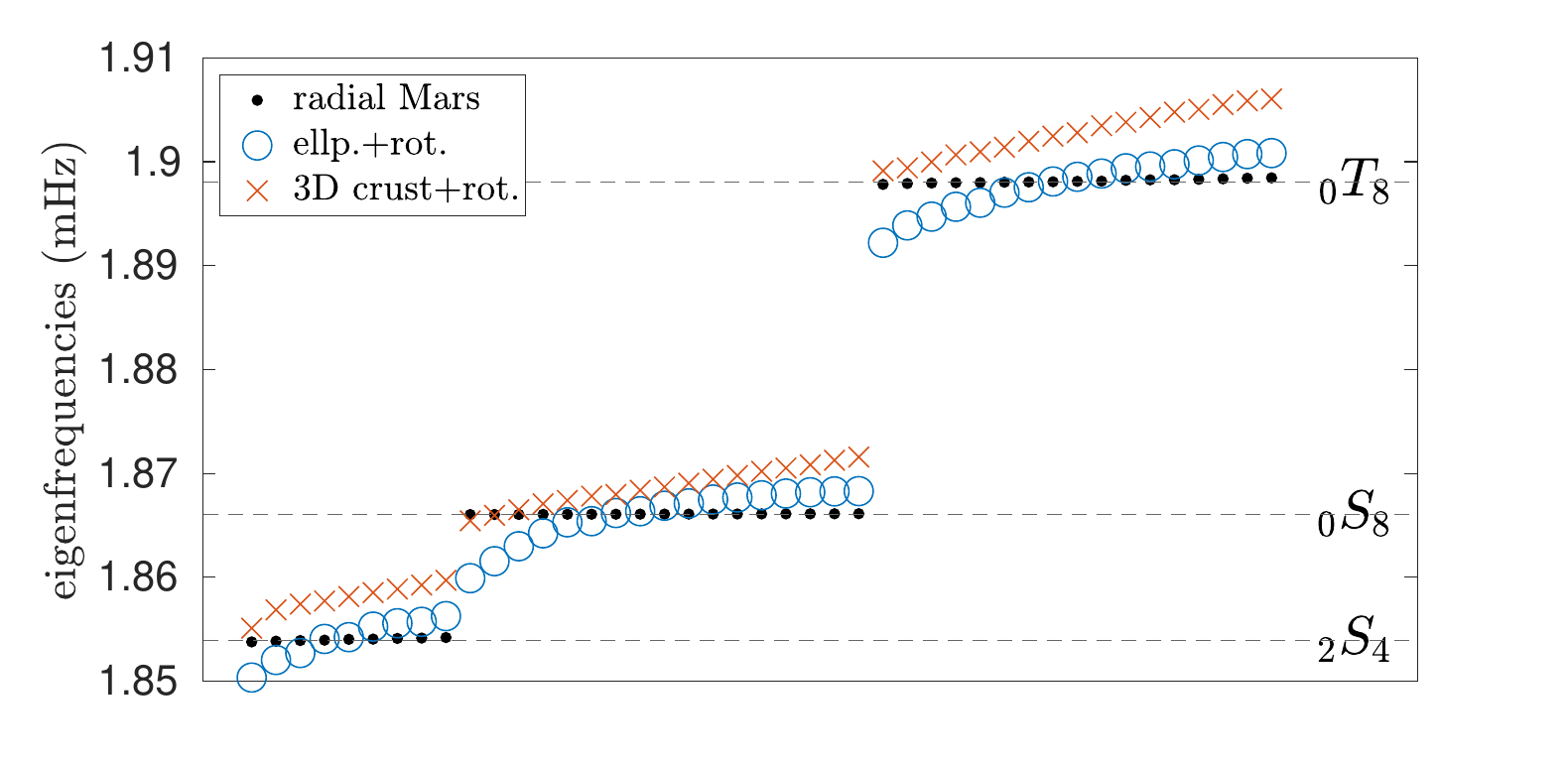}
\caption{Illustration of a subinterval in Fig.~\ref{fig:marsresults0}
  (d).  The x-axis indicates the indexes of eigenfrequencies with
  ascending order.  Splitting of modes ${}_2S_4$, ${}_0S_8$, and
  ${}_0T_8$ due to the three-dimensional crust. The maximum difference
  between the eigenfrequencies is 5.2 $\mu$Hz.}
\label{fig:highlight}
\end{figure}

In Fig.~\ref{fig:marsresults0}, we show eigenfrequencies computed in
different Mars models listed in Table~\ref{tab:marstests}. Symbols
$\bullet$, $\circ$ and $\times$ represent the eigenfrequencies
computed in Mars models M2Mp2, EM2Mp2 and TM2Mp2
(cf.~Table~\ref{tab:marstests}). The horizontal dashed lines represent
the eigenfrequencies of a spherically symmetric Mars model computed
with a one-dimensional solver
\cite{masters2011mineos,jingchen2018revisiting}. Mode splitting is
apparent due to ellipticity, rotation and heterogeneity in three
dimensions. The three-dimensional crust does not have a clear
influence on the lowest eigenfrequencies associated with ${}_0S_2$,
${}_0T_2$, ${}_1S_1$, ${}_0S_3$, ${}_0T_3$, ${}_1S_2$ and ${}_0S_4$ in
Fig.~\ref{fig:marsresults0} (a). The three-dimensional crust has a
noticeable effect on the surface wave modes, such as ${}_0T_6$,
${}_0T_7$, ${}_0T_8$, ${}_0S_6$, ${}_0S_7$ and ${}_0S_8$, as
expected. In Fig.~\ref{fig:highlight}, we show the eigenfrequencies in
a subinterval of the interval used in Fig.~\ref{fig:marsresults0}
(d). Here, we note the splitting of modes ${}_2S_4$, ${}_0S_8$ and
${}_0T_8$ and highlight the effects of the three-dimensional
crust. The maximum difference between the eigenfrequencies in
Fig.~\ref{fig:highlight} is 5.2 $\mu$Hz, which, in principle, can be
detected. There is no mode-coupling observed in these experiments.  

\begin{figure}[ht!]
\centering
\begin{tabular}{ccc}
\includegraphics[trim= 5cm 0cm 0cm 0cm,clip=true,width=0.30\linewidth]{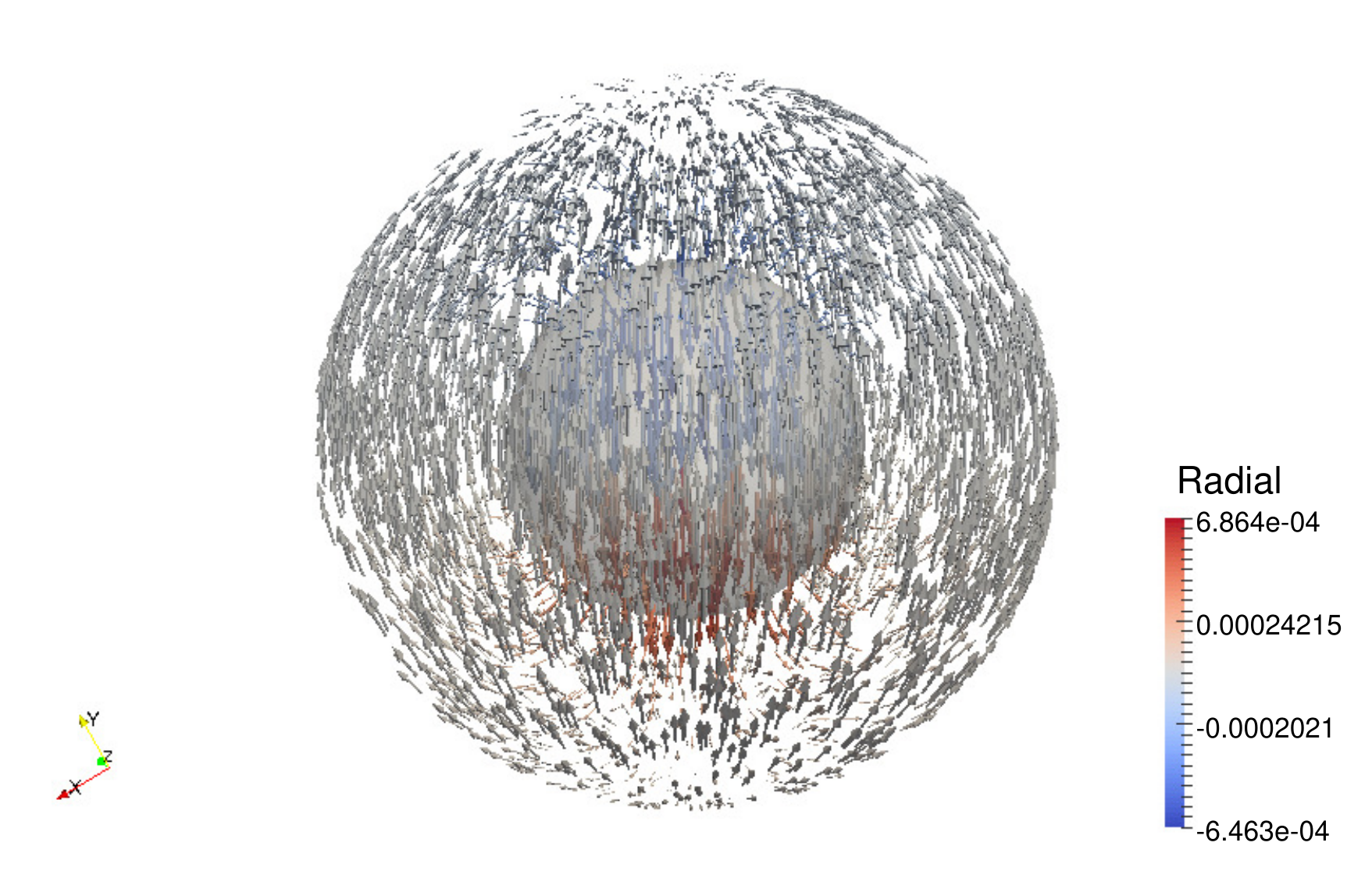} &
\includegraphics[trim= 5cm 0cm 0cm 0cm,clip=true,width=0.30\linewidth]{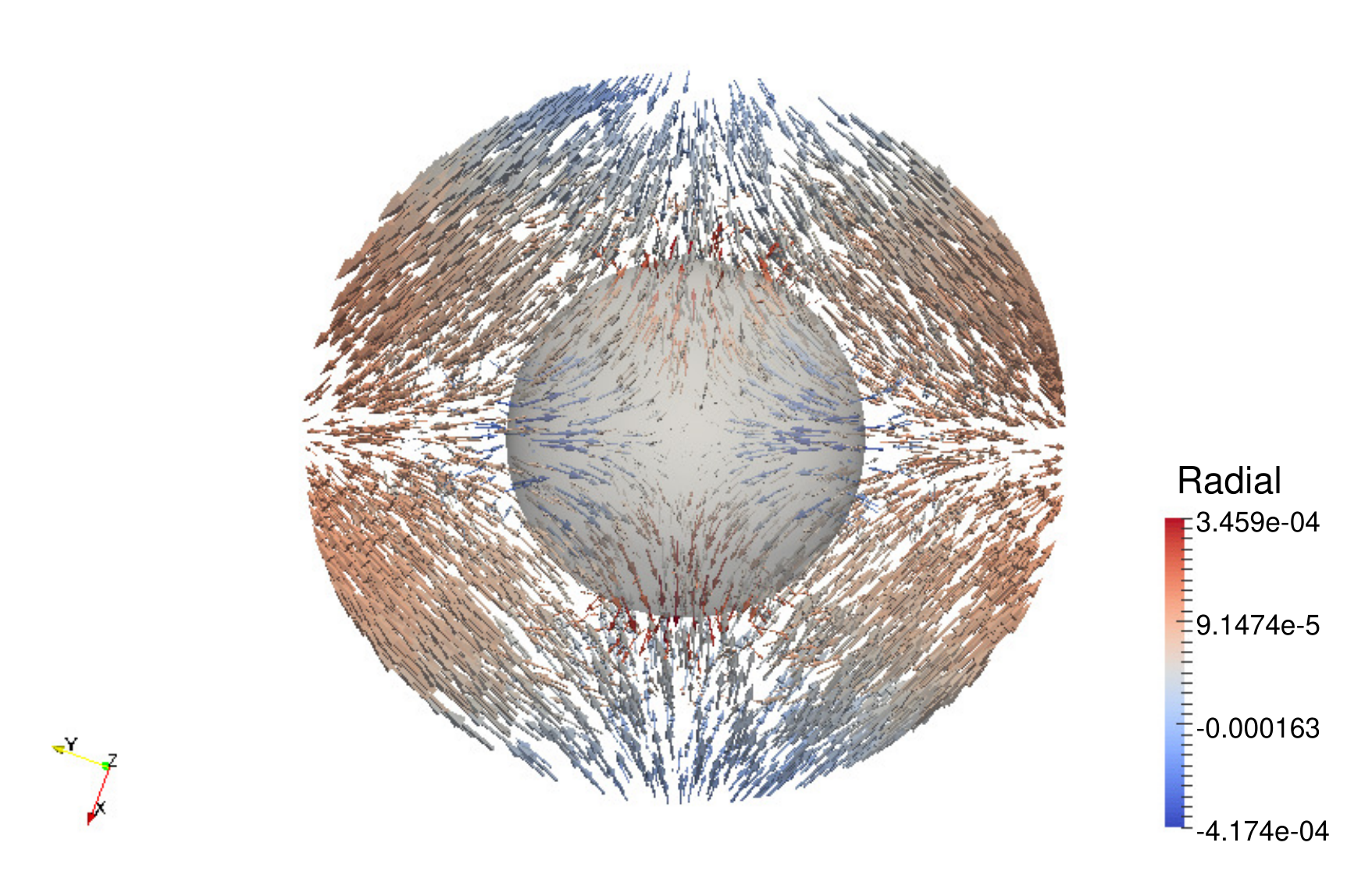} &
\includegraphics[trim= 5cm 0cm 0cm 0cm,clip=true,width=0.30\linewidth]{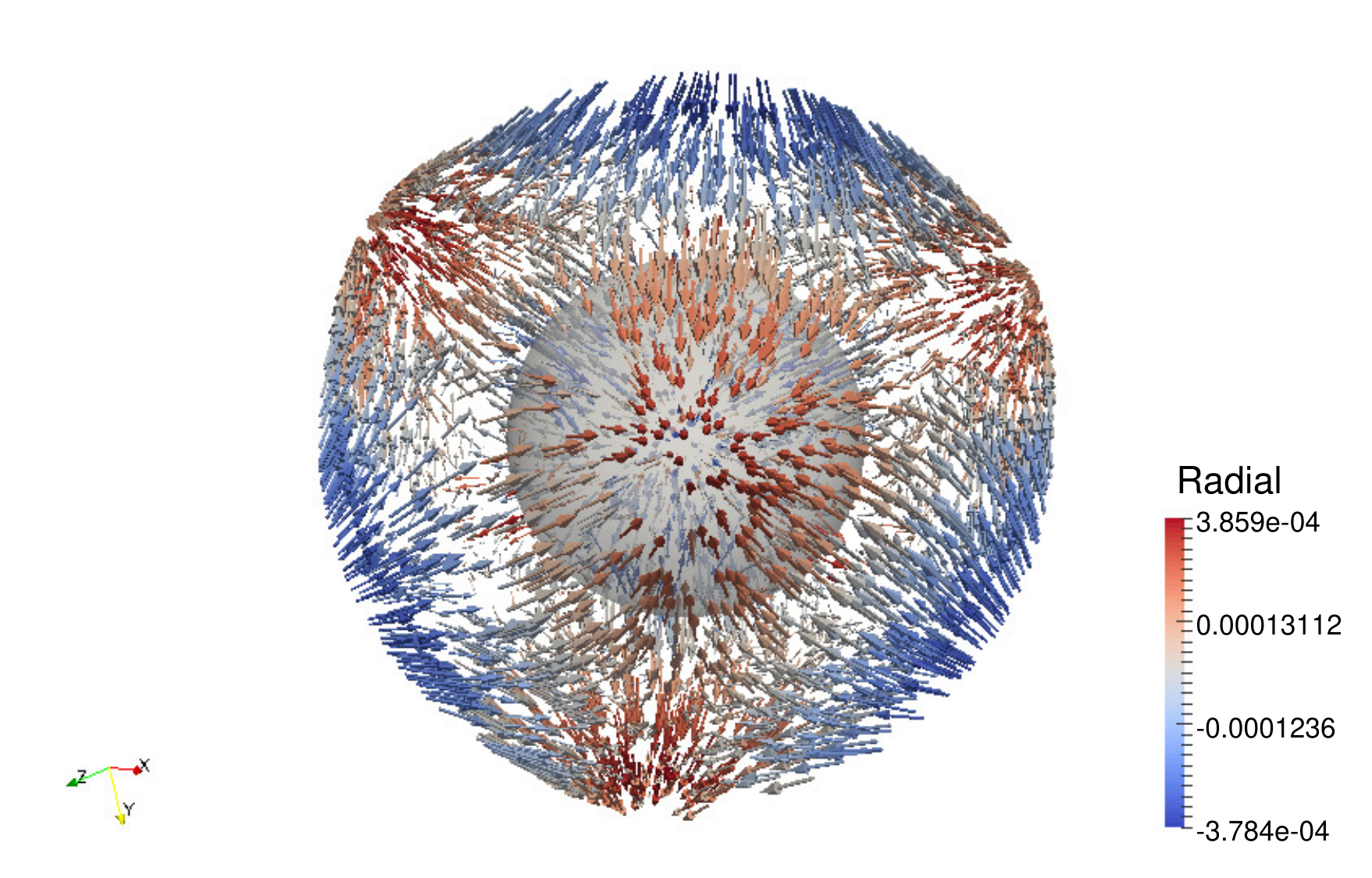} \\
(a1) ${}_1S_1$ & (a2) ${}_1S_2$ & (a3) ${}_1S_3$ \\
\includegraphics[trim= 5cm 0cm 0cm 0cm,clip=true,width=0.30\linewidth]{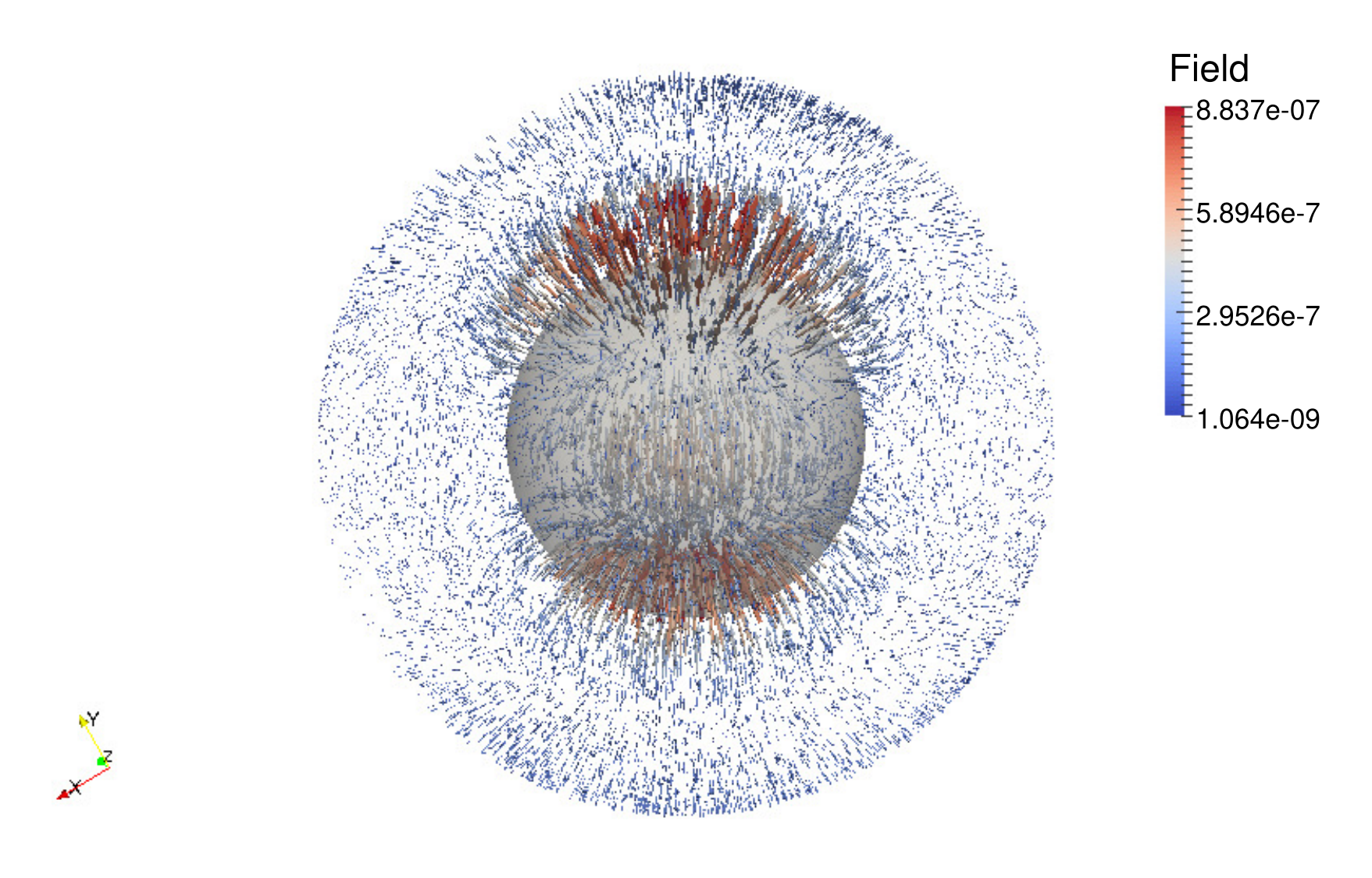} &
\includegraphics[trim= 5cm 0cm 0cm 0cm,clip=true,width=0.30\linewidth]{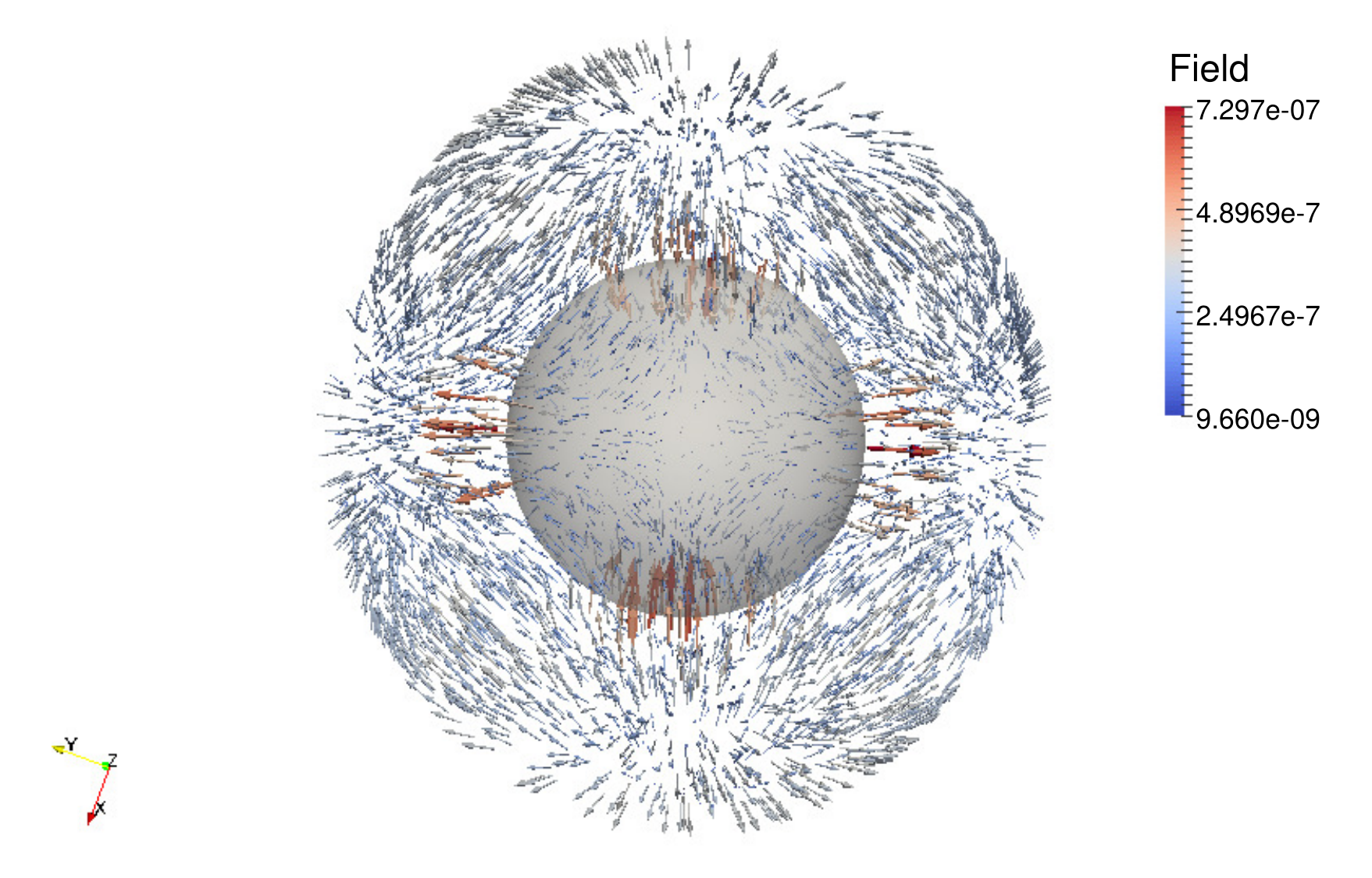} &
\includegraphics[trim= 5cm 0cm 0cm 0cm,clip=true,width=0.30\linewidth]{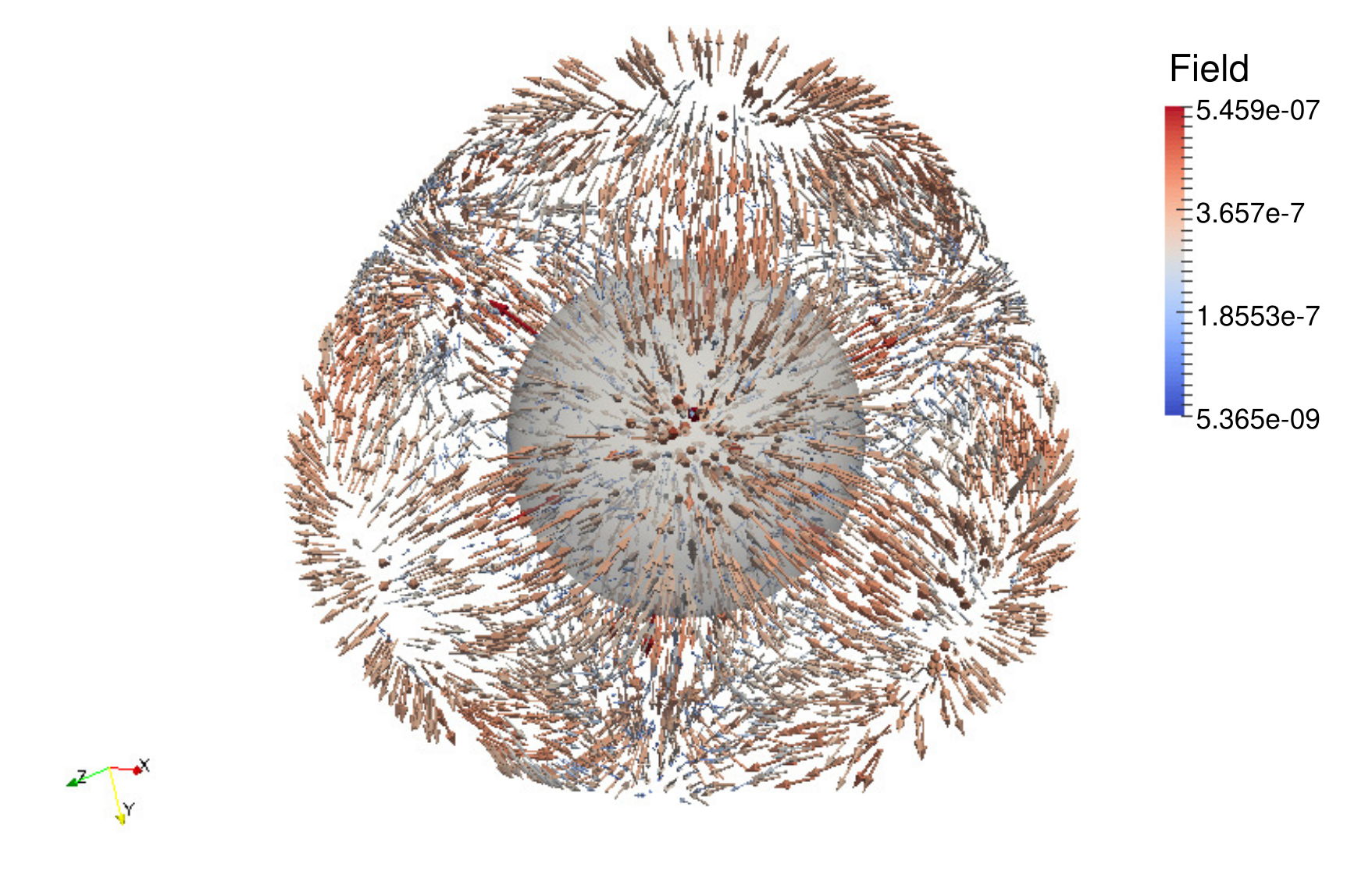} \\
(b1) $\nabla S(u)$ of ${}_1S_1$ & (b2) $\nabla S(u)$ of ${}_1S_2$ & (b3) $\nabla S(u)$ of ${}_1S_3$ \\
\includegraphics[trim= 5cm 0cm 0cm 0cm,clip=true,width=0.30\linewidth]{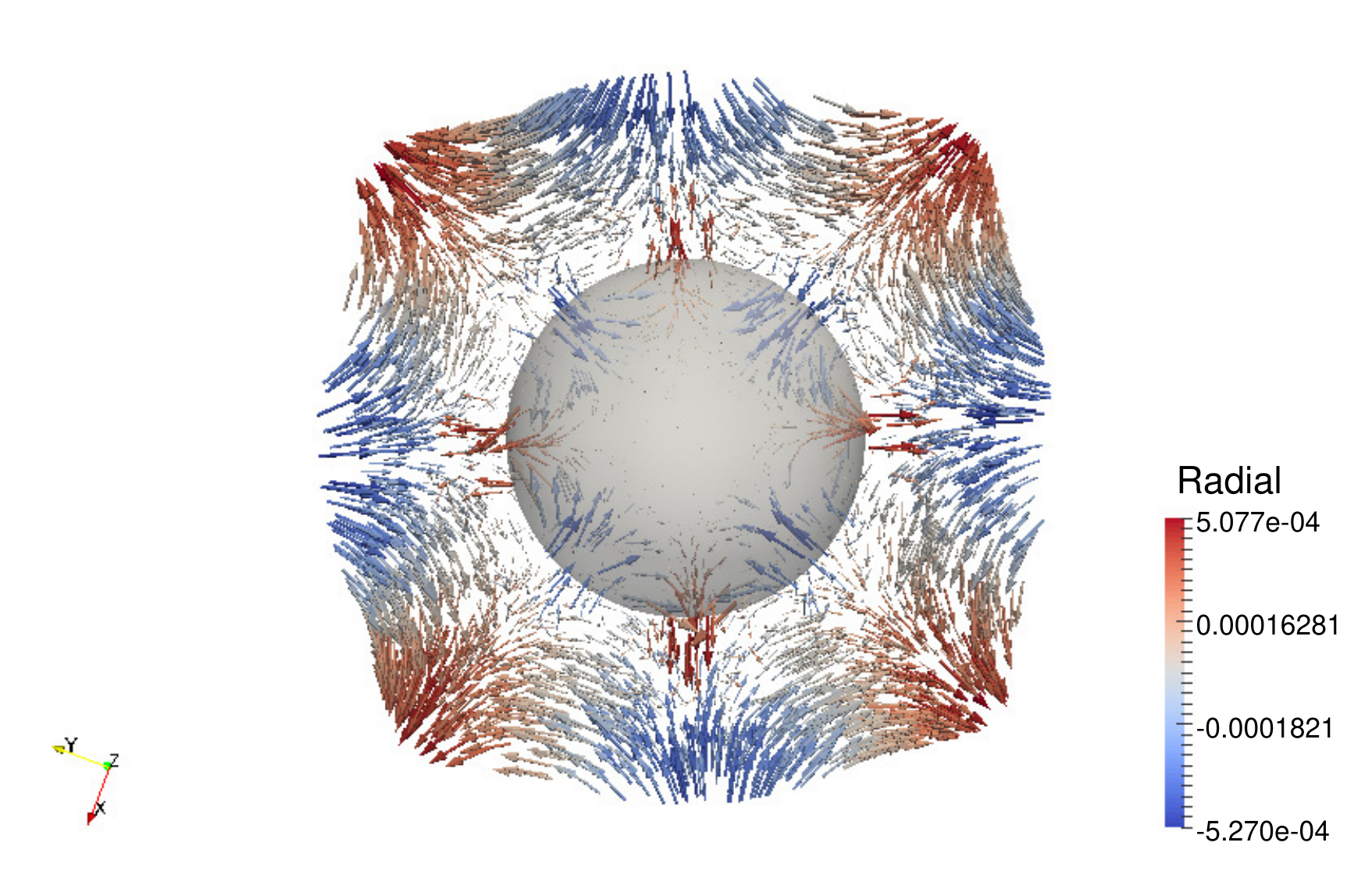} &
\includegraphics[trim= 5cm 0cm 0cm 0cm,clip=true,width=0.30\linewidth]{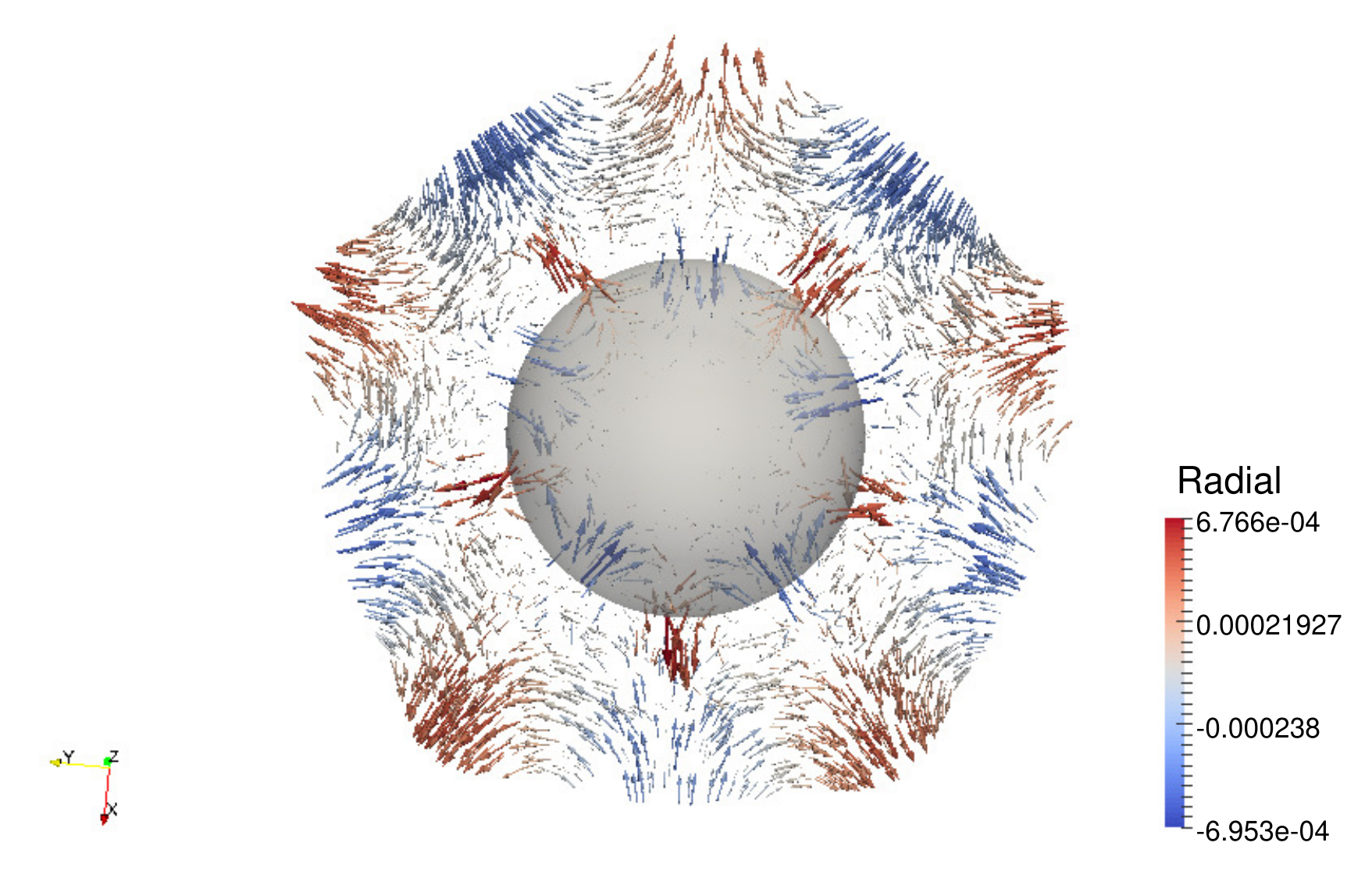} &
\includegraphics[trim= 5cm 0cm 0cm 0cm,clip=true,width=0.30\linewidth]{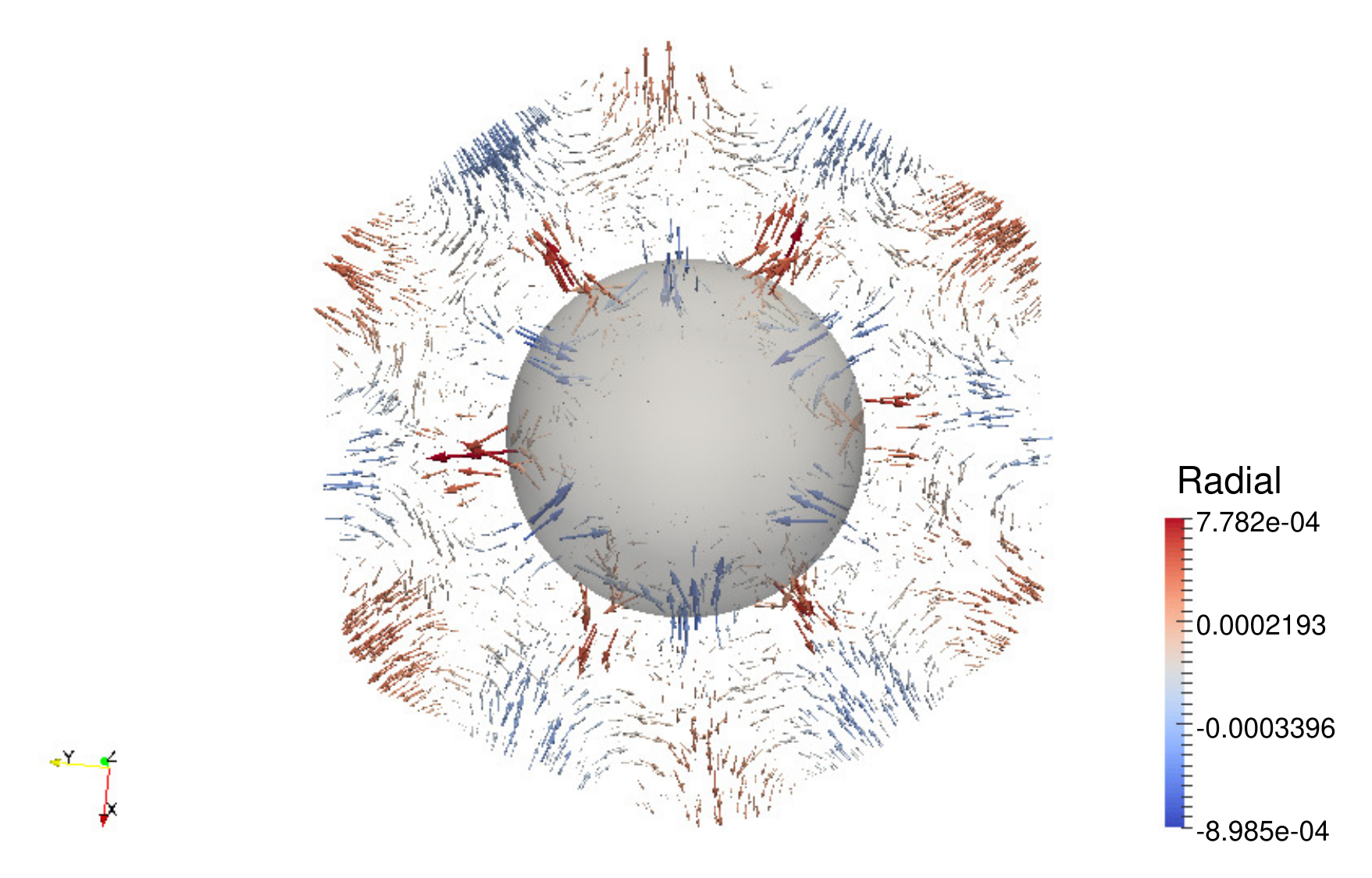} \\
(a4) ${}_1S_4$ & (a5) ${}_1S_5$ & (a6) ${}_1S_6$ \\
\includegraphics[trim= 5cm 0cm 0cm 0cm,clip=true,width=0.30\linewidth]{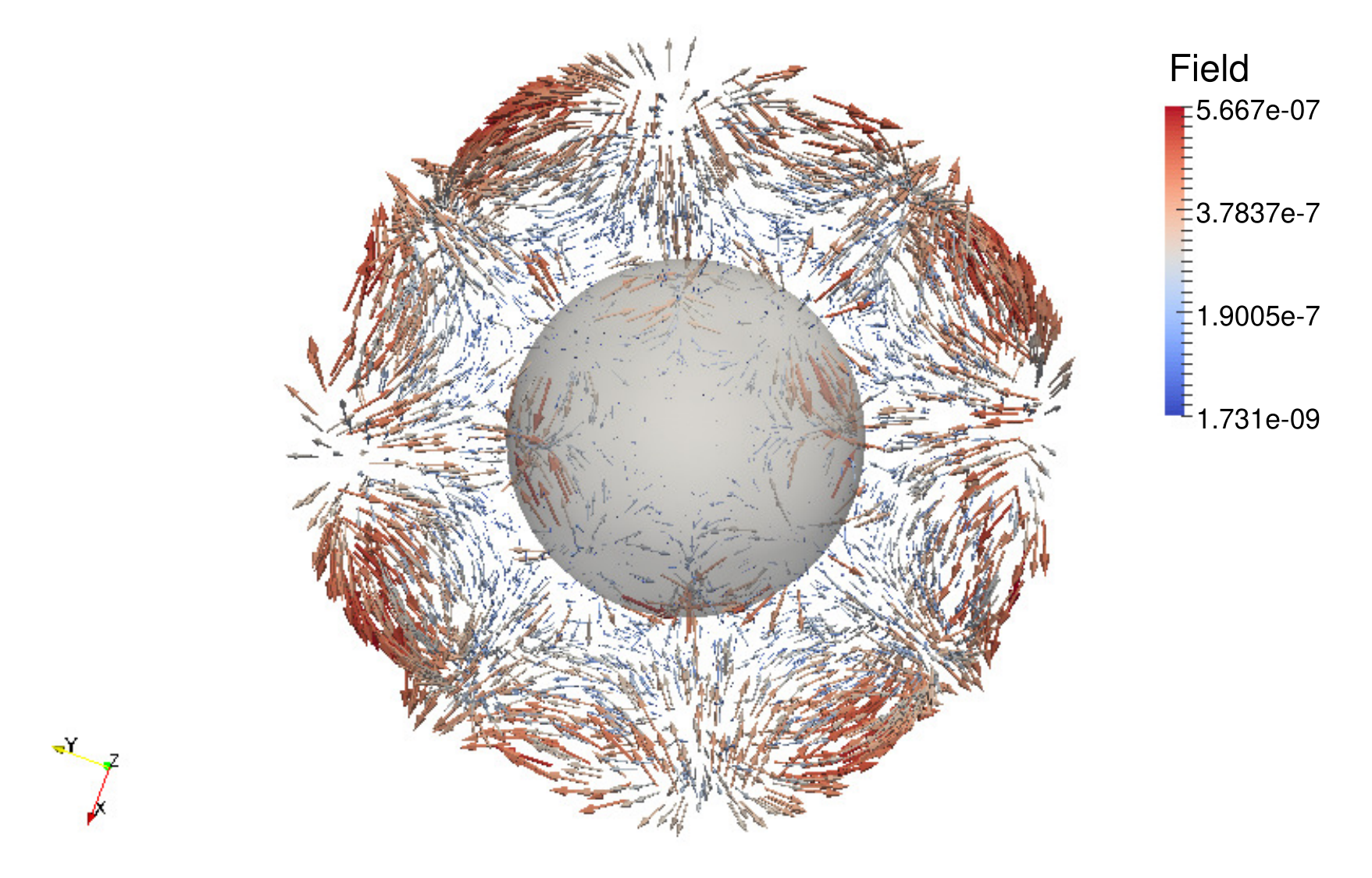} &
\includegraphics[trim= 5cm 0cm 0cm 0cm,clip=true,width=0.30\linewidth]{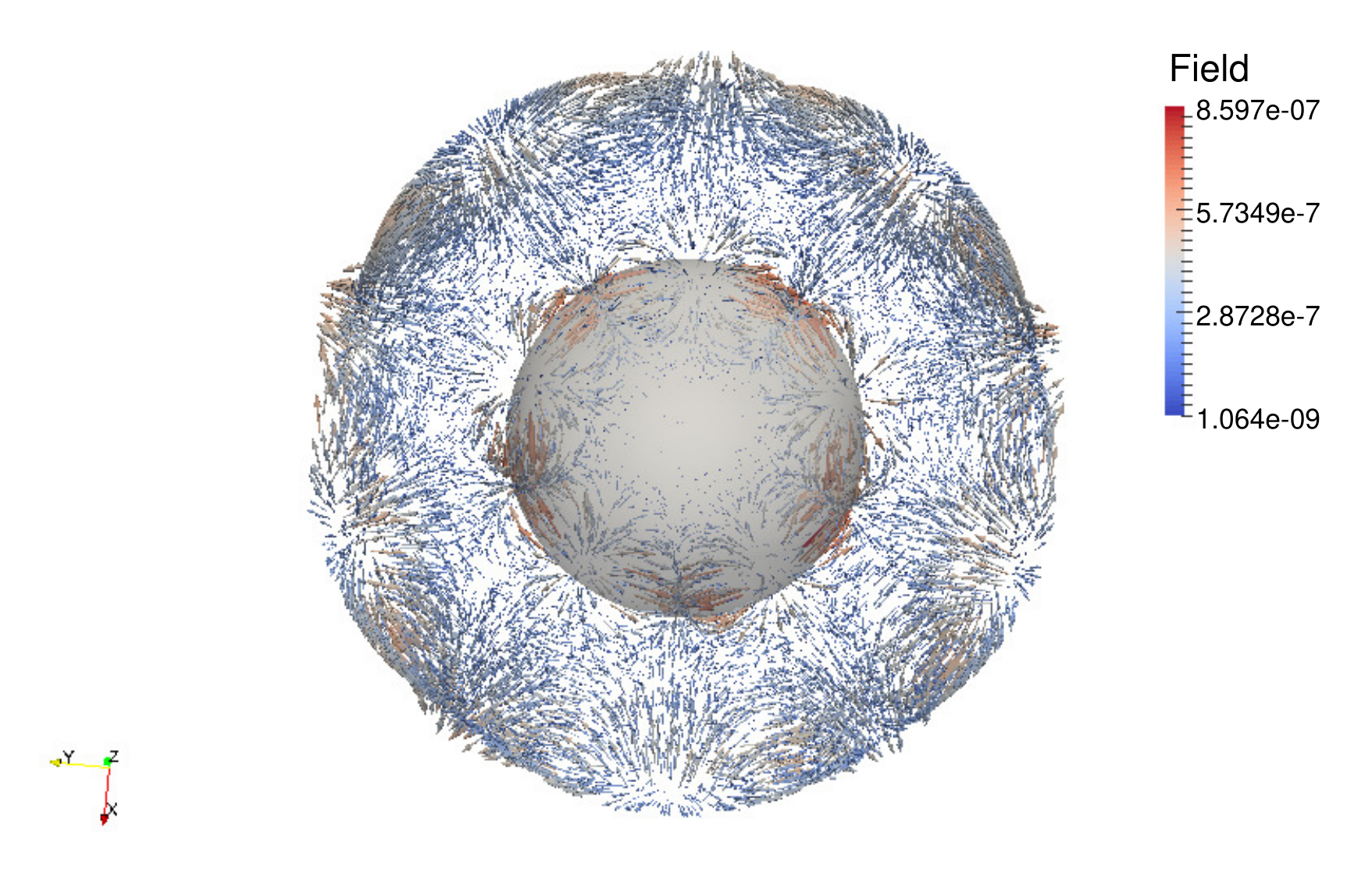} &
\includegraphics[trim= 5cm 0cm 0cm 0cm,clip=true,width=0.30\linewidth]{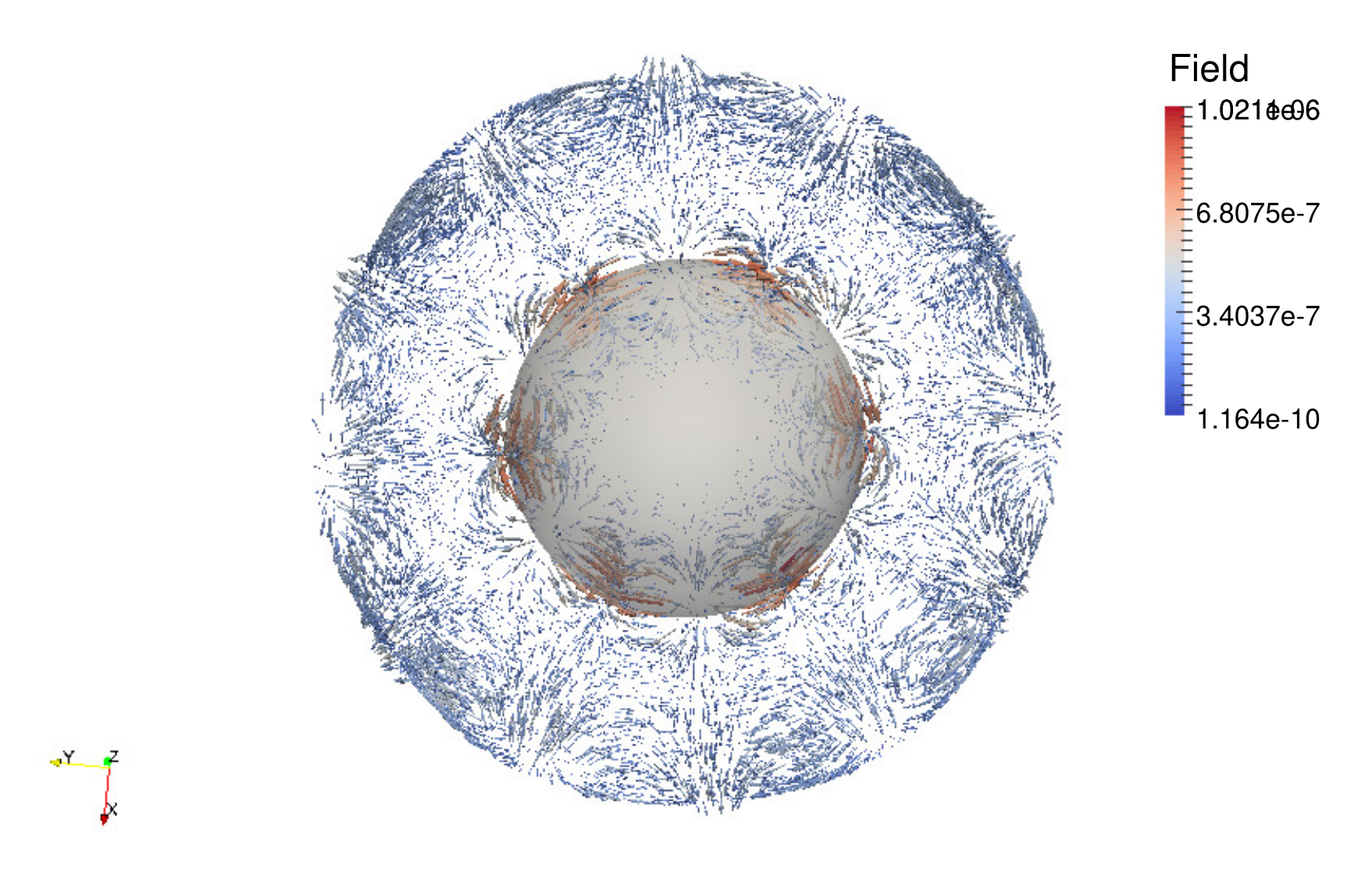} \\
(b4) $\nabla S(u)$ of ${}_1S_4$ & (b5) $\nabla S(u)$ of ${}_1S_5$ & (b6) $\nabla S(u)$ of ${}_1S_6$ 
\end{tabular}
\caption{Visualization of ${}_1S_l$ branch of a Mars model with a
  three-dimensional crust and rotation from TM2Mp2 experiment. The
  light ball indicates the position of the core-mantle
  boundary. (a1)--(a6) illustrate the modes ${}_1S_1$ to ${}_1S_6$,
  respectively. The unit in the color of (a1) - (a6) is
  meter. (b1)--(b6) illustrate the perturbed gravitational field
  $\nabla S(u)$ of the modes ${}_1S_1$ to ${}_1S_6$, respectively. The
  unit in the colorbar of (b1)--(b6) is millimeter.}
\label{fig:tm2m1Sl}
\end{figure}

In Fig.~\ref{fig:tm2m1Sl}, we plot the branch ${}_1S_l$ as well as the
corresponding incremental gravitational fields $\nabla S(u)$. The
superconducting gravimeters are expected to contribute to normal mode
seismology 
\cite{crossley1999network,van1999measuring,widmer2003can,rosat2003search,hafner2012signature}. We anticipate that both the
seismic and gravity measurements of these modes could help to estimate
the size of the Martian core.

\section{Conclusion}
\label{sec:conclusion}

In this work, we propose a method to compute the normal modes of a
fully heterogeneous rotating planet.  We apply the mixed
finite-element method to the elastic-gravitational system of a
rotating planet and utilize the FMM to calculate the self-gravitation.
We successfully separate out the essential spectrum 
by using a polynomial filtering eigensolver and thus, are able to compute the
normal modes associated with seismic point spectrum. To solve the
relevant QEP, we utilize extended Lanczos vectors computed in a
non-rotating planet -- with the shape of boundaries of a rotating
planet and accounting for the centrifugal potential -- spanning a
subspace to reduce the dimension of an equivalent linear form of the
QEP. The reduced system can be solved with a standard eigensolver. We
demonstrate our ability to compute the seismic normal modes with and
without rotation accurately. We then study the computational accuracy
and use a standard Earth model to perform a benchmark test against a
perturbation calculation.  We carry out computational experiments on
various Mars models and illustrate mode splitting due to rotation,
ellipticity and heterogeneity of the crust. The use of modern
supercomputers enables us to capture normal modes associated with the
seismic point spectrum of a fully heterogeneous planet accurately. The
computational efficiency can be further improved by using acceleration
techniques. The extension to include viscoelastic relaxation (for a
review, see \cite{romanowicz20071}), in particular Maxwell and Burger models, 
leads to a nonlinear rational eigenvalue problem, which 
is tractable at current subject of research. 

\section*{Acknowledgement}
We would like to thank Bernard Valette for his thoughtful comments. 
J.S. would like to thank Petroleum Geo-Services for using their supercomputer
Abel, and Danny Sorensen, Ruichao Ye, and Harry Matchette-Downes for helpful discussions.


\appendix

\section{Construction of orthonormal bases and submatrices}
\label{sec:localmatrices}

Here, we introduce three-dimensional polynomial bases $\{ \psi_n^s
\}_{n=1}^{N_{p^s}}$, $\{ \psi_n^f \}_{n=1}^{N_{p^f}}$ and $\{ \psi_n^p
\}_{n=1}^{N_{p^p}}$ while addressing the fact that the Lagrange
polynomials are not orthogonal to one another. 
We suppress superscripts $s$, $f$, $p$ in the notation in the remainder of 
this subsection. 
To simplify the computations, we introduce reference
volume and boundary elements. That is, we introduce a mapping that
connects any element $K$ to the reference tetrahedron defined by
\[
   \mathbf{I} = \{ r = (r_1,r_2,r_3)\ :\
       r_1 \geq -1 ,\ r_2 \geq -1 ,\ r_3 \geq -1,\
                   r_1 + r_2 + r_3 \leq -1 \} .
\]
Likewise, we introduce a mapping that connects any boundary element
$E$ to the reference triangle defined by
\[
   \mathbf{I}_{2D} = \{ t = (t_1,t_2)\ :\
       t_1 \geq -1 ,\ t_2 \geq -1 ,\ t_1 + t_2 \leq 0 \} . 
\]
We note that any two tetrahedra are connected through an affine
transformation, $x \to r$, with a constant Jacobian, $J$, which is
the determinant of $(\partial_r x)$. For the local approximation on
the reference element $\mathbf{I}$, we have
\[
   u_j(r) = \sum_{n=1}^{N_p} (\hat{u}_j)_n \psi_n(r) = \sum_{i=1}^{N_p} u_j(r_i) \ell_i(r) .
\]
The vector fields are treated component-wise in our discretization. 
This yields the expression $\mathcal{V} \hat{u}_j = u_j$, where the
generalized Vandermonde matrix takes the form of $\mathcal{V}_{in} =
\psi_n(r_i)$ with $i,n$ as indices of nodal points. Here, $\{\psi_n
\}$ is a polynomial basis that is orthonormal on $\mathbf{I}$. 
We later introduce submatrices of $\mathcal{V}$. 
We then evaluate derivatives and mass matrices according to
\[
   \partial_{x_i} = (\partial_{x_i} r_j) \mathcal{D}_{j} ,
\quad
   \mathcal{D}_{j} = (\partial_{r_j} \mathcal{V}) \mathcal{V}^{-1} ,
\quad
   \mathcal{M} = \mathcal{V}^{-T} \mathcal{V}^{-1} ,
\]
where $\mathcal{D}_j$ and $\mathcal{M}$ are the derivative matrix and
the mass matrix on the reference tetrahedron. More details of the
constructions of $J$, $\mathcal{V}$, $\mathcal{D}_j$ and $\mathcal{M}$
can be found in \cite[Chapter 10.1]{hesthaven2007nodal}. Thus, we
introduce
\[
   \mathcal{V}_s ,\ \mathcal{V}_f ,\ \mathcal{V}_p, \quad 
   \mathcal{M}_s ,\ \mathcal{M}_f ,\ \mathcal{M}_p 
\quad\text{and}\quad
   \mathcal{D}_j^s ,\ \mathcal{D}_j^f ,\ \mathcal{D}_j^p.
\]
We employ the notation
\[ 
   \mathsf{D}^s_i = (\partial_{x_i} r_j) \mathcal{D}^s_j ,\quad
   \mathsf{D}^f_i = (\partial_{x_i} r_j) \mathcal{D}^f_j ,\quad
   \mathsf{D}^p_i = (\partial_{x_i} r_j) \mathcal{D}^p_j ,
\] 
reflecting the mapping of the derivatives from the reference tetrahedron to the target
element. 
We follow a similar approach for boundary elements and introduce 
\[
\mathcal{M}_s^{2D}, \ \mathcal{M}_f^{2D} \quad \text{and}\quad  J^{2D}, 
\]
where $\mathcal{M}_s^{2D}$ and $\mathcal{M}_f^{2D}$
are the mass matrices for solid and fluid 
boundary elements, respectively; $J^{2D}$ denotes the Jacobian, which is 
the determinant of $(\partial_t x)$ on the boundary element. 
The construction of the mass matrices
$\mathcal{M}^{2D}_s$ and $\mathcal{M}^{2D}_f$ on the reference triangle $\mathbf{I}_{2D}$ is
similar to the construction of $\mathcal{M}$ \cite[Chapter 6.1]{hesthaven2007nodal}.

\subsection{Submatrices: $A_{sg}$, $A_f$, $A_p$, $M_s$, $M_f$, $R_s$ and $R_f$}

We extract $\tilde{u}^s |_{K_k}$, $\tilde{u}^f |_{K_k}$ and $\tilde{p}
|_{K_k}$ from $\tilde{u}^s$, $\tilde{u}^f$ and $\tilde{p}$,
respectively, by restricting the nodes to the ones of element $K_k$. In a
similar fashion, we extract $\tilde{v}^s |_{K_k}$, $\tilde{v}^f
|_{K_k}$ and $\tilde{v}^p |_{K_k}$ on any element $K_k$.  
For the evaluation of matrix $A_{sg}$ in
Table~\ref{table:matrixcomponentsquadratic} we need to evaluate the 
submatrices on element $K_k$ through
\begin{align}
\int_{K_k^{\text{S}}} \partial_{x_i} (\overline{v}^s_h)_j (c_{ijmn} \partial_{x_m} (u^s_h)_n) \dd x & = 
(\tilde{v}^s_j|_{K_k})\ctrans [ J_k (\mathsf{D}_i^s)\trans c^{k}_{ijmn} \mathcal{M}_s \mathsf{D}_m^s ] \tilde{u}^s_n|_{K_k} , \\
\int_{K_k^{\text{S}}} \partial_{x_i} (\overline{v}^s_h)_i  g'_j (u^s_h)_j   \rho^0 \dd x & = 
(\tilde{v}^s_i|_{K_k})\ctrans  [ J_k (\mathsf{D}_i^s)\trans  \rho^0_k \mathcal{M}_s D_{g'_j} ] \tilde{u}^s_j|_{K_k} , \\
\int_{K_k^{\text{S}}} -  (u^s_h)_i \partial_{x_i} g'_j  (\overline{v}^s_h)_j \rho^0 \dd x & = 
(\tilde{v}^s_i|_{K_k})\ctrans [-  J_k \rho^0_k D_{\partial_{x_i} g'_j} \mathcal{M}_s ] \tilde{u}^s_j|_{K_k}  ,
\\
\int_{K_k^{\text{S}}} -  (u^s_h)_j (\partial_{x_j} (\overline{v}^s_h)_i) g'_i \rho^0 \dd x & = 
(\tilde{v}^s_i|_{K_k})\ctrans  [- J_k \mathsf{D}^s_j \mathcal{M}_s  \rho^0_k D_{g'_i} ] \tilde{u}^s_j|_{K_k}  ,
\end{align}
where $c^{k}_{ijmn}$, $\rho^0_k$ and $J_k$ denote the stiffness
tensor, density and the Jacobian on element $K_k$, respectively; $D_{g'_i}$
and $D_{\partial_{x_i} g'_j}$ denote the diagonal matrices whose
diagonal entries are $g'_i$ and $\partial_{x_i} g'_j$, respectively. For
the evaluation of the boundary integration in $A_{sg}$, we need to
evaluate the submatrix on element $E_l^{\text{FS}}$ through
\begin{equation}
\int_{E_l^{\text{FS}}} (\overline{v}^s_h)_i g'_i  \nu^{s \rightarrow f}_j (u^s_h)_j  [\rho^0]^f \dd \Sigma  = 
(\tilde{v}^s_i|_{E_l})\ctrans [ J^{2D}_l \rho^0_l D_{g'_i} \mathcal{M}_s^{2D} \nu^{s \rightarrow f}_j|_{E_l} ] \tilde{u}^s_j|_{E_l} , 
\end{equation}
where $\rho^0_l$ and $\nu^{s \rightarrow f}_j|_{E_l}$ denote 
the density and normal vector on the boundary element $E_l^{\text{FS}}$, respectively,
upon extracting $\tilde{v}^s_i |_{E_l}$ and $\tilde{u}^s_i|_{E_l}$. 
We can deal with the integral over $\Sigma^{\text{FF}}$ similarly. 

We then evaluate the submatrices for $A_f$, $A_p$, $M_s$, $M_f$
in Table~\ref{table:matrixcomponentsquadratic} and obtain
\begin{align}
\int_{K_k^{\text{F}}} \rho^0 N^2 \frac{  g'_i (\overline{v}^f_h)_i g'_j  (u^f_h)_j }{\| g' \|^2} \dd x &=
 (\tilde{v}^f_i|_{K_k})\ctrans  [ J_k D_{g'_i/\|g'\|}  \rho_k^0 N_k^2 \mathcal{M}_f   D_{g'_j/\|g'\|} ] \tilde{u}^f_j|_{K_k} , \\
\int_{K_k^{\text{F}}}  -  \overline{v}^p_h  p_h \kappa^{-1} \dd x &= (\tilde{v}^p|_{K_k})\ctrans [ - J_k \kappa^{-1}_k \mathcal{M}_p ] \tilde{p}|_{K_k} , \\
\int_{K_k^{\text{S}}} (\overline{v}^s_h)_i   (u^s_h)_i\rho^0 \dd x  & = (\tilde{v}^s_i|_{K_k})\ctrans  [ J_k \rho^0_k \mathcal{M}_s ] \tilde{u}^s_i|_{K_k} , \\
\int_{K_k^{\text{F}}}  (\overline{v}^f_h)_i (u^f_h)_i \rho^0 \dd x & = (\tilde{v}^f_i|_{K_k})\ctrans  [ J_k \rho^0_k \mathcal{M}_f ] \tilde{u}^f_i|_{K_k} ,
\end{align}
where $ D_{g'_j/\|g'\|}$ denotes a diagonal matrix whose diagonal
entries are $g'_j/\|g'\|$ and $N_k^2$ denotes the square of the Brunt-V\"{a}is\"{a}l\"{a}
frequency on element $K_k$.
We also obtain the rotation components $R_s$ and $R_f$,
\begin{align}
\int_{K_k^{\text{S}}} \epsilon_{ijm}  (\overline{v}^s_h)_i  (u^s_h)_j  \rho^0 \dd x  & = (\tilde{v}^s_i|_{K_k})\ctrans  [ \epsilon_{ijm}  J_k \rho^0_k \mathcal{M}_s ] \tilde{u}^s_j|_{K_k} , \\
\int_{K_k^{\text{F}}} \epsilon_{ijm}  (\overline{v}^f_h)_i (u^f_h)_j \rho^0 \dd x & = (\tilde{v}^f_i|_{K_k})\ctrans  [ \epsilon_{ijm}  J_k \rho^0_k \mathcal{M}_f ] \tilde{u}^f_j|_{K_k} , 
\end{align}
where $\epsilon_{ilm}$ denotes the Levi-Civita symbol.

\subsection{Submatrices: $A_{\text{dg}}$ and $A_{\text{dg}}\trans$}\label{sec:mixbody}

Here, we discuss the integration between the different variables. For the
inner products between $u^f_h$ and $p_h$ for $A_{\text{dg}}$ and
$A_{\text{dg}}\trans$ in Table~\ref{table:matrixcomponentsquadratic}, we
evaluate the mass matrices $\mathcal{M}_{pf}$ and $\mathcal{M}_{fp}$,
\[
   \mathcal{M}_{pf} = (\mathcal{V}_p^{-1}(I_{f}))^T
                                  \mathcal{V}_{f}^{-1}(I_{p}) ,
\quad
   \mathcal{M}_{fp} = (\mathcal{V}_{f}^{-1}(I_{p}))^T
                                  \mathcal{V}_p^{-1}(I_{f}) ,
\]
where we refine the notation to indicate submatrices of $\mathcal{V}$; 
$\mathcal{V}(I)$ denotes the submatrix of $\mathcal{V}$ formed by columns indexed by  
$I \subseteq \{1,\ldots, N_{p}\}$. 
The selection of submatrices is based on the polynomial construction
\cite[(10.6)]{hesthaven2007nodal}. 
For instance, if the polynomial orders used 
for both $u^f_h$ and $p_h$ are the same, i.e., $p^f=p^p$, 
$I_{f}=I_{p}=\{1,\ldots, N_{p^f}\}$; 
if $p^p=1$ and $p^f=2$, we have $N_{p^p}=4$, $N_{p^f}=10$ and $I_{f}=\{1,2,3,4\}$, 
$I_{p}=\{1,2,4,7\}$.
It is apparent that $\mathcal{M}_{pf} = \mathcal{M}_{fp}^T$.

Evaluating $A_{\text{dg}}$ in Table~\ref{table:matrixcomponentsquadratic}
requires the evaluation of the submatrices on element $K_k$ through
\begin{align}
\int_{K_k^{\text{F}}} (\overline{v}^f_h)_j (\partial_{x_j} p_h ) \dd x & =  (\tilde{v}^f_j|_{K_k})\ctrans [ J_k \mathcal{M}_{fp} \mathsf{D}_{j}^p ] \tilde{p}|_{K_k} , \\
\int_{K_k^{\text{F}}}   (\overline{v}^f_h)_j g'_j   p_h  \rho^0  \kappa^{-1}\dd x & 
=  (\tilde{v}^f_j|_{K_k})\ctrans [ J_k D_{g'_j}  \rho^0_k  \kappa^{-1}_k \mathcal{M}_{fp}  ] \tilde{p}|_{K_k} , 
\end{align}
where $\kappa^{-1}_k$ denotes the inverse of the bulk modulus on
element $K_k$. To evaluate $A_{\text{dg}}\trans$ in
Table~\ref{table:matrixcomponentsquadratic}, we also need to evaluate 
the submatrices on element $K_k$ through
\begin{align}
\int_{K_k^{\text{F}}} (\partial_{x_j} \overline{v}^p_h)  (u^f_h)_j \dd x &=  (\tilde{v}^p|_{K_k})\ctrans  [ J_k (\mathsf{D}_{j}^p)^{T}  \mathcal{M}_{pf} ] \tilde{u}^f_j|_{K_k} ,
\\
\int_{K_k^{\text{F}}} \overline{v}^p_h g'_j  (u^f_h)_j   \rho^0 \kappa^{-1}  \dd x & = (\tilde{v}^p|_{K_k})\ctrans [ J_k \rho^0_k \kappa^{-1}_k \mathcal{M}_{pf} D_{g'_j} ] \tilde{u}^f_j|_{K_k} .
\end{align}

\subsection{Submatrices: $E_{\text{FS}}$ and $E_{\text{FS}}\trans$}
For $E_{\text{FS}}$ and $E_{\text{FS}}\trans$ , similar to Section~\ref{sec:mixbody}, 
we introduce two new indices to 
construct $\mathcal{M}_{ps}^{2D}$ and $\mathcal{M}_{sp}^{2D}$ on the boundary elements 
associated with the fluid-solid boundary. 
The selection of the submatrix is based on \cite[Chapter 6]{hesthaven2007nodal}. 
$\mathcal{M}_{ps}^{2D} = {\mathcal{M}_{sp}^{2D}}\trans$ holds true as well. 
To evaluate $E_{\text{FS}}\trans$ in Table~\ref{table:matrixcomponentsquadratic},
we need to compute the submatrix on boundary element
$E^{\text{FS}}_l$ through
\begin{equation}
\int_{E_l^{{\text{FS}}}}   (\overline{v}^s_h)_j  \nu^{s \rightarrow f}_j  p_h \dd \Sigma = ( \tilde{v}^s_j|_{E_l} ) \ctrans  [ J^{2D}_l \nu^{s \rightarrow f}_j \mathcal{M}_{sp}^{2D}  ] \tilde{p}|_{E_l} ,
\end{equation}
upon extracting $\tilde{p}|_{E_l}$ on boundary element
$E^{\text{FS}}_l$. To evaluate $E_{\text{FS}}$ in
Table~\ref{table:matrixcomponentsquadratic}, we need to evaluate 
the submatrix on boundary element $E^{\text{FS}}_l$ through
\begin{equation}
\int_{E^{{\text{FS}}}_l} \overline{v}^p_{h}  \nu^{f \rightarrow s}_j (u^s_{h})_j \dd \Sigma = 
 (\tilde{v}^p|_{E_l} )\ctrans [ J^{2D}_l \nu^{f \rightarrow s}_j \mathcal{M}_{ps}^{2D}  ] \tilde{u}^s_j|_{E_l} ,
\end{equation}
upon extracting $\tilde{v}^p|_{E_l}$ on $E^{\text{FS}}_l$.

\medskip\medskip

\noindent
We are now able to build all the submatrices for the evaluation of
the integrals in Table~\ref{table:matrixcomponentsquadratic}.
We then assemble the global matrices from all these submatrices using 
standard techniques similar to those in \cite{bathe2006finite,hughes2012finite}.

\subsection{Construction of the submatrices for the perturbation of the gravitational potential}
\label{sec:matrixselfG}
Similar to the previous subsections, we construct the submatrices in $C_s$ in Table~\ref{table:matrixcomponentsselfG}, 
\begin{align}
\int_{K_k^{\text{S}}}  \partial_{x_i} ( \rho^0 (u^s_h)_i )  \dd x &=
 (\mathbf{1}|_{K_k})\ctrans  [J_k \mathcal{M}_s \mathsf{D}_i^s \rho^0_k ] \tilde{u}^s_i|_{K_k} , \\
\int_{E_l^{\text{FS}}} \nu^{f\rightarrow s}_i  (u^s_h)_i \left[\rho^0\right]^s \dd \Sigma &= 
(\mathbf{1}|_{E_l})\ctrans  [ J^{2D}_l \nu^{f\rightarrow s}_i  [\rho^0]^s_l \mathcal{M}_s^{2D} ] \tilde{u}^s_i|_{E_l} ,\\
\int_{E_l^{\text{S}}} \nu_i  (u^s_h)_i \left[\rho^0\right]^+_- \dd \Sigma & = 
(\mathbf{1}|_{E_l})\ctrans  [J^{2D}_l \nu_i  ([\rho^0]^+_-)_l \mathcal{M}_s^{2D} ] \tilde{u}^s_i|_{E_l}  , 
\end{align}
and the submatrices in $C_s\trans$, 
\begin{align}
\int_{K_k^{\text{S}}}  [\partial_{x_i} ( \rho^0 (\overline{v}^s_h)_i )] S_k(u_h) \dd x &=
 (\tilde{v}^s_i|_{K_k})\ctrans  [J_k  \rho^0_k (\mathsf{D}_i^s)\trans \mathcal{M}_s  S_k(\tilde{u}) ] \mathbf{1}|_{K_k}, \\
\int_{E_l^{\text{FS}}} \nu^{f\rightarrow s}_i  (\overline{v}^s_h)_i  S_l(u_h) \left[\rho^0\right]^s \dd \Sigma &= 
(\tilde{v}^s_i|_{E_l})\ctrans  [ J^{2D}_l \nu^{f\rightarrow s}_i \mathcal{M}_s^{2D}  [\rho^0]^s_l S_l(\tilde{u}) ] \mathbf{1}|_{E_l} ,\\
\int_{E_l^{\text{S}}} \nu_i  (\overline{v}^s_h)_i S_l(u_h) \left[\rho^0\right]^+_- \dd \Sigma & = 
(\tilde{v}^s_i|_{E_l})\ctrans  [J^{2D}_l \nu_i  \mathcal{M}_s^{2D} ([\rho^0]^+_-)_l  S_l(\tilde{u}) ] \mathbf{1}|_{E_l}  , 
\end{align}
where $\mathbf{1}$ denotes a vector of all ones. 
The construction of the submatrices in $C_f$ and $C_f\trans$ is the same. 
We are now able to build all the submatrices for the evaluation of
the integrals in Table~\ref{table:matrixcomponentsselfG}.

\section{Full mode coupling}\label{app:sphharmonics}

Concerning the Galerkin approximation, we can use different, nonlocal bases of
functions in the appropriate energy space, for example, the spectral-Galerkin method \cite{shen1994efficient}. 
In this appendix, we consider the use of the eigenfunctions of a spherically symmetric, non-rotating,
perfectly elastic and isotropic (SNREI) reference model as a basis in this method. 
This has been implemented by 
\cite{woodhouse1978effect,woodhouse1980coupling,deuss2001theoretical,deuss2004iteration}, 
and named the full mode coupling approach.
An immediate drawback of using this basis,
however, is that the fluid-solid boundaries need to be spherically
symmetric, as these are encoded in these basis functions.

We let $u_{km}$ represent the eigenfunctions associated with
eigenfrequencies, $\omega_k$, in terms of spherical
harmonics, $Y_l^m$, that is, 
\[
   u_{km} = U_{km} \mathbf{P}_{lm}
      +  V_{km}  \mathbf{B}_{lm}
         +  W_{km} \mathbf{C}_{lm}  \quad \, \text{(no summation over $m$)}, 
\]
where $k$ is the multi-index for the eigenfrequency; 
$m=-l,-l+1\ldots,l-1,l$ is the index corresponding with the degeneracy with $l$ denoting the
spherical harmonic degree; 
$U_{km}, V_{km}$ and $W_{km}$ are the three components of
eigenfunctions and are functions of the radial coordinate; 
$\mathbf{P}_{lm}$, $\mathbf{B}_{lm}$ and $\mathbf{C}_{lm}$ 
are the vector spherical harmonics, see \cite[(8.36)]{dahlen1998theoretical} for their definition. 
In addition, $p_{km}$ needs to be introduced to constrain the solution, cf. \eqref{eq:fluidconstraint} \cite[Subsection 3.3]{de2019note}.  
Since $\nabla \cdot u_{km}(x)$ can be expanded using $Y_l^m(x)$ 
\cite[(8.38)]{dahlen1998theoretical} and $u_{km}(x)  \cdot g_{(r)}$ can also be expanded using $Y_l^m(x)$ for the radial models, 
we let $p_{km} = P_{km} Y_l^m$ with
\[
P_{km} = - \kappa_{(r)} \left[ \partial_r U_{km} + r^{-1} (2 U_{km} - \sqrt{l(l+1)}V_{km}) \right] + \rho^0_{(r)} g_{(r)} U_{km}, 
\]
where $\rho_{(r)}^0$,  $\kappa_{(r)}$ and $g_{(r)}$ denote the radial profiles of the density, bulk modulus and reference gravitational field of a radial model, respectively.
Similarly, the incremental gravitational potential of the radial models takes the form, 
$s_{km} = S_{km} Y_l^m$, 
where $S_{km}$ is also a function in the radial coordinate. 
In the following, $l$ and $m$ are fixed. 
 
In a SNREI model, for the computation of the toroidal modes, 
we only need to consider a solid annulus comprising the mantle and the crust.  
We exemplify the computations with the spheroidal modes and let $U'_{km}$, $P'_{km}$ and 
$S'_{km}$ be test functions for $U_{km}$, $P_{km}$ and 
$S_{km}$ following the Galerkin method. 
We let the $\tilde{X}_{(r)}$ be the 1D interval of the radial planet and have $\tilde{X}_{(r)} = 
\Omega_{(r)}^{\text{S}} \cup \Omega_{(r)}^{\text{F}}$, 
where $\Omega_{(r)}^{\text{S}}$ and $\Omega_{(r)}^{\text{F}}$ denote 
the 1D intervals for the solid and fluid regions, respectively. 
Given a regular finite-element partitioning $\mathcal{T}_h^{(r)}$ of the interval $\tilde{X}_{(r)}$, 
we denote an element of the mesh by $L_q \in \mathcal{T}_h^{(r)}$ 
and have $\tilde{X}_{(r)} =  \bigcup_{q=1}^{N_L} L_q$, 
where $N_L$ denotes the total number of 1D elements. 
Furthermore, we let $L_q^{\text{S}}$ and $L_q^{\text{F}}$ specifically be elements in the solid 
and fluid regions and have 
\[
\Omega_{(r)}^{\text{S}} = \bigcup_{q=1}^{N_L^{\text{S}}} L_q^{\text{S}}, \quad
\Omega_{(r)}^{\text{F}} = \bigcup_{q=1}^{N_L^{\text{F}}} L_q^{\text{F}}, 
\]
where $N_L^{\text{S}}$ and $N_L^{\text{F}}$ denote the numbers of 1D elements in the 
solid and fluid regions, respectively. 
We let $\Sigma^{\text{FS}}_{(r)}$ denote the fluid-solid boundary points in the radial interval. 
We introduce the finite-element solutions, $U_{km;h}^s$, $U_{km;h}^f$, $V_{km;h}^s$, $V_{km;h}^f$, $P_{km;h}$ and $S_{km;h}$, and test
functions, $U_{km;h}^{s'}$, $U_{km;h}^{f'}$, $V_{km;h}^{s'}$, $V_{km;h}^{f'}$, $P'_{km;h}$ and $S'_{km;h}$. 
We set $N_{p^U} =(p^U+1)/2$, where $N_{p^U}$ is the number of nodes on
a 1D element for the $p^U$-th order polynomial approximation. We have
likewise expressions for $N_{p^V}$, $N_{p^P}$ and $N_{p^S}$. 
As in Subsection~\ref{sec:matrixform}, we introduce nodal-based Lagrange polynomials, 
$\ell_i^{U}$, $\ell_i^{V}$, $\ell_i^{P}$, $\ell_i^S$, on the respective 1D elements $L \in \mathcal{T}_h^{(r)}$, 
and write 
\begin{align}
   U_{km;h}^s(x) &= \sum_{i=1}^{N_{p^U}} U_{km;h}^s(x_i) \ell^U_i(x) , \quad 
  U_{km;h}^f(x) = \sum_{i=1}^{N_{p^U}} U_{km;h}^f(x_i) \ell^U_i(x) ,
\\
   V_{km;h}^s(x) &= \sum_{i=1}^{N_{p^V}} V_{km;h}^s(x_i) \ell^V_i(x) , \quad 
  V_{km;h}^f(x) = \sum_{i=1}^{N_{p^V}} V_{km;h}^f(x_i) \ell^V_i(x) ,
\\
   P_{km}(x) &= \sum_{i=1}^{N_{p^P}} P_{km}(x_i) \ell^P_i(x), \quad 
   S_{km}(x) = \sum_{i=1}^{N_{p^S}} S_{km}(x_i) \ell^S_i(x), 
\end{align}
for $x\in L^{\text{S}}$ and $x \in L^{\text{F}}$, respectively; similar representations hold for 
$U_{km;h}^{s'}$, $U_{km;h}^{f'}$, $V_{km;h}^{s'}$, $V_{km;h}^{f'}$, $P'_{km;h}$ and $S'_{km;h}$, 
respectively. 
We note that the fluid-solid boundary points coincide with nodes. 

As in Subsection~\ref{sec:mixedFEM} and Section~\ref{sec:selfG}, 
we collect the ``values" of $U_{km;h}^s$, $U_{km;h}^f$, $V_{km;h}^s$, $V_{km;h}^f$, $P_{km;h}$ and $S_{km;h}$ at all the nodes, 
in vectors $\tilde{U}_{km}^s$, $\tilde{U}_{km}^f$, $\tilde{V}_{km}^s$, $\tilde{V}_{km}^f$, $\tilde{V}_{km}$ and $\tilde{S}_{km}$, respectively, 
and collect the values of $U_{km;h}^{s'}$, $U_{km;h}^{f'}$, $V_{km;h}^{s'}$, $V_{km;h}^{f'}$, $P'_{km;h}$ and $S'_{km;h}$ at all the nodes, 
in ``vectors" $\tilde{U}_{km}^{s'}$, $\tilde{U}_{km}^{f'}$, $\tilde{V}_{km}^{s'}$, $\tilde{V}_{km}^{f'}$, $\tilde{P}'_{km}$ and $\tilde{S}'_{km}$, respectively. 
We let 
\begin{align*}
\tilde{u}_{km}^{(r)} &= ((\tilde{U}_{km}^s)\trans, (\tilde{V}_{km}^s)\trans, 
(\tilde{U}_{km}^f)\trans, (\tilde{V}_{km}^f)\trans)\trans,  \\
\tilde{u}^s_{km} &= ((\tilde{U}^s_{km})\trans, (\tilde{V}^s_{km})\trans)\trans, \quad 
\tilde{u}^f_{km} = ((\tilde{U}^f_{km})\trans, (\tilde{V}^f_{km})\trans)\trans, 
\end{align*}
and obtain the resulting eigenvalue problem (cf. \eqref{eq:eigfull})
\begin{equation}\label{eq:1Dform}
   (A_{G}^{(r)} - E_{G}^{(r)} {A_p^{(r)}}^{-1} {E_{G}^{(r)}}\trans
       - {C^{(r)}}\trans (S^{(r)})^{-1} C^{(r)}) \tilde{u}_{km}^{(r)}
                   = \omega_k^2  M^{(r)} \tilde{u}_{km}^{(r)} ,
\end{equation}
where 
\begin{align*}
A_{G}^{(r)} = \left( \begin{array}{cc}
A_{sg}^{(r)}         & 0           \\
0                & A_f^{(r)}                 
\end{array} \right) , \, 
E_G^{(r)} &= \left( \begin{array}{c}
E_{\text{FS}}^{(r)}         \\ 
A_{\text{dg}}^{(r)}                                   
\end{array} \right) , \,
{C^{(r)}}\trans = \left( \begin{array}{c}
{C_s^{(r)}}\trans      \\
{C_f^{(r)}}\trans                      
\end{array} \right)
, \\
M^{(r)} = \left( \begin{array}{cc}
M_s^{(r)}      & 0           \\
0                 & M_f^{(r)}                           
\end{array} \right) , \,
{E_G^{(r)}}\trans &= \left( \begin{array}{cc}
{E_{\text{FS}}^{(r)}}\trans      & {A_{\text{dg}}^{(r)}}\trans                                   
\end{array} \right) , \,
C^{(r)} = \left( \begin{array}{cc}
C_s^{(r)}      & C_f^{(r)}                      
\end{array} \right) , 
\end{align*}
in which
$A_{sg}^{(r)}$, $A_f^{(r)}$, $A_p^{(r)}$, $E_{\text{FS}}^{(r)}$, 
${E_{\text{FS}}^{(r)}}\trans$, $A_{\text{dg}}^{(r)}$, ${A_{\text{dg}}^{(r)}}\trans$, 
$M_s^{(r)} $, $M_f^{(r)} $, 
${C_s^{(r)}}\trans$, ${C_f^{(r)}}\trans$, $S^{(r)}$, $C_s^{(r)}$ and $C_f^{(r)}$, 
are given in Tables~\ref{tab:radialresults} and \ref{tab:radialresults1}. 
We note that the matrices in \eqref{eq:1Dform} are
obtained using separation of variables with spherical harmonics
in \eqref{eq:eigfull}. 
We substitute
\[
\tilde{P}_{km} =- {A_p^{(r)}}^{-1} {E_{G}^{(r)}}\trans \tilde{u}_{km}^{(r)}
\]
upon solving \eqref{eq:weakfluidconstraint} and 
\[
\tilde{S}_{km} = (S^{(r)})^{-1} C^{(r)} \tilde{u}_{km}^{(r)}
\] 
upon solving \eqref{eq:perturbedpotentialQ}. 
We only need to invoke a finite-element basis
in the radial coordinate. We note that the resulting system can be
solved via a standard eigensolver, such as \texttt{LAPACK} \cite{anderson1999lapack}.

\begin{table}[ht!]
	\centering
	\resizebox{\columnwidth}{!}{
	\begin{tabular}{ c  c  r}
		\hline
		operations  & physical meanings   & corresponding formulae \\ \hhline{===}
        \(\displaystyle  (\tilde{U}_{km}^{s'})\trans A_{sg}^{(r)} \tilde{U}_{km}^s \)  &     solid stiffness matrix  &   \cite[(3.1)]{jingchen2018revisiting}\\    \hdashline
         &  & \(\displaystyle \int_{\Omega_{(r)}^{\text{F}}}  U_{km;h}^{f'} U_{km;h}^f N^2_{(r)} \rho^0_{(r)} r^2 \dd r  \) \\  
          \(\displaystyle (\tilde{U}_{km}^{f'})\trans A_{f}^{(r)} \tilde{U}_{km}^{f} \)  &  buoyancy term   & \(\displaystyle + [\rho^0_{(r)}]_-^+g_{(r)} U_{km;h}^{f} U_{km;h}^{f'} r^2|_{\Sigma_{(r)}^{\text{FF}}}  \) \\  \hdashline
          \(\displaystyle (\tilde{P}'_{km})\trans A_p^{(r)} \tilde{P}_{km} \) & fluid potential  &   \(\displaystyle \int_{\Omega_{(r)}^{\text{F}}} P'_{km;h} P_{km;h} \kappa_{(r)}^{-1} r^2 \dd r  \) \\  \hdashline 
        &  &   \(\displaystyle 
		\int_{\Omega_{(r)}^{\text{F}}} U_{km;h}^{f'}  ( \partial_r P_{km;h} + \rho^0_{(r)} g_{(r)} \kappa_{(r)}^{-1}  P_{km;h} ) r^2 \dd r \) \\ 
          \(\displaystyle (\tilde{u}_{km}^{f'})\trans A_{\text{dg}}^{(r)} \tilde{P}_{km} \) & fluid stiffness matrix  &   \(\displaystyle 
		+ \int_{\Omega_{(r)}^{\text{F}}}  \sqrt{l(l+1)} P_{km;h}  V_{km;h}^{f'} r \dd r \) \\ \hdashline 
		&  &   \(\displaystyle 
		\int_{\Omega_{(r)}^{\text{F}}}  \left( \partial_r P'_{km;h} + \rho^0_{(r)} g_{(r)} \kappa_{(r)}^{-1}  P'_{km;h} \right) U^f_{km;h} r^2 \dd r \) \\         
		\(\displaystyle (\tilde{P}'_{km})\trans {A_{\text{dg}}^{(r)}}\trans \tilde{u}_{km}^{f} \) & constraint  &   \(\displaystyle 
		+ \int_{\Omega_{(r)}^{\text{F}}}  \sqrt{l(l+1)} P'_{km;h} V_{km;h}^{f}  r \dd r\) \\     \hdashline 
		\(\displaystyle (\tilde{U}_{km}^{s'})\trans E_{\text{FS}}^{(r)} \tilde{P}_{km} \) & fluid-solid boundary condition & 
		\(\displaystyle - P_{km;h} U_{km;h}^{s'} r^2|_{\Sigma^{\text{FS}}_{(r)}}  \) \\ 
	   \(\displaystyle (\tilde{P}'_{km})\trans {E_{\text{FS}}^{(r)}}\trans \tilde{U}_{km}^{s} \) & fluid-solid boundary condition & 
		\(\displaystyle  - P'_{km;h} U_{km;h}^s r^2|_{\Sigma^{\text{FS}}_{(r)}} \) \\ 
        \(\displaystyle (\tilde{U}_{km}^{s'})\trans M_s^{(r)} \tilde{U}_{km}^{s} \)  & solid mass matrix & \(\displaystyle  \int_{\Omega_{(r)}^{\text{S}}}  \left( U_{km;h}^{s'} U_{km;h}^{s} + V_{km;h}^{s'} V_{km;h}^{s}  \right) \rho^0_{(r)} r^2 \dd r \) \\ 
          \(\displaystyle (\tilde{U}_{km}^{f'})\trans M_f^{(r)} \tilde{U}_{km}^{f} \)  & fluid mass matrix & \(\displaystyle  \int_{\Omega_{(r)}^{\text{F}}}  \left( U_{km;h}^{f'} U_{km;h}^{f} + V_{km;h}^{f'} V_{km;h}^{f}   \right) \rho^0_{(r)} r^2 \dd r \) \\ 
          \hline
          \end{tabular}
          }
   \captionof{table}{Implicit definition of the matrices in \eqref{eq:1Dform} (no summations over $k$ and $m$). Since the construction of $A_{sg}^{(r)}$ is standard, we refer to \cite[(8.43) \& (8.44)]{dahlen1998theoretical} and  \cite[(3.1)]{jingchen2018revisiting}. 
   In the above, $\int_{\Omega_{(r)}^{\text{S}}}=\sum_{q=1}^{N_L^{\text{S}}} \int_{L_q^{\text{S}}}$ and $\int_{\Omega_{(r)}^{\text{F}}}=\sum_{q=1}^{N_L^{\text{F}}} \int_{L_q^{\text{F}}}$.} 
   \label{tab:radialresults}
\end{table}
          
   \begin{table}[ht!]
	\centering
	\resizebox{\columnwidth}{!}{
	\begin{tabular}{ c  c  r}
		\hline  
		operations  & physical meanings   & corresponding formulae \\ \hhline{===}
        &  & \(\displaystyle  \int_{\Omega_{(r)}^{\text{S}}}  (\partial_r S'_{km;h})  U_{km;h}^s \rho^0_{(r)} r^2 \dd r    \) \\
         \(\displaystyle (\tilde{S}'_{km})\trans C_s^{(r)} \tilde{u}_{km}^{s} \)   & density changes in $\overline{\Omega^{\text{S}}_{(r)}}$  & \(\displaystyle  + \int_{\Omega_{(r)}^{\text{S}}} \sqrt{l(l+1)}  S'_{km;h} V_{km;h}^s \rho^0_{(r)}  r \dd r  \) \\ \hdashline 
        &  & \(\displaystyle  \int_{\Omega_{(r)}^{\text{F}}} (\partial_r S'_{km;h}) U_{km;h}^f  \rho^0_{(r)} r^2 \dd r   \) \\ 
          \(\displaystyle (\tilde{S}'_{km})\trans C_f^{(r)} \tilde{u}_{km}^{f} \)   & density changes in $\overline{\Omega^{\text{F}}_{(r)}}$  & \(\displaystyle   + \int_{\Omega_{(r)}^{\text{F}}} \sqrt{l(l+1)}  S'_{km;h} V_{km;h}^f  \rho^0_{(r)}  r \dd r  \) \\ \hdashline 
         & & \(\displaystyle (4 \pi G)^{-1} \int_0^{\infty} ( \partial_r S'_{km;h} \partial_r S_{km;h}   r^2 \) \\ 
         \(\displaystyle (\tilde{S}'_{km})\trans S^{(r)} \tilde{S}_{km} \)   & Poisson's equation & \(\displaystyle  + l(l+1) S'_{km;h} S_{km;h} ) \dd r  \) \\ \hdashline 
         & incremental gravitational field  & \(\displaystyle  \int_{\Omega_{(r)}^{\text{S}}} U_{km;h}^{s'} (\partial_r S_{km;h})  \rho^0_{(r)} r^2 \dd r   \) \\ 
          \(\displaystyle (\tilde{u}_{km}^{s'})\trans {C_s^{(r)}}\trans \tilde{S}_{km} \)   & in $\overline{\Omega^{\text{S}}_{(r)}}$  & \(\displaystyle  + \int_{\Omega_{(r)}^{\text{S}}} \sqrt{l(l+1)} V_{km;h}^{s'} S_{km;h} \rho^0_{(r)}  r \dd r  \) \\  \hdashline 
         & incremental gravitational field & \(\displaystyle  \int_{\Omega_{(r)}^{\text{F}}} U_{km;h}^{f'} (\partial_r S_{km;h})  \rho^0_{(r)} r^2 \dd r    \) \\ 
             \(\displaystyle (\tilde{u}_{km}^{f'})\trans {C_f^{(r)}}\trans \tilde{S}_{km} \)   &  in $\overline{\Omega^{\text{F}}_{(r)}}$  & \(\displaystyle    + \int_{\Omega_{(r)}^{\text{F}}} \sqrt{l(l+1)} V_{km;h}^{f'} S_{km;h} \rho^0_{(r)}  r \dd r  \) \\
       \hline
	 \end{tabular}
	 }
 \captionof{table}{Implicit definition of the matrices in
\eqref{eq:1Dform} (no summation over $k$ and $m$). 
In the above, $\int_{\Omega_{(r)}^{\text{S}}}=\sum_{q=1}^{N_L^{\text{S}}} \int_{L_q^{\text{S}}}$ and $\int_{\Omega_{(r)}^{\text{F}}}=\sum_{q=1}^{N_L^{\text{F}}} \int_{L_q^{\text{F}}}$. 
In the Poisson's equation, the computation of the integral $\int_0^{\infty}$ requires special treatment, 
see \cite[Chapter 3.2.2]{jingchen2018revisiting}. 
}
\label{tab:radialresults1}
\end{table}

As mentioned above, we may consider the finite-element solution
denoted as $\{u_{km;h}\}$ as an alternative basis.  Since
$\{u_{km;h}\}$ is a global basis for the general problem, we have no
separation in the solid and fluid components and no longer have the
fluid-solid boundary terms in the system.  Following the Galerkin
method, we then consider an expansion for the general solution $u_c =
\sum_{km} y_{km} u_{km;h}$ and the corresponding test functions $v_c =
\sum_{k'm'} y'_{k'm'} u_{k'm';h}$.  We introduce $s_c$ and its
corresponding test functions $v^{s_c}$ for self-gravitation.  We have
$s_c = \sum_{km} z_{km} S_{km;h}$ and $v^{s_c} = \sum_{k'm'} z'_{k'm'}
S_{k'm';h}$. Assuming that all the discontinuities in a fully
heterogeneous model coincide with the ones in the reference radial
model and the fluid outer core, the eigenfuncions represented by the
mentioned expansions lie in $H_1 \subset E$ (cf. \eqref{eq:h1h2space})
for the fully heterogeneous problem while the constraint equation
disappears.  We let $y$, $y'$, $z$ and $z'$ be the ``vectors" with
components $y_{km}$, $y'_{k'm'}$, $z_{km}$ and $z'_{k'm'}$,
respectively, and obtain
\begin{equation}\label{eq:gep1D}
   (A_{G}^{(c)} - {C^{(c)}}\trans {S^{(c)}}^{-1}  C^{(c)}) y
                = \omega^2  M^{(c)} y,
\end{equation}
as the counterpart of \eqref{eq:eigfull}. 
Here, $A_{G}^{(c)}$, $M^{(c)}$, ${C^{(c)}}\trans$, 
$S^{(c)}$ and $C^{(c)}$, obtained via substituting the above-mentioned expansion of $u_c$
in \eqref{eq:eigfull}, are given
in Tables~\ref{table:couplingmatrixcomponents} and \ref{table:couplingmatrixcomponents1}. 

If all the discontinuities in a fully heterogeneous model with a fixed fluid outer core
coincide with the reference radial model, we note that the matrix elements in \eqref{eq:gep1D}, 
Tables~\ref{table:couplingmatrixcomponents} and \ref{table:couplingmatrixcomponents1} 
are similar to \cite[(A1)]{woodhouse1980coupling}, which describe mode coupling in non-radial models. 
However, Woodhouse \cite[(A1)]{woodhouse1980coupling} includes additional terms accounting for 
changes in the fluid-solid boundaries while in the previous work \cite[(42)]{woodhouse1978effect}, 
perturbation theory is used to compute the eigenfrequency changes in terms of the unperturbed eigenfunctions; 
both calculations violate the condition that normal modes need to remain in $E$ and in $H_1$.

\begin{table}[ht!]
	\centering
	\resizebox{\columnwidth}{!}{
	\begin{tabular}{ c  c  r}
		\hline
		operations  & physical meanings   & corresponding formulae \\ \hhline{===}
		&    & \(\displaystyle \sum_{km}  \sum_{k'm'} y'_{k'm'} \bigg\{  \int_{\Omega^{\text{S}}}  \nabla u_{k'm';h} : \left(c : \nabla u_{km;h} \right) \dd x \) \\
		&  & \(\displaystyle + \int_{\Sigma^{\text{FS}}} \mathfrak{S}\left\{\left(g \cdot u_{k'm';h} \right)   \left(\nu^{s \rightarrow f} \cdot u_{km;h} \right)  [\rho^0]^f \right\} \dd \Sigma \) \\
		  &  
		& \(\displaystyle  + \int_{\Omega^{\text{S}}} \mathfrak{S} \bigg\{  \left( \nabla \cdot u_{k'm';h} \right) \left(\rho^0 u_{km;h} \cdot g \right)  - \rho^0 u_{k'm';h}  \cdot (\nabla g) \cdot u_{km;h}  \) \\
	&  & \(\displaystyle  - \rho^0 u_{km;h} \cdot \left(\nabla u_{k'm';h} \right)  \cdot g \bigg\} \dd x  \) \\ 
		  &  & \(\displaystyle  + \int_{\Omega^{\text{F}}} \rho^0 N^2 \frac{  \left(g \cdot u_{k'm';h} \right)\left(g \cdot u_{km;h} \right) }{\| g \|^2} \dd x \) \\   
		  &  & \( \displaystyle + \int_{\Sigma^{\text{FF}}} (g\cdot \nu) (u_{km;h} \cdot \nu) (u_{k'm';h} \cdot \nu) [\rho^0]^+_- \dd \Sigma   \) \\
		&  & \(\displaystyle + \int_{\Omega^{\text{F}}}  \kappa  \left( \nabla u_{k'm';h} + \rho^0 \kappa^{-1} u_{k'm';h} \cdot g \right)  \) \\  
		    \(\displaystyle (y')\trans A_G^{(c)} y \)  & stiffness matrix   & \(\displaystyle  \left( \nabla u_{km;h} + \rho^0 \kappa^{-1} u_{km;h} \cdot g \right) \dd x  \bigg\} y_{km} \) \\    \hdashline
		&  & \(\displaystyle \sum_{km}  \sum_{k'm'} y'_{k'm'}  \bigg\{  \int_{\Omega^{\text{S}}}  u_{k'm';h}  \cdot u_{km;h} \rho^0  \dd x \) \\ 
		\(\displaystyle (y')\trans M^{(c)} y \)  &  mass matrix & \(\displaystyle  +  \int_{\Omega^{\text{F}}}  u_{k'm';h}  \cdot u_{km;h} \rho^0 \dd x \bigg\}  y_{km}\) \\  
		\hline
		\end{tabular}
		}
	\captionof{table}{Implicit definition of the matrices in \eqref{eq:gep1D} for the Cowling approximation. }	
\label{table:couplingmatrixcomponents}
		\end{table}

  \begin{table}[ht!]
	\renewcommand{\arraystretch}{1.0}
	\setlength{\tabcolsep}{1pt}
	\centering
	\resizebox{\columnwidth}{!}{%
	\begin{tabular}{ c  c  r}
		\hline
		operations  & physical meanings   & corresponding formulae \\ \hhline{===}
		  &  & \(\displaystyle \sum_{km}  \sum_{k'm'} z'_{k'm'}  \bigg\{ \int_{\Omega^{\text{S}}} s_{k'm';h} \nabla \cdot \left(\rho^0 u_{km;h} \right) \dd x \)  \\
		&  & \(\displaystyle  +\int_{ \Sigma^{\text{SS} \cup \partial \tilde{X}^{\text{S}}} }  s_{k'm';h} \nu \cdot u_{km;h} \left[\rho^0\right]^+_- \dd \Sigma     \) \\
		  & & \(\displaystyle    + \int_{\Sigma^{\text{FS}}}  s_{k'm';h}  \nu^{f\rightarrow s} \cdot u_{km;h} \left[\rho^0\right]^s \dd \Sigma   \) \\
		    &  & \(\displaystyle  + \int_{ \Sigma^{\text{FF} \cup \partial \tilde{X}^{\text{F}}} } s_{k'm';h}   \nu \cdot u_{km;h}  \left[\rho^0\right]^+_- \dd \Sigma  \)  \\
		 & & \(\displaystyle +  
		  \int_{\Omega^{\text{F}}}  s_{k'm';h}  \nabla \cdot (\rho^0 u_{km;h} )\dd x \) \\ 
		   \(\displaystyle (z')\trans C^{(c)} y\)   & density changes in $\tilde{X}$  & \(\displaystyle + \int_{\Sigma^{\text{FS}}} s_{k'm';h}  \nu^{s\rightarrow f} \cdot u_{km;h} \left[\rho^0\right]^f \dd \Sigma   \bigg\} y_{km} \) \\ 
		  \hdashline 
		   \(\displaystyle (z')\trans S^{(c)} z \)   & Poisson's equation & \(\displaystyle  \sum_{km}  \sum_{k'm'} z'_{k'm'}  \bigg\{ \int_{\R^3} (\nabla s_{k'm';h})  \cdot (\nabla s_{km;h} )  \dd x \bigg\} z_{km}\) \\ \hdashline 
		  & & \(\displaystyle \sum_{km} \sum_{k'm'} y'_{k'm'}  \bigg\{ \int_{\Omega^{\text{S}}}  \nabla \cdot (\rho^0 u_{k'm';h} )  s_{km;h}  \dd x    \) \\
		 &  & \(\displaystyle + \int_{ \Sigma^{\text{SS}}  \cup \partial \tilde{X}^{\text{S}} } [\rho^0]^+_-  \nu \cdot u_{k'm';h} s_{km;h} \dd \Sigma \)  \\
		 &   & \(\displaystyle +   \int_{\Sigma^{\text{FS}}} [\rho^0]^s \nu^{f\rightarrow s} \cdot u_{k'm';h}  s_{km;h}  \dd \Sigma \)  \\
		   & & \(\displaystyle + \int_{ \Sigma^{\text{FF}}  \cup \partial \tilde{X}^{\text{F}}  } [\rho^0]^+_-  \nu \cdot u_{k'm';h} s_{km;h}  \dd \Sigma  \) \\
		 & incremental gravitational field  &  \(\displaystyle +  \int_{\Omega^{\text{F}}} \nabla \cdot (\rho^0 u_{k'm';h} ) s_{km;h}  \dd x \)  \\
		    \(\displaystyle (y')\trans {C^{(c)}}\trans z \)  & in $\tilde{X}$  &  \(\displaystyle  +  \int_{\Sigma^{\text{FS}}} [\rho^0]^f \nu^{s\rightarrow f} \cdot u_{k'm';h} s_{km;h}  \dd \Sigma  \bigg\} z_{km}  \)  \\
		  \hline
	\end{tabular}
	}
\captionof{table}{Implicit definition of the matrices in \eqref{eq:gep1D}. 
} \label{table:couplingmatrixcomponents1}
\end{table}

%
%

\bibliographystyle{spmpsci}      
\bibliography{JS_NM_JSC}   

\end{document}